%% file: main.tex
\documentclass[11pt]{article}
\usepackage{amsmath}

\usepackage[numbers, sort&compress ]{natbib}
\usepackage[margin=1in]{geometry}

\usepackage[T1]{fontenc}
\usepackage[latin9]{inputenc}
\usepackage{xcolor}
\usepackage{mathtools}
\usepackage{amstext}
\usepackage{amsthm}
\usepackage{amssymb}
\usepackage{amsfonts,bm}
\usepackage{enumitem}
\usepackage{wasysym}
\usepackage{stmaryrd}
\usepackage{bbm}
\usepackage{bm}
\usepackage{thmtools}

\usepackage{tikz}
\usetikzlibrary{patterns.meta}

\usepackage[ruled,vlined]{algorithm2e} \PassOptionsToPackage{algo2e,ruled}{algorithm2e}
\numberwithin{equation}{section}

\newtheorem{theorem}{Theorem}
\newtheorem{lemma}{Lemma}
  
  \newtheorem{proposition}{Proposition}
  \newtheorem{remark}{Remark}
  \newtheorem{corollary}{Corollary}
  \newtheorem{definition}{Definition}

\newtheorem{assumption}[theorem]{Assumption}

\newcounter{gamecounter}
\newcounter{savedalgocf}

\newenvironment{game}[1][htbp]{%
  \setcounter{savedalgocf}{\value{algocf}}%
  \setcounter{algocf}{\value{gamecounter}}%
  \begin{algorithm}[#1]%
}{%
  \end{algorithm}%
  \setcounter{gamecounter}{\value{algocf}}%
  \setcounter{algocf}{\value{savedalgocf}}%
}

\usepackage{subcaption}
\usepackage[colorlinks=true]{hyperref}
\usepackage[capitalize]{cleveref}\hypersetup{colorlinks=true,linkcolor=blue,citecolor=blue}
\usepackage{etoolbox}
\makeatletter
\patchcmd{\algocf@caption@algo}
  {}
  {}
  {}{\PackageError{game-links}{Could not patch algorithm2e hyperlink names}{Check the algorithm2e caption definition.}}
\makeatother
\crefname{lemma}{Lemma}{Lemmas}
\Crefname{lemma}{Lemma}{Lemmas}
\crefname{assumption}{Assumption}{Assumptions}
\Crefname{assumption}{Assumption}{Assumptions}

\makeatother

\usepackage{babel}
\newcommand{\todo}[1]{{\color{red}TODO: #1}}

\newcommand{\jnote}[1]{{\color{purple}Jialin: #1}}
\newcommand{\comment}[1]{}

\title{Exponential memory/query-complexity tradeoffs in high-accuracy feasibility problems}
\author{%
  \textbf{Jialin Yu}\thanks{H. Milton Stewart School of Industrial and Systems Engineering, Georgia Institute of Technology. Emails: \texttt{\{yujl1,mblanchard41\}@gatech.edu}.}
  \and
  \textbf{Mo\"ise Blanchard}\footnotemark[1]
}

\input{shortcuts}
\usepackage{tikz,graphicx}
\usetikzlibrary{arrows.meta,calc,intersections}
\begin{document}
\maketitle
\begin{abstract}
    We prove memory/query-complexity tradeoff lower bounds for the \emph{feasibility problem}, where the goal is to find a point within an $\epsilon$-ball hidden in the unit $d$-dimensional ball, with access to a separation oracle.
    This problem admits three regimes: the low-accuracy regime $\epsilon\geq 1/\sqrt d$ where memory does not play a role, the medium-accuracy regime $1/\sqrt d\geq \epsilon\geq e^{-\Omega(d\ln d)}$ where prior works identified strong memory/query-complexity separations, and the \emph{high-accuracy regime} $\epsilon\leq e^{-\Omega(d\ln d)}$ for which much less is known. In this regime, low-memory algorithms with significantly faster convergence rates than gradient descent have been developed in \cite{blanchard2024memory}: for accuracy $\epsilon$ and any $\delta\in[0,1]$, these require $d^{1+\delta}\ln \frac1\epsilon$ memory bits and use $\left(d\ln\frac1{\epsilon}\right)^{O(d^{1-\delta})}$ oracle queries. In comparison, gradient descent uses the same memory for $\delta=0$ but requires $1/\epsilon^2$ oracle queries. On the lower-bound side, no non-trivial memory/query-complexity tradeoff was known: the best lower bound remained the historical memory-independent bound $\Omega(d\ln \frac1\epsilon)$ from \cite{Nemirovsky1983}.
    We make progress towards characterizing the memory landscape in this high-accuracy regime by showing that for any $\delta\in[0,1]$, any algorithm with $d^{1+\delta}\ln\frac1{\epsilon}$ bits of memory uses at least $\left(d\ln\frac1{\epsilon}\right)^{\widetilde{\Omega} (d^{(1-\delta)/8})}$ oracle calls. This result justifies the seemingly peculiar exponential-in-$d$ dependence of oracle complexity for the aforementioned low-memory algorithms. In particular, our result shows that for high-accuracy feasibility problems, insufficient memory does result in an exponential price in the dimension for the query-complexity.
\end{abstract}

\thispagestyle{empty}
\newpage
\setcounter{page}{1}

\section{Introduction}
We consider the \emph{feasibility problem} where a learner aims to find a small convex set $Q$ hidden somewhere in the unit Euclidean ball in $\Rbb^d$, with access to a separation oracle that hints at the location of $Q$ using separating hyperplanes. For the feasibility problem with accuracy $\epsilon$, we assume $Q$ contains a small Euclidean ball of radius $\epsilon$. As a fundamental problem in optimization, the feasibility problem is a standard generalization of non-smooth convex optimization \cite{Nemirovsky1983,nesterov2003introductory}, in the sense that each convex gradient provides a separating hyperplane from the set of minimizers. Both setups have served as key components for various other problems in optimization \cite{lee2015faster}, machine learning \cite{joachims2009cutting} and physics \cite{Yu2021QuantumMarginal}.

From an information-theoretic perspective, the performance of an algorithm for the feasibility problem can be measured by its oracle/query complexity, which corresponds to the number of times it queries the separation oracle for information before finding a solution. The feasibility problem has been extensively studied under this model. For different choices of accuracy $\epsilon$, the corresponding optimal oracle complexity is known to be attained by different algorithms. In particular, in the \emph{low-accuracy regime} where $\epsilon> 1/\sqrt{d}$, gradient descent achieves the optimal oracle complexity.
Going beyond the low-accuracy regime and with $\epsilon\leq 1/\sqrt{d}$, it is known that $\Theta(d\ln 1/\epsilon)$ queries are necessary for success \cite{Nemirovsky1983}, which can be attained by the broad family of cutting plane methods that stems from the celebrated ellipsoid method \cite{khachiyan1980polynomial,yudin76evaluation}, including the inscribed ellipsoid method \cite{nesterov1989self,tarasov1988method}, Vaidya's method \cite{anstreicher1997vaidya,atkinson1995cutting,vaidya1996new}, and random-walk-based methods \cite{bertsimas2004solving,levin1965algorithm}. Importantly, for those methods, the dependency in accuracy is $\Theta(\ln 1/\epsilon)$, which substantially improves on the $\Theta(1/\epsilon^2)$ rate of gradient descent on problems with higher accuracies.

Despite having suboptimal oracle complexity for higher accuracy problems, gradient descent remains the dominant method in practice, even in the $\epsilon\leq 1/\sqrt{d}$ regime. Cutting planes, on the other hand, are often considered impractical due to their excessive usage of computational resources for each oracle query. Notably, they all require at least $\Omega(d^3\ln 1/\epsilon)$ total runtime even with recent improvements \cite{jiang2020improved,lee2015faster}. Another important reason is that these methods typically involve storing a large number of previous cutting planes, or at least require the storage and manipulation of a summary matrix that uses $\Omega(d^2\ln1/\epsilon)$ bits of memory. In contrast, gradient descent only needs to keep track of a few vectors, which corresponds to $O(d\ln1/\epsilon)$ memory usage. This observation, along with other practical concerns, motivates the study of tradeoffs between oracle complexity and other computational resources such as memory \cite{blanchard2023quadratic,blanchard2024memory,chen2023memory,marsden22efficient,woodworth2019open} or communication 
\cite{lan_communication-efficient_2020,Mota2013,Reddi2016AIDEFA,shamir14,Smith2017,wang17aMemo,Wangni2018,Zhang2012Communication}.

\vspace{4pt}
\noindent\textbf{Memory/query-complexity tradeoffs for medium accuracies.} The question of understanding the tradeoffs between memory usage and query complexity in the $\epsilon\leq 1/\sqrt{d}$ regime was posed as a COLT 2019 open problem in \cite{woodworth2019open} under the convex optimization setup. 
In the \emph{medium accuracy regime} where $e^{-\Omega(d\ln d)}< \epsilon\leq 1/\sqrt{d}$, gradient descent and cutting planes are the only known algorithms for the memory/query complexity tradeoff in both convex optimization and feasibility problems.
Though memory-efficient algorithms have also been studied under other optimization settings \cite{liu_limited_1989,Nocedal1980,fletcher1964function,hager2006survey,pilanci2017newton,roosta2019sub,altschuler2023acceleration,das2023branch}, they do not improve over the two aforementioned algorithms in terms of the memory/query-complexity landscape. 

In the absence of better algorithms, there has been substantial interest on the lower-bound side, aiming to understand impossibility results within this landscape. Starting from the breakthrough of \cite{marsden22efficient} for convex optimization, which demonstrates the existence of such a tradeoff by showing that any algorithm either uses $d^{1+\delta}$ bits of memory or makes $\widetilde{\Omega}(d^{4(1-\delta)/3})$ queries, subsequent works improved this polynomial tradeoff for deterministic \cite{blanchard2023quadratic} and randomized \cite{chen2023memory} algorithms, and the recent work by \cite{menart2026stronger} even shows a sharp polynomial phase transition when the memory differs from quadratic by only a logarithmic factor. For the harder feasibility problem, exponential memory/query-complexity separations were proved in \cite{blanchard24gradient}, which shows that any algorithm with $O(d\ln 1/\epsilon)$ memory needs $\Omega(1/\epsilon^2)$ queries to succeed, matching the performance of gradient-descent methods. In summary, for the feasibility problem, no algorithm can Pareto-improve over either gradient descent or cutting planes in this medium accuracy regime.

\vspace{4pt} \noindent\textbf{The high-accuracy regime.} While these prior works on medium accuracies proved impossibility results, somewhat surprisingly, gradient descent can be improved upon for the \emph{high-accuracy regime} where $\epsilon\leq e^{-\Omega(d\ln d)}$ \cite{blanchard2024memory}. They provide a family of algorithms based on cutting-planes which use $O(d^{1+\delta}\log 1/\epsilon)$ bits of memory for $O(d \ln 1/\epsilon)^{d^{1-\delta}}$ oracle queries, for any $\delta\in[0,1]$. 
In particular, for the same memory as gradient descent, these algorithms only use $(d\ln1/\epsilon)^{O(d)} \ll 1/\epsilon^2$ queries.\footnote{As a side-note, these algorithms require an additional \emph{time-consistency assumption}: the separation oracle must output the same answer if queried on the same vector on different rounds.}

This suggests a qualitatively different memory/query-complexity landscape in this new accuracy regime: while low-memory algorithms such as gradient descent required a polynomial convergence rate, beyond the high-accuracy threshold, the convergence becomes exponential with some peculiar exponential dependency in the dimension $d$. Importantly, it remains unknown whether this exponential dependency in $d$ is necessary or an artifact of their algorithms. This leads to the following question:
\begin{center}
    \textit{What is the qualitative picture of the memory/query-complexity tradeoff for high-accuracy regimes?}
\end{center}
In particular, is linear convergence $O_d(\ln 1/\epsilon)$ achievable, or is the exponential dependence on $d$ observed in prior works necessary for low-memory algorithms? As a reminder, the only lower bound known for this regime remains \cite{Nemirovsky1983} which states that $\Omega(d\ln 1/\epsilon)$ queries are necessary irrespective of memory constraints.

\comment{
All such works, however, are only significant in the medium-accuracy regime. Specifically, the aforementioned works on convex optimization only involves showing at most a polynomial dependency in $d$, which does not capture the $(d\ln 1/\epsilon)^d$ rate in the $\epsilon\leq e^{-\Omega(d\ln d)}$ regime of high accuracy, and the simple extension of \cite{blanchard24gradient} to the same regime does not provide any beyond-linear dependency in $\ln(1/\epsilon)$ either.

As we can see, for both problems the memory/query complexity tradeoffs largely remain unknown when $\epsilon\leq e^{-\Omega(d\ln d)}$. In particular, we are uncertain about the right dependency in $\epsilon$, which probably dominates the rate in this regime of very small accuracy. Under the linear $O(d\ln1/\epsilon)$ memory, the best known lower bound dependency in $\epsilon$ remains to be $\ln(1/\epsilon)$, which creates a substantial gap from the $O((d\ln 1/\epsilon)^d)$ algorithm in \cite{blanchard2024memory}. This leads to the following question:

\emph{What is the right dependency in $\epsilon$ for memory/query complexity tradeoffs in the $\epsilon\leq e^{-\Omega(d\ln d)}$ regime?}
}

\vspace{4pt} \noindent\textbf{Our contribution.}
In this paper, we provide the first memory/query complexity tradeoff lower bound for high-accuracy feasibility problems. Essentially, we show that the exponential dependency in some polynomial in $d$ is \emph{unavoidable} whenever an algorithm faces non-trivial memory constraints. 

\begin{theorem}\label{thm:main_lower_bound_result}
Let $d$ be sufficiently large and $0<\epsilon\leq e^{-d}$. Then, for any $\delta\in[0,1]$, any algorithm that solves feasibility problems up to accuracy $\epsilon$ with probability $\frac23$ either uses $M=d^{1+\delta}\log\frac1\epsilon$ bits of memory or makes at least $\left(d\ln\frac1\epsilon\right)^{\widetilde{\Omega}(d^{(1-\delta)/8})}$ 
queries, where $\tilde\Omega$ only hides $\polylog \,d$ factors. Here, the considered feasibility oracles may be \emph{adaptive} in the learner's queries. For deterministic algorithms, the oracle can be non-adaptive. 
\end{theorem}
In particular, any low-memory algorithm with linear memory $O(d\ln 1/\epsilon)$ must make at least $(d\ln(1/\epsilon))^{\widetilde{\Omega}(d^{1/8})}$ oracle queries on some feasibility problem with accuracy $\epsilon$. Further, the polynomial exponent within our lower-bound tradeoff decays as $d^{(1-\delta)/8}$ as the algorithm's memory $d^{1+\delta}\log 1/\epsilon$ grows, which essentially matches the performance of the recursive-cutting planes algorithms in \cite{blanchard2024memory} up to a factor $1/8$ in the double-exponent. To the best of our knowledge, this is the second example of exponential memory/query-complexity tradeoffs in continuous optimization besides \cite{blanchard24gradient} which focuses on medium-accuracy regimes.

This completes a qualitative picture of the memory/query-complexity tradeoff for the feasibility problem: (1) in low-accuracy regimes, no tradeoff exists; (2) in medium-accuracy regimes, there is a strong separation between low-memory algorithms which converge polynomially $O(1/\epsilon^2)$ and high-memory algorithms which converge exponentially $O(d\ln 1/\epsilon)$; and (3) in high-accuracy regimes, there is still an exponential separation, but it is milder: low-memory algorithms only pay an exponent $d$ in query-complexity, independently of $\epsilon$.

For other related works and potential applications, we refer the reader to Appendix~\ref{app:additional-related-work}.

\paragraph{Problem setup.} We start by formally defining the \emph{feasibility problem} over the unit ball $B_d(0,1)$ for a known accuracy $\epsilon>0$. In this problem, the goal is to find a point in an unknown convex set $Q\subseteq B_d(0,1)$---referred to as the \emph{feasible set}---that contains a Euclidean ball of radius
$\epsilon$, having access to $Q$ through a \emph{separation oracle}. Specifically, given a query $x_t\in B_d(0,1)$, the oracle returns
$\texttt{Success}$ if $x_t\in Q$, and otherwise returns a separating hyperplane with its normal 
$g_t\in\Rbb^d$ (sometimes called a gradient) satisfying
$g_t^\top(y-x_t)<0$ for all $y\in Q$.
The oracle may be time-dependent, randomized and \emph{adaptive}: its response
$g_t$ may depend on the current query and the previous history,\footnote{Note that if the algorithm is deterministic, non-adaptive (oblivious) oracles are as powerful as adaptive oracles.} as long as it remains a valid separating hyperplane for the feasible set $Q$.
\comment{
As a remark, we assume here that the feasible set $Q$ contains a ball of known radius $\epsilon>0$. Since we focus on lower bounds, our results will also apply to more general settings in which the algorithm must either find a feasible point $x\in Q$ or certify that $Q$ does not contain an $\epsilon$-ball \cite{lee2015faster,jiang2020improved}.
}

In this work, we study the performance of \emph{memory-constrained} algorithms for the feasibility problem, using the same memory model from \cite{marsden22efficient}. An $M$-bit memory-constrained algorithm maintains an $M$-bit register $\texttt{Memory}_t\in\{0,1\}^M$ which represents its memory state at the beginning of each round $t\geq 1$. Formally, it is specified by some query map $\psi_{\mathrm{query}}:\{0,1\}^M\to B_d(0,1)$ and memory update function $\psi_{\mathrm{update}}:\{0,1\}^M\times B_d(0,1)\times (\Rbb^d\cup\{\texttt{Success}\})\to\{0,1\}^M$. These may be randomized, in which case the algorithm may use i.i.d.\ fresh randomness on each round.

These functions specify the behavior of the algorithm: its memory state is initialized at $\texttt{Memory}_1=0$, and on each round it queries
$x_t=\psi_{\mathrm{query}}(\texttt{Memory}_t)$,
receives $g_t\in \Rbb^d\cup\{\texttt{Success}\}$, and then updates its memory state via $\texttt{Memory}_{t+1}
    =\psi_{\mathrm{update}}
      (\texttt{Memory}_t,x_t,g_t)$. 
      This is a fairly mild memory-constrained model: we restrict the algorithm's memory \emph{between} rounds; however, the algorithm may use infinite memory and computation time within rounds to perform its query and memory update. 

      \comment{
\begin{remark}
    While the query and update functions do not depend explicitly on the round $t$, this can be captured by storing the round number $t$ within the memory state $\texttt{Memory}_t$, using only $\ceil{\log t}$ additional bits.
\end{remark}
}
\vspace{4pt}
\noindent\textbf{Organization.} The remainder of the main paper is organized as follows. Section \ref{sec:technical_overview} provides a technical overview of our construction and the proof ideas. Section \ref{sec:proof_main_result} presents the hard instance construction and the proof strategy for our main result, including the blocking property and its propagation across layers. Section \ref{sec:missing_proofs} provides the detailed proofs of the main technical results. 

\section{Technical overview}
\label{sec:technical_overview}

We start by giving the main ideas about the construction of our hard class of feasibility problems. For concreteness, we suppose that the learner has memory $M = d^{1+\delta}\ln(1/\epsilon)$. To achieve exponential lower bounds, we use a multi-layer construction akin to \cite{blanchard24gradient}. Specifically, the construction will include $P\in[d]$ layers and for each layer we sample independently a Haar uniform $\tilde d$-dimensional subspace $E_p\sim\mathrm{Unif}(\mathrm{Gr}(\tilde d,d))$, where $\tilde d\leq \frac{d}{2P}$. Let $\bm e:=(1,0,...,0)\in\Rbb^d$. The target feasible region is then defined via
\begin{equation*}
    Q_{E_1,...,E_P}:=B_d(0,1)\cap\set{x:\bm e^\top x\leq -\frac{1}{2}}\cap\bigcap_{p\in[P]}\set{x:\|P_{E_p}(x)\|\leq\epsilon},
\end{equation*} 
where $\epsilon>0$ is the fixed accuracy of the problem. Hence, the learner aims to query vectors approximately perpendicular to all subspaces $E_1,...,E_P$. The first constraint $B_d(0,1)$ simply restricts the search space to the unit ball, and the second constraint $\bm e^\top x\leq -1/2$ ensures that small-norm vectors---e.g., $x=0$---are not trivially feasible. We can easily check that with high probability, the region $Q_{E_1,...,E_P}$ indeed contains a ball of radius $\epsilon$: it suffices to center it at some norm-$3/4$ vector in $\Span(E_p,p\in[P])^\perp$.

It remains to define the separation oracle for this feasible region. To motivate our final construction, we introduce the main elements separately in the following paragraphs, in the specific $P=2$-layer case. This will already contain most of the intuitions for our result. We remark that in the high-accuracy regime, the source of hardness comes from multiple accuracy levels for $\epsilon$, which requires new analysis compared to prior works on medium-accuracy regimes. In particular, we introduce a special form of \emph{anisotropic noise} which could be of interest for other high-accuracy regimes. 

\vspace{4pt} \noindent\textbf{Multi-phase subspace accuracies.} For now, let $P=1$ and focus only on the last constraint $\{x:\|P_{E_1}(x)\|\leq \epsilon\}$. Our goal involves obtaining query-complexity lower bounds that grow as $\Omega(\log(1/\epsilon))^{O(d)}$ against any algorithm using $O(d\log(1/\epsilon))$ memory. For the milder goal of achieving $\Omega(\log(1/\epsilon))$ query-complexity lower bounds, we can use a relatively standard \emph{blocking oracle}: Instead of revealing exactly $E_1$ to the learner, we may iteratively reveal a sequence of $O(\log(1/\epsilon))$ versions of this subspace corresponding to increasing accuracies, and reveal these only as necessary depending on the learner's progress.

For concreteness, we construct a sequence of increasing-accuracy versions $(E_1^{(i)})_{i\in[N]}$ from the initial subspace $E_1$, by adding increasing-level noise:
\begin{equation}\label{eq:rough_noiseaddition}
    E_1^{(N+1)} := E_1\quad \text{and}\quad E_1^{(i)}\sim \mathrm{Unif}(B(E_1^{(i+1)},\delta^{(i)})),\quad i=N,N-1,\ldots,1,
\end{equation}
for $\delta^{(i)}=\poly(d)^{-i}$.
Initially, the blocking oracle may reveal the lowest-precision subspace $E_1^{(1)}$ to the learner, until the learner queries some vector approximately perpendicular to $E_1^{(1)}$ with the same order of precision: $\|P_{E_1^{(1)}}(x)\| \lesssim \delta^{(1)}$. With such a query, $E_1^{(1)}$ no longer provides a legitimate separation from $\set{x:\|P_{E_1}(x)\|\leq \epsilon}$, so the oracle reveals the next-accuracy subspace $E_1^{(2)}$, until a query reaches the corresponding precision $\|P_{E_1^{(2)}}(x)\| \lesssim \delta^{(2)}$, etc. In words, the oracle returns separation hyperplanes of the same precision-order as the learner's query, revealing $E_1$ only at the necessary accuracy level. We refer to the (layer-$1$) phase $i$ as the interval of time when the learner has observed $E_1^{(i)}$ but not yet $E_1^{(i+1)}$. 
As a remark, it is possible that the learner \emph{skips} phase $i$ if they make a query $x$ that forces the blocking oracle to directly reveal $E_1^{(j)}$ for $j>i$ without revealing $E_1^{(i)}$ on prior queries.

Usual arguments then show that no phase $i\in[N]$ can be skipped with significant probability by the learner, leading to $\Omega(N) = \Omega(\log(1/\epsilon)/\log d)$ queries.

\vspace{4pt} \noindent\textbf{Naive 2-layer construction.} To achieve an $\Omega((\log(1/\epsilon))^2)$ oracle-complexity lower bound against linear-memory algorithms, we aim to repeat the argument for the second subspace $E_2$. 
Precisely, during each layer-$1$ phase $i_1$, we let the learner interact with an increasing-accuracy sequence $(E_2^{(i_2)})_{i_2\in[N]}$ for $E_2$ and move on to the next layer-$1$ phase once the learner reaches the full-accuracy subspace $E_2^{(N)}$. This gives the following \emph{nested} oracle given a query $x$:
\begin{enumerate}
    \item If $\bm e^\top x > -1/2$, answer trivially with $\bm e$.
    \item\label{item:naive_construction-2} Otherwise, during layer-$1$ phase $i_1$, check if $\|P_{E_1^{(i_1)}}(x)\| \gg \delta^{(i_1)}$. If so, answer with $P_{E_1^{(i_1)}}(x)$.
    \item\label{item:naive_construction-3} Otherwise, answer with the blocking-oracle for $(E_2^{(i_2)})_{i_2\in[N]}$. If the query has accuracy corresponding to the highest accuracy: $\|P_{E_2^{(N)}}(x)\|<\delta^{(N)}$, the oracle moves to the next layer-$1$ phase $i_1\gets i_1+1$.
\end{enumerate}

Ideally, on each layer-$1$ phase $i_1$, the blocking oracle for $E_2$ forces the learner to make $\Omega(N)$ queries, as described in the previous paragraph. If this holds for all phases $i_1\in[N]$, the desired query lower bound $\Omega(N^2) = \Omega((\log(1/\epsilon)/\log d)^2)$ follows. Similarly, using $P\geq 2$ layers, one can hope to prove a query lower bound of the order $\Omega(N^P)$ with this nested oracle, as desired in \cref{thm:main_lower_bound_result}.

\vspace{4pt} \noindent\textbf{Failure of the naive construction.} 
Unfortunately, the above construction cannot enforce an $\Omega(N^2)$ query lower bound. Specifically, there is a strategy for the learner which stores only one vector and manages to skip most of the second-layer phases. The above construction would only enforce $\Omega(N\log N)$ query complexity; using more layers would at best prove query lower-bounds of the form $\Omega(N(\log N)^d)$. 
\comment{
The failure case in the previous $2$-layer construction is illustrated in \cref{fig:naive-construction-geometry}: consider a layer-$1$ phase $i_1$ and suppose that the learner has already stored some vector $x^\star$ perpendicular to $E_2$ and the previous-accuracy version $E_1^{(i_1-1)}$ of $E_1$---both are observed by the player in the previous layer-$1$ phase. 
}

For intuition, this strategy utilizes the fact that the uncertainty of a layer-2 phase $E_2^{(i)}$ is largely reduced by a restriction to the set $\{x:\|P_{E_1^{(i_1)}}(x)\|\lesssim \delta^{(i_1)}\}$ as in line~\ref{item:naive_construction-2} of the construction. 
Indeed, during the layer-$1$ phase $i_1$, the player observes $E_1^{(i_1)}$ and the highest-accuracy version of $E_2$. The player may therefore choose to store some vector perpendicular to both, denoted as $x^\star\in E_2^\perp\cap (E_1^{(i_1)})^\perp$. In the next layer-$1$ phase, $i_1+1$, by construction of the oracle above, the learner is essentially tasked with querying a point approximately in $E_2^\perp \cap (E_1^{(i_1+1)})^\perp$. Since $E_1^{(i_1)}$ is an $O(\delta^{(i_1)})$-accuracy version of $E_1^{(i_1+1)}$, the stored vector $x^\star$ is already $\delta^{(i_1)}$-close to the target region $E_2^\perp \cap (E_1^{(i_1+1)})^\perp$.
\comment{
Indeed, at layer-1 phase $i_1$, one can only observe a layer-2 phase with a query $x$ satisfying $\|P_{E_1^{(i_1)}}(x)\|\lesssim \delta^{(i_1)}$, and this restriction actually helps the learner to accelerate in the next phase $i_1+1$, where the learner essentially wants to query a point $x\in E_2^\perp \cap (E_1^{(i_1+1)})^\perp$. At the end of phase $i_1$, the learner has observed the highest-accuracy version of $E_2$, and chooses to store a vector $x^*\in \set{\|P_{E_1^{(i_1)}}(x)\|\lesssim \delta^{(i_1)}}$ that is extremely close to being perpendicular to $E_2$, then essentially, throughout the next layer-1 phase, we know that $E_2^\perp$ passes through this known point $x^*$. 
And given that both $x^*$ and $(E_1^{(i_1+1)})^\perp$ belong to $\set{\|P_{E_1^{(i_1)}}(x)\|\lesssim \delta^{(i_1)}}$, we have the intersection $E_2^\perp \cap (E_1^{(i_1+1)})^\perp$ being $\delta^{(i_1)}$-close to $x^*$, provided that $E_2$ and $E_1^{(i_1+1)}$ are in general positions. }
In turn, the uncertainty in $E_2^\perp \cap (E_1^{(i_1+1)})^\perp$ given the current phase $E_2^{(i_2)}$ that we have observed will be multiplied by a $\delta^{(i_1)}$ factor, which means that the learner can skip $\log(1/\delta^{(i_1)})=\Omega (i_1)$ layer-2 phases at a time. See Appendix~\ref{app:transform_necessity} for a pictorial explanation.

\vspace{4pt} \noindent\textbf{An adjusted construction with transformations.} The failure of this naive oracle construction, however, motivates a stronger construction that eventually leads to \cref{thm:main_lower_bound_result}, which we now describe. To ensure feasibility in the above construction, we required
\begin{equation}\label{eq:transform_necessity-crux}
    \set{x\in B_d(0,1): \|P_{E_2}(x)\|\leq\epsilon}\subseteq \{x:\|P_{E_2^{(i_2)}}(x)\|\leq \delta^{(i_2)}\}.
\end{equation}
Intuitively, this forces $(E_2^{(i_2)})^\perp$ to be close to $E_2^\perp$ isotropically in all directions. This is depicted in Figure~\ref{fig:illustration}: the green region represents the left-hand side of \eqref{eq:transform_necessity-crux}---a widened $E_2^\perp$---and purple regions within Figure~\ref{fig:illustration}(a) represent acceptable choices for the right-hand side of \eqref{eq:transform_necessity-crux}---a widened $(E_2^{(i_2)})^\perp$. However, in the 2-layer construction, queries that lie outside of the set $\{x:\|P_{E_1^{(i_1)}}(x)\|\lesssim \delta^{(i_1)}\}$ (shaded in pink in Figure~\ref{fig:illustration}) are answered directly using $E_1^{(i_1)}$. Therefore, it suffices to ensure
\begin{equation*}\label{eq:transform_necessity-answer}
    \set{x\in B_d(0,1): \|P_{E_2}(x)\|\leq\epsilon}\cap \{x:\|P_{E_1^{(i_1)}}(x)\|\lesssim \delta^{(i_1)}\}\subseteq \{x:\|P_{E_2^{(i_2)}}(x)\|\leq \delta^{(i_2)}\}.
\end{equation*}
This left-hand side region is depicted in red within Figure~\ref{fig:illustration}.
This observation allows us to add significantly more noise to $E_2^{(i_1)}$ in the direction of $(E_1^{(i_1)})^\perp$: all the purple regions in Figure~\ref{fig:illustration}(b) are still allowed since the corresponding separation oracle remains feasible. This anisotropic noise is crucial for achieving our desired lower bound tradeoffs. For convenience, we may view this anisotropic noise as isotropic noise after applying the \emph{transformation} of the space which stretches the space along $E_1^{(i_1)}$ by a factor $1/\delta^{(i_1)}$. This is depicted in Figure~\ref{fig:illustration}(c).

\comment{
Intuitively, \eqref{eq:transform_necessity-crux} means that the blocking oracle must be responsible for the entire set of points in $B_d(0,1)\cap E_2^\perp$ that corresponds to the red region in Figure~\ref{fig:symmetric-a}. However, as depicted in the same figure, a large proportion of this set, shadowed in green, lies outside of $\set{\|P_{E_1^{(i_1)}}(x)\|\leq\delta^{(i_1)}}$, which can already be answered by the layer-1 phases. In the end, rather than blocking every query away from $B_d(0,1)\cap E_2^\perp$, the oracle for second layer only needs to be responsible for points in
\begin{equation}\label{eq:transform_necessity-answer}
    B_d(0,1)\cap E_2^\perp \cap\set{\|P_{E_1^{(i_1)}}(x)\|\leq\delta^{(i_1)}},
\end{equation}

which allows us to add even more noise to $E_2$ in the direction of $E_1^{(i_1)}$. For intuition, we compare between Figure~\ref{fig:symmetric-b}, which corresponds to the isotropic noise by the blocking oracle, and Figure~\ref{fig:symmetric-c}, which depicts in purple the enlarged heterogeneous noise that is actually allowed for us.}
\definecolor{basegreen}{RGB}{97,197,111}
\definecolor{basered}{RGB}{235,106,106}
\definecolor{overlaypurple}{RGB}{134,91,202}

\begin{figure}[t]
\centering
\definecolor{basegreen}{RGB}{97,197,111}
\definecolor{basered}{RGB}{220,48,57}
\definecolor{overlaypurple}{RGB}{134,91,202}

\resizebox{0.95\linewidth}{!}{%
\begin{tikzpicture}[
  x=1cm,y=1cm,
  >=Stealth,
  base boundary/.style={draw=black,line width=0.65pt},
  purple band/.style={fill=overlaypurple,fill opacity=0.23,draw=none},
  pink annotation/.style={draw=basered!75!black,line width=0.55pt},
  green annotation/.style={draw=basegreen!45!black,line width=0.55pt}
]
\def\W{2.15}
\def\H{1.95}
\def\V{0.36}  %
\def\B{0.085} %
\def\P{0.27}  %
\def\GreenAngle{12} %
\pgfmathsetmacro{\GreenSlope}{tan(\GreenAngle)}
\pgfmathsetmacro{\GreenHalfY}{\B/cos(\GreenAngle)}
\def\PinkFraction{0.80} %
\pgfmathsetmacro{\Stretch}{\PinkFraction*\W/\V}

\foreach \panel/\offset/\letter in
  {1/0/a,2/5.05/b,3/10.10/c}{
  \begin{scope}[xshift=\offset cm]
    \def\PanelStretch{1}
    \ifnum\panel=3
      \edef\PanelStretch{\Stretch}
    \fi

    \path (-\W,-\H) rectangle (\W,\H);
    \begin{scope}[overlay]
      \clip (-\W,-\H) rectangle (\W,\H);
      \begin{scope}[xscale=\PanelStretch]
        \ifnum\panel=1
          \foreach \deviation in {-4,4}{
            \pgfmathsetmacro{\angle}{\GreenAngle+\deviation}
            \begin{scope}[rotate=\angle]
              \path[purple band] (-4,-\P) rectangle (4,\P);
            \end{scope}
          }
        \fi
        \ifnum\panel>1
          \foreach \deviation in {-30,0,30}{
            \pgfmathsetmacro{\angle}{\GreenAngle+\deviation}
            \begin{scope}[rotate=\angle]
              \path[purple band] (-4,-\P) rectangle (4,\P);
            \end{scope}
          }
        \fi

        \path[fill=basered,fill opacity=0.14,draw=none]
          (-\V,-\H) rectangle (\V,\H);

        \foreach \Left/\Right in {-\W/-\V,\V/\W}{
          \path[fill=basegreen,fill opacity=0.72,draw=none]
            (\Left, {\GreenSlope*(\Left)-\GreenHalfY}) --
            (\Right,{\GreenSlope*(\Right)-\GreenHalfY}) --
            (\Right,{\GreenSlope*(\Right)+\GreenHalfY}) --
            (\Left, {\GreenSlope*(\Left)+\GreenHalfY}) -- cycle;
        }
        \path[fill=basered,fill opacity=0.94,draw=none]
          (-\V,{-\GreenSlope*\V-\GreenHalfY}) --
          ( \V,{ \GreenSlope*\V-\GreenHalfY}) --
          ( \V,{ \GreenSlope*\V+\GreenHalfY}) --
          (-\V,{-\GreenSlope*\V+\GreenHalfY}) -- cycle;

        \draw[base boundary] (-\V,-\H) -- (-\V,\H);
        \draw[base boundary] ( \V,-\H) -- ( \V,\H);
        \draw[base boundary]
          (-\W,{-\GreenSlope*\W-\GreenHalfY}) --
          ( \W,{ \GreenSlope*\W-\GreenHalfY});
        \draw[base boundary]
          (-\W,{-\GreenSlope*\W+\GreenHalfY}) --
          ( \W,{ \GreenSlope*\W+\GreenHalfY});
      \end{scope}
    \end{scope}

    \ifnum\panel=1
      \begin{scope}[rotate=\GreenAngle]
        \draw[green annotation,line width=0.5pt,<->,
              >={Stealth[length=2pt,width=2pt]}]
          ({-\W-0.1},-\B) -- ({-\W-0.1},\B)
          node[midway,left,text=basegreen!45!black,
               font=\scriptsize] {$2\epsilon$};
      \end{scope}

      \node[anchor=north east,inner sep=2pt,font=\scriptsize,
            text=basegreen!45!black] (greenlabel) at ({-\V-0.12},-1.10)
        {$\{x:\|P_{E_2}(x)\|
          \leq\epsilon\}$};
      \draw[green annotation,->]
        (greenlabel.north) -- (-1.30,{-1.30*\GreenSlope-\GreenHalfY});

      \node[anchor=east,inner sep=2pt,font=\scriptsize,
            text=basered!75!black] (pinklabel) at ({-\V-0.13},1.10)
        {$\{x:\|P_{E_1^{(i_1)}}(x)\|
          \leq\delta^{(i_1)}\}$};
      \draw[pink annotation,->]
        (pinklabel.east) -- (-0.10,1.10);

      \draw[pink annotation,<->,
            >={Stealth[length=1.8pt,width=2.8pt]}]
        (-\V,{\H+0.1}) -- (\V,{\H+0.1})
        node[midway,above,text=basered!75!black,
             font=\scriptsize] {$2\delta^{(i_1)}$};
    \fi

    \node[anchor=north,font=\normalsize]
      at (0,{-\H-0.18}) {(\letter)};
  \end{scope}
}
\draw[->,draw=black!75,line width=0.7pt]
  (5.45,{\H+0.28}) -- (9.70,{\H+0.28})
  node[midway,above=-1pt,font=\scriptsize,text=black!85]
    {stretch along $E_1^{(i_1)}$ by $1/\delta^{(i_1)}$};
\end{tikzpicture}%
}
\captionsetup{font=footnotesize}
\caption{Anisotropic noise in the $2$-layer construction and corresponding transformation. The green strip represents the reference layer-$2$ feasible set, the pink strip represents the current layer-$1$ constraint set, and their intersection is shown in red. Purple strips represent candidate layer-$2$ constraint sets $\{x:\|P_{E_2^{(i_2)}}(x)\|\leq\delta^{(i_2)}\}$. (a) Requiring each purple strip to contain the entire green strip restricts the permissible perturbations. (b) Since queries outside the pink strip are answered by the first-layer oracle, feasibility requires only containment of the red intersection, allowing substantially larger anisotropic perturbations. (c) After stretching along $E_1^{(i_1)}$ by a factor $1/\delta^{(i_1)}$, perturbations become isotropic in the transformed space.}
\label{fig:symmetric}
\label{fig:illustration}
\end{figure}
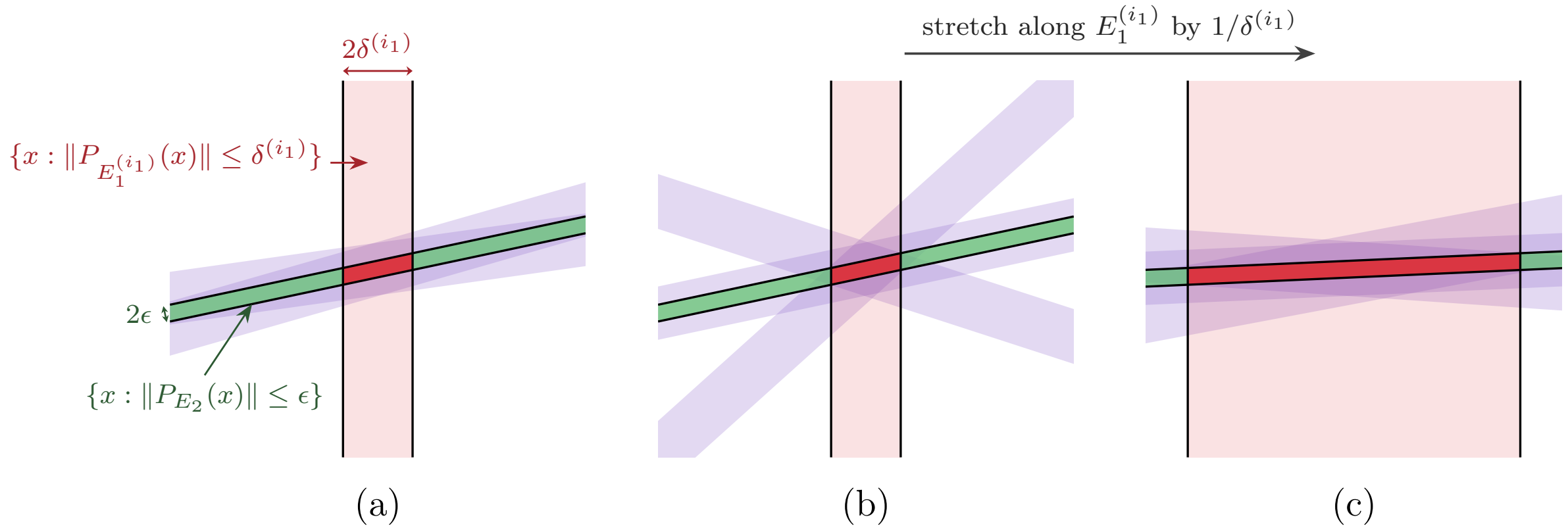

\comment{

As an equivalent way of viewing this, given that we only need to provide separations for the red region, we can stretch the set in \eqref{eq:transform_necessity-answer} along $E_1^{(i_1)}$ by $1/\delta^{(i_1)}$ times, so that the red region in Figure~\ref{fig:symmetric-a} is no longer thin in the direction of $E_1^{(i_1)}$. And naturally, isotropic noise can be applied in this transformed space.
}

As suggested above, we can fix the issue from the previous construction by stretching the space along $E_1^{(i_1-1)}$ for each phase $i_1$.
This gives our final $2$-layer construction:
\begin{enumerate}
    \item At the beginning of each new layer-$1$ phase $i_1$, resample increasing-accuracy versions $(E_2^{(i_2)})_{i_2\in N}$ of $T^{-\top}(E_2)$, where $T$ is the linear map that stretches $\Rbb^d$ by a factor $1/\delta_1^{(i_1-1)}$ along $E_1^{(i_1-1)}$. 
    \item For a new query $x$ during layer-$1$ phase $i_1$, if $\bm e^\top x > -1/2$, answer trivially with $\bm e$.
    \item Otherwise, check if $\|P_{E_1^{(i_1)}}(x)\| \gg \delta^{(i_1)}$. If so, answer directly with $P_{E_1^{(i_1)}}(x)$. \label{item:first_check}
    \item Otherwise, answer with the blocking-oracle for $(E_2^{(i_2)})_{i_2\in[N]}$ using appropriate mappings between the original and transformed spaces.\footnote{Specifically, $x$ is mapped to $T(x)$ in the transformed space, and a separating hyperplane with normal $g$ in the transformed space is mapped to a separating hyperplane with normal $T^\top(g)$ in the original space. 
    } 
    If the query has accuracy corresponding to the highest-accuracy subspace $E_2^{(N)}$, the oracle moves on to the next layer-$1$ phase $i_1\gets i_1+1$.
\end{enumerate}

With this construction at hand, the core part of the proof is to show that during each layer-$1$ phase $i_1\in[N]$, the player cannot skip a majority of layer-$2$ phases corresponding to the increasing-accuracy versions $(E_2^{(i_2)})_{i_2\in[N]}$ of $E_2$. We give below some of the main intuitions for this property, and defer the formal structure of our proof to the next section.

\vspace{4pt} \noindent\textbf{High-level information-theoretic argument.} First, denote by $\texttt{Memory}_{i_1}$ the memory state of the algorithm just before the beginning of layer-$1$ phase $i_1$, i.e., just before observing $E_1^{(i_1)}$. By the chain rule:
\begin{equation*}
    M \geq H(\texttt{Memory}_{i_1}) \geq I(\texttt{Memory}_{i_1};E_2^{(1)},\ldots,E_2^{(N)}) = \sum_{i\in[N]} I(\texttt{Memory}_{i_1};E_2^{(i)} \mid E_2^{(1)},\ldots,E_2^{(i-1)}).
\end{equation*}
Therefore, for a majority of the layer-$2$ phases $i\in[N]$, the learner retains only
\begin{equation*}
    I(\texttt{Memory}_{i_1};E_2^{(i)} \mid E_2^{(1)},\ldots,E_2^{(i-1)}) \leq \frac{M}{N/2} = d^{1+\delta}\poly\log d,
\end{equation*}
information about this layer $i$. 
Intuitively, this shows that the player may retain at most a $k$-dimensional subspace $W_i$ approximately perpendicular to $E_2^{(i)}$ in the transformed space, where $k=\tilde O(d^{\delta})\ll d$. For simplicity, suppose that the learner proceeds in this manner, i.e., only memorizes some subspace $W_i$ in the transformed space that is perpendicular to $E_2^{(i)}$ with precision $\delta^{(i)}$. This is a reasonable strategy for the learner: querying any vector $x\in W_i$ would skip the layer-$2$ phase corresponding to $E_2^{(i)}$, provided it passed the first check $\|P_{E_1}^{(i_1)}(x)\|\lesssim \delta^{(i_1)}$ (see step~\ref{item:first_check}).

Importantly, the learner must commit to this subspace $W_i$ \emph{before} observing the new layer-$1$ subspace $E_1^{(i_1)}$. Further, the transformed space was designed so that $E_1^{(i_1)}$ has high variability in it. Conceptually, we may consider that $E_1^{(i_1)}$ is a Haar uniform $\tilde d$-dimensional subspace, independent of $W_i$. Since the memorized subspace $W_i$ only has dimension $k\ll d$, we can show that with reasonable probability, no vector in $W_i$ can successfully pass the first check $\|P_{E_1}^{(i_1)}(x)\|\lesssim \delta^{(i_1)}$ (see step~\ref{item:first_check}). Intuitively, the information memorized about $E_2^{(i)}$ by the learner is ``blocked'' by the new layer-$1$ subspace $E_1^{(i_1)}$. In turn, this prevents the learner from skipping the subspace $E_2^{(i)}$ in the blocking oracle, with reasonable probability.

\comment{
\paragraph{From $2$ layers to $P$ layers.} 
The construction for $2$ layers can be naturally generalized to $P$ layers. As a note, this requires implementing a transformation of the space for each layer $p=1,2,\ldots,P-1$. Roughly speaking, after applying the combined transformation from the first $p-1$ layers, everything from layers $p,p+1,...,P$ in this transformed space would look similar to the corresponding quantities from layers $1,2,...,P-p+1$ in the original space. 
}

\comment{
The construction for $2$ layers can be naturally generalized to $P$ layers. As a note, this requires implementing a transformation of the space for each layer $p=1,2,\ldots,P-1$: at the beginning of a layer-$p$ phase corresponding to a new-accuracy subspace $E_p^{(i_p)}$ we apply a linear transformation that stretches the space along $(E_p^{(i_p-1)})^\perp$ by $1/\delta_p^{(i_p-1)}$, then sample increasing-accuracy versions $(E_{p+1}^{(i)})_{i\in[N]}$ of $E_{p+1}$ in that transformed space.
}

\vspace{4pt}
\noindent\textbf{Notations.} Throughout, $\log$ denotes the base-$2$ logarithm and $\ln$ denotes the base-$e$ logarithm. We denote by $\Gr(k,d)$ the collection of $k$-dimensional subspaces of $\Rbb^d$. For $E\in\Gr(k,d)$, we denote by $P_E$ the orthogonal projection onto $E$.
Throughout the paper, we use balls in the Grassmannian with respect to the \emph{largest angle distance}---i.e., the infinity norm of principal angles: for any center $F\in\mathrm{Gr}(k,d)$ and radius $\rho\in[0,\tfrac{\pi}{2}]$ we denote
$B(F,\rho):=\set{E\in\mathrm{Gr}(k,d):\theta_{\max}(E,F)\leq\rho}$, where the principal angle is $\theta_{\max}(E,F)=\arcsin\|P_E-P_F\|_{op}$.
For a linear map $\Tcal:\Rbb^d\to\Rbb^d$, we denote by $\lambda_{\min}(\Tcal),\lambda_{\max}(\Tcal)$ its minimum and maximum singular values. We then denote by $\kappa(\Tcal):= \frac{\lambda_{\max}(\Tcal)}{\lambda_{\min}(\Tcal)}$ its condition number.

We will also use the shorthand $u^{(\leq N)}$ to refer to a given sequence $(u^{(0)},u^{(1)},\ldots,u^{(N)})$. Next, for any two sequences $u,v$ we denote by $u\parallel v$ their concatenation. By an abuse of notation, we denote by $u\parallel x := u\parallel (x)$ the sequence obtained by appending $x$ to $u$. Finally, the multiplication of a sequence $u$ by a scalar $a\in\Rbb$ is meant coordinate-wise: $a \cdot u^{(\leq N)} = (au^{(0)},\ldots,au^{(n)})$.

\section{Proof of the main result}
\label{sec:proof_main_result}
\subsection{Construction of the hard class of feasibility problems}\label{subsec:construction_of_oracle}

In this section, we start by formally constructing the hard class of feasibility problems with accuracy $\epsilon\leq e^{-d}$, for which we prove our query-complexity lower bound.
As discussed in \cref{sec:technical_overview}, the construction depends on some depth parameter $P$. We then assume $P\leq d^{1/4}$ and sample independently
\begin{equation*}
    E_1,\ldots,E_P\overset{iid}{\sim}\mathrm{Unif}(\mathrm{Gr}(\tilde d,d)),\quad \text{where}\quad \tilde d:=\floor{\frac{d}{36P}},
\end{equation*}
a sequence of $\tilde d$-dimensional subspaces of $\Rbb^d$. We then define the feasible region via
\begin{equation*}
    Q_{E_1,...,E_P}:=B_d(0,1)\cap\set{x:\bm e^\top x\leq -\tfrac{1}{2}}\cap\bigcap_{p\in[P]}\set{x:\|P_{E_p}(x)\|\leq\epsilon},
\end{equation*} 
Hence, the algorithm aims to query vectors approximately perpendicular to all subspaces $E_1,...,E_P$. The first constraint $\bm e^\top x\leq -1/2$ is included only to ensure that the algorithm does not query small-norm vectors.

To formally define the corresponding separation oracle, we need to introduce a few notions. 
First, we define a noise-adding Markov process which takes as input a subspace $F\in\mathrm{Gr}(\tilde d,d)$, a sequence of non-negative, decreasing accuracies $\bm\delta:=(\delta^{(0)}, \ldots,\delta^{(N)})$ and constructs a sequence $F^{(0)},\ldots,F^{(N)}\in\mathrm{Gr}(\tilde d,d)$ of noisy versions of $F$ with the corresponding accuracies.
These are constructed backwards via a simple random walk on the Grassmannian initialized at $F$, where each step is taken uniformly on the local ball of radius $\delta^{(N)},\ldots,\delta^{(0)}$ in that order. The corresponding procedure $\NoiseAddition$ is summarized in \cref{alg:noise_addition}. This procedure will be used to construct different precision / noisy versions of the subspaces $E_p$ which will be revealed by the oracle as frugally as possible, starting with the lowest-precision / highest-noise version.

\begin{algorithm}[h]
\caption{$\NoiseAddition(F,\bm\delta)$}\label{alg:noise_addition}
\LinesNumbered
\everypar={\nl}
\KwIn{subspace $F\in\mathrm{Gr}(k,d)$, accuracies $\bm\delta=(\delta^{(0)},\ldots,\delta^{(N)})$}
Initialize $F^{(N+1)}\gets F$
and \lFor{$i=N,N-1,\dots,0$}{
Sample $F^{(i)}\sim \text{Unif}(B(F^{(i+1)},\delta^{(i)}))$}
Return $F^{(\leq N)}$.
\end{algorithm}

The separation oracle proceeds by \emph{nested phases} for each layer $p\in[P]$: each layer-$(p-1)$ phase contains exactly $N_p$ layer-$p$ phases of precisions $\delta_p^{(1)},\dots,\delta_p^{(N_p)}$, where 
\begin{equation*}
    N_p=\floor{\frac{\log(1/\epsilon)}{2Pp\log d}},\quad \lambda_p:=d^{-p},\quad \delta_p^{(i)} = d^{-1}\epsilon^{\frac{p-1}{2P}} \lambda_p^i,\quad\forall i\in\set{0,1,\dots,N_p}.
\end{equation*}
This nested procedure is most easily described through its \emph{skeleton tree}, which corresponds
to the full $(N_1,\ldots,N_P)$-ary tree with vertices $\Vcal:=\bigsqcup_{p=0}^P  \Vcal_p$ where $\Vcal_p:=[N_1]\times\cdots\times [N_p]$ corresponds to the set of layer-$p$ phases. The children of a layer-$p$ phase $(i_1,\ldots,i_p)\in\Vcal_p$ are precisely all layer-$(p+1)$ phases $(i_1,\ldots,i_p,i_{p+1})$ for $i_{p+1}\in[N_{p+1}]$.

We associate each phase $v\in\Vcal_p$ with (1) a version $E_{v}$ of the subspace $E_p$ corresponding to some accuracy $\delta_{v}$ in the corresponding transformed space, and (2) a linear transformation $\Tcal_v$ corresponding to the concatenation of all transformations leading up to this phase. For convenience, we introduce the following notation for stretching the space by a factor $\alpha$ along $F$:
\begin{equation*}
    T_{F,\alpha}:=\alpha P_F+P_{F^\perp}.
\end{equation*}
The construction proceeds as follows: for the root phase $\emptyset$ we initialize $\Tcal_\emptyset = I$.
Having defined these quantities for a given layer-$(p-1)$ phase $v'\in\Vcal_{p-1}$, we first sample
\begin{equation*}
    E_{v',p}^{(\leq N_p)}:=\NoiseAddition(\Hcal_{v'}(E_p),\delta_p^{(\leq N_p)}/\sqrt{3P})
\end{equation*}
independently from all other variables, then set, for any child phase $v=v'\parallel i_p$ for $i_p\in[N_p]$:
\begin{equation*}
    (E_{v},\delta_{v}):=(E_{v',p}^{(i_p)},\delta_p^{(i_p)})\quad\text{and}\quad \Tcal_v:=  T_{E_{v',p}^{(i_p-1)},1/ \delta_p^{(i_p-1)}}\circ \Tcal_{v'},\quad \Hcal_v:=\Tcal_v^{-\top}.
\end{equation*}
Here, we introduce the adjoint transform $\Hcal_v$ which intuitively applies to the subspaces $E$: note that $\Tcal_v(E^\perp) =(\Hcal_v(E))^\perp$.
In words, to construct the layer-$p$ phases nested within $v'$, we run the noise addition process on $E_p$ in the $\Tcal_{v'}$-transformed space: each different version $E_{v',p}^{(i_p)}$ corresponds to a different layer-$p$ phase. We then apply a new transformation which stretches the space along the lower-accuracy version $E_{v',p}^{(i_p-1)}$. Intuitively, after applying this transformation, the current subspace $E_{v',p}^{(i_p)}$ is ``well-conditioned'': it can be shown to have variability up to $1/d$ principal angle (this is formalized in \cref{lemma:reduce_to_uniform}). This construction is summarized in \cref{alg:construct_tree_structure}.

\begin{algorithm}[h]
\caption{Skeleton tree construction}\label{alg:construct_tree_structure}
\LinesNumbered
\everypar={\nl}
\KwIn{dimensions $d>\tilde d$, depth $P$, parameters $N_1,\ldots,N_P,\delta_1^{(\leq N_1)},\ldots,\delta_P^{(\leq N_P)}$}
Sample $E_1,\dots,E_P\overset{iid}{\sim}\mathrm{Unif}(\mathrm{Gr}(\tilde d,d))$
and initialize $\Tcal_\emptyset=I$

\For{$p=1,\ldots,P$}{
    \For{$v'\in \Vcal_{p-1}$}{
        Sample independently $E_{v',p}^{(\leq N_p)}:=\NoiseAddition (\Hcal_{v'}(E_p),\tfrac{1}{\sqrt{3P}}\delta_p^{(\leq N_p)})$ (see \cref{alg:noise_addition})

        \For{$i_p\in [N_p]$}{
             Let $v=v'\parallel i_p$, set $(E_v,\delta_v):=(E_{v',p}^{(i_p)},\delta_p^{(i_p)})$, $\Tcal_v:=T_{E_{v',p}^{(i_p-1)},1/\delta_p^{(i_p-1)}}\circ \Tcal_{v'}$,\, $\Hcal_v:=\Tcal_v^{-\top}$
        }
    }
}

\end{algorithm}

We now focus on leaves $\Vcal_P$ which we order lexicographically. For convenience, we denote by $\Succ(l)=\min_{\leq_{lex}}\set{l'\in\Vcal_P: l<_{lex} l'}$ the successor of a given leaf $l\in\Vcal_P$. For the last leaf $l=(N_1,\ldots,N_P)$ we arbitrarily set $\Succ(l)>_{lex} l$.
The separation oracle proceeds by following leaves of the skeleton tree $\ell\in \Vcal_P$ in increasing order. More precisely, each leaf is associated with some tentative path separation oracle $\Ocal_\ell$ for the corresponding-accuracy subspaces $E_{\ell,1},\ldots,E_{\ell,P}$, and the oracle will move from one leaf to its successor as needed by the learner's queries. 

We start by defining the path separation oracle $\Ocal_\ell$ for a given leaf $\ell\in\Vcal_P$. Denote by $v_p=(\ell_1,\ldots,\ell_p)$ the layer-$p$ phase corresponding to this leaf. We recall that $v_p$ is associated with a version $E_{v_p}$ of $E_p$ with accuracy $\delta_{v_p}$. The path oracle first uses $E_{v_1}$ as a proxy to test perpendicularity of the query to $E_1$ up to its accuracy: $\|P_{E_{v_1}}(x)\|\leq\delta_{v_1}$. If not, the oracle may use $E_{v_1}$ to construct a separating hyperplane. Otherwise, the oracle continues by testing $\delta_{v_2}$-perpendicularity to the next proxy subspace $E_{v_2}$, taking into account the fact that these were sampled in a space transformed by $\Tcal_{v_1}$. The path oracle continues testing perpendicularity to the proxy subspaces $E_{v_3},\ldots,E_{v_P}$ in that order, and finally returns $\texttt{success}$ if the query $x$ successfully passed all these tests. This path separation oracle $\Ocal_\ell$ is summarized in \cref{alg:leaf_oracle}.

\begin{algorithm}[h]
    \caption{Path separation oracle $\Ocal_\ell(x)$ for a given leaf $\ell\in\Vcal_P$}\label{alg:leaf_oracle}
    \LinesNumbered
    \everypar={\nl}
    \KwIn{query $x$}
    \lIf{$\boldsymbol{e}^\top x > -1/2$}{Answer $x$ with $\boldsymbol{e}$}

    Denote by $v_p=(\ell_1,\ldots,\ell_p)$ the layer-$p$ ancestor of $\ell$
    
    \For{$p=1,\ldots ,P$}{
        \lIf{\label{eq:alg:hard_oracle_explicit_single-4}$\|P_{E_{v_p}}(\Tcal_{v_{p-1}}(x))\|>\delta_{v_p}$}{\Return $\Tcal_{v_{p-1}}^{\top} P_{E_{v_p}}(\Tcal_{v_{p-1}}(x))$}
    }
    \Return $\texttt{Success}$.
\end{algorithm}

We are now ready to formally define the final \emph{transform separation oracle} for the feasible region $Q_{E_1,\ldots,E_P}$. Throughout the learning process the oracle keeps track of its current leaf $\ell\in\Vcal_P$, initialized to the first leaf $\ell=(1,\ldots,1)$. For a new query $x$, the oracle tentatively uses the response from $\Ocal_\ell(x)$. If this path oracle replies $\texttt{Success}$, the oracle needs to increase the accuracy of its proxy subspaces by moving to the next leaf $\ell\gets \Succ(\ell)$. The process continues until the last leaf $\ell=(N_1,\ldots,N_P)$ has been exhausted, in which case the oracle may default to some arbitrary separation oracle for $Q_{E_1,\ldots,E_P}$. This is summarized in \cref{alg:final_oracle}.

\begin{algorithm}
\caption{The transform separation oracle $g$}\label{alg:final_oracle}
\LinesNumbered
\everypar={\nl}
\KwIn{dimensions $\tilde d,d$, depth $P$, parameters $N_1,\ldots,N_P,\delta_1^{(\leq N_1)},\ldots,\delta_P^{(\leq N_P)}$}

Construct the skeleton tree using \cref{alg:construct_tree_structure} then initialize the leaf $\ell\gets (1,\ldots,1)$

\For{$t\geq 1$}{
    \lWhile{$\Ocal_\ell(x_t)=\texttt{Success}$ and $\ell\leq_{lex}(N_1,\ldots,N_P)$}{
        Update $\ell\gets \Succ(\ell)$
    }
    \lIf{\label{line:transform-4}$\ell\leq_{lex}(N_1,\ldots,N_P)$}{
        Answer $x_t$ with $\Ocal_\ell(x_t)$
    }
    \lElse{\label{line:transform-5}Answer $x_t$ with any arbitrary separation oracle for $Q_{E_1,\ldots,E_P}$}
}
\end{algorithm}

As an important remark, the skeleton tree and path separation oracles for all leaves are constructed independently of the learner's decisions. The only source of adaptivity in the final separation oracle is through the selection of the current leaf $\ell$ that will be used for each round: as the leaf $\ell$ increases, the oracle is revealing more information (lexicographically) about the subspaces $E_1,\ldots,E_P$ and the oracle does so only when forced by the queries of the learner.

We can check that the constructed instance is indeed a valid feasibility problem for accuracy $\epsilon$.
\begin{lemma}\label{lem:oracle-consistency}
    For $d$ sufficiently large, with probability at least $1-e^{-d/300}$, the set $Q_{E_1,...,E_P}$ contains a ball of radius $\epsilon$ and all oracle responses by $g$ are valid separating hyperplanes for $Q_{E_1,...,E_P}$.
\end{lemma}

\subsection{General proof strategy}

By construction, the skeleton tree has $N_1 N_2\cdots N_P$ leaves, each corresponding to a specific path separation oracle that the learner may interact with at some point of the learning procedure. 
We aim to show that a significant fraction of these leaves cannot be ``skipped'' by the learner; this would immediately imply an exponential query-complexity lower bound in the depth $P$, as desired.

We will assume the learner is deterministic for this proof. We first introduce a few definitions about the initial memory of the learner for each phase of the transform oracle.
\begin{definition}[Start/end times and initial memory of phases]\label{def:phases_and_memory}
    Let $p\in[P]$ and fix any layer-$p$ phase $v\in\Vcal_p$. The \textbf{start} (resp. \textbf{end}) \textbf{time} of phase $v$, denoted $t_{v,0}$ (resp. $t_{v,1}$) is the first (resp. last) round $t$ when the transform separation oracle had its internal leaf be a descendant leaf of $v$, i.e. the first round when $\ell=v\parallel(1,\ldots,1)$ (resp. last round when $\ell=v\parallel(N_{p+1},\ldots,N_P)$) if it exists.

    The \textbf{initial memory} of phase $v$, denoted $\texttt{Mem}_v$ is the memory state of the learner at the beginning of phase $v$, i.e., $\texttt{Mem}_v := \texttt{Memory}_{t_{v,0}}$.
\end{definition}

As discussed in \cref{sec:technical_overview}, the core difficulty stems from this initial memory $\texttt{Mem}_v$ that the learner may have at the beginning of each phase $v$. Specifically, in the $2$-layer case, we argued that at the beginning of any layer-$1$ phase $(i_1)$, the learner may have memorized within $\texttt{Mem}_{(i_1)}$ a $k$-dimensional subspace $W$ perpendicular to $E_2$ in the corresponding transformed space after applying $\Tcal_{(i_1)}$. Further, if any vector $x\in W$ passes the two tests $\bm e^\top x\leq -1/2$ and $\|P_{E_{i_1}}(x)\|\leq \delta_{(i_1)}$, in the transformed space it would skip all layer-$2$ phases $(i_1,i_2)$ for $i_2\in[N_2]$. Fortunately, since $E_{(i_1)}$ has $\tilde d\gg k$ dimensions and has high variability even after observing the lower-accuracy subspaces $E_{(1)},\ldots,E_{(i_1-1)}$, no vector in a $(E_{(1)},\ldots,E_{(i_1-1)})$-measurable subspace $W$ of dimension $k$ is likely to pass these tests. Intuitively, these two tests ``block'' subspaces of dimension $k$.

We formalize this property through the notion of \emph{blocking region} defined below.

\begin{definition}[Blocking property]
    Fix $\Delta>0$ and $q\in[0,1]$. Let $R\subset \Rbb^d$ be a random region and let $C$ be any potentially correlated random variable. 
    We say that $R$ has the \textbf{blocking property} $\Delta$-$\Block_k(q|C)$ if for any $C$-measurable subspace $W\in\Gr(k,d)$,
    \[\Pbb\left(R\cap (W+\Delta B)= \emptyset\right)\geq q.\]
\end{definition}
In words, a region $R$ is $\Delta$-$\Block_k(q|C)$ if even after observing some \emph{context} $C$, one cannot construct a $k$-dimensional subspace $W$ such that its widening $W+\Delta B$ intersects $R$ with probability more than $1-q$. 
As a remark, the blocking property is harder to achieve when adding more variables to the context $C$, but we can add a random variable $X$ to $C$ for free provided $X$ and $R$ are conditionally independent given $C$.

Following the intuition from the $2$-layer case, natural candidates for blocking regions correspond to the regions which satisfy the perpendicularity tests corresponding to some path on the skeleton graph. More specifically, for any layer-$p$ phase $v\in\Vcal_p$, we define its region $R_v$ as the set of potential queries in the corresponding transformed space, that would pass all proxy perpendicularity tests for $E_1,\ldots,E_p$ in the constructed path separation oracles on phase $v$ (see \cref{alg:leaf_oracle}):
\begin{definition}[Phase regions and contexts]
\label{def:regions_contexts}
    Fix $p\in[P]$ and a layer-$p$ phase $v=(i_1,\ldots,i_p)\in\Vcal_p$. Let $v_s=(i_1,\ldots,i_s)$ be the layer-$s$ ancestor of $v$ for $s\leq p$. We define the \textbf{phase-$v$ region} via
    \begin{equation*}
        R_v:=\Tcal_v\paren{\set{x\in B_d(0,1): \bm e^\top x\leq -\tfrac12 \text{ and } \forall s\in[p], \|P_{E_{v_s}}(\Tcal_{v_{s-1}}(x))\| \leq \delta_{v_s}}}.
    \end{equation*}
    Next, letting $v'$ be the parent of $v$, we recall that $E_v=E_{v',p}^{(i_p)}$ (see \cref{alg:construct_tree_structure}). We then define the \textbf{phase-$v$ context} via
    \begin{equation*}
        C_v:=(E_{v_0,1}^{(\leq i_1)},\ldots,E_{v_{p-2},p-1}^{(\leq i_{p-1})}, E_{v',p}^{(<i_p)}).
    \end{equation*}
\end{definition}
In words, the context $C_v$ for a layer-$p$ phase includes information about the different-accuracy versions of $E_{p'}$ leading to $E_{v_{p'}}$ for shallower layers $p'<p$, but for the current layer $p$, we only include strictly-lower accuracy versions of $E_p$. As a remark, the transformation $\Tcal_v$ is $C_v$-measurable.

With these definitions at hand, our goal is to show that for each $p\in[P]$, for a significant fraction of layer-$p$ phases $v$, their region enjoys the blocking property for the context $C_v$. 
Specifically, we aim to propagate the blocking property for phase regions from shallow layers $p=1$ to the deepest layers $p=P$ by induction. 

As a remark, the blocking property cannot necessarily be propagated to \emph{all} phases. Indeed, the total memory of the algorithm is $M=O(d^{1+\delta}\log(1/\epsilon))$ and since we consider exponentially-small accuracy $\epsilon$ regimes, this total memory may be significantly larger than $d^2$. In particular, the memory $\texttt{Mem}_v$ is more than enough to store fully an approximate encoding of some child subspace $E_{v,p}^{(i_p)}$ given $E_{v,p}^{(<i_p)}$, in which case the phase $(v\parallel i_p)$ may be easily skipped by the learner. However, we will show that the player cannot do so for most child phases $i_p\in[N_p]$.
Accordingly, we focus on \emph{well-behaved} phases which satisfy the following information constraint:
\begin{definition}[Well-behaved phases]
    Fix $p\in[P]$ and a layer-$p$ phase $w=(v\parallel i_p)\in\Vcal_p$, where $v$ denotes its parent. We recall that by construction, $E_w:=E_{v,p}^{(i_p)}$. We say that $w$ is well-behaved if
    \begin{equation*}
        I(E_w;\texttt{Mem}_{v}\mid E_{> p}, C_{w})
        \leq 4M/N_p.
    \end{equation*}
\end{definition}
In particular, if $w$ is well-behaved, the previous memory includes $M/N_p \ll kd$ information about the subspace $E_w$, which is not enough to encode it fully.

We are now ready to state a propagation result for the blocking property.
This is one of the core steps of our query-complexity lower bound argument.

\begin{theorem}[Blocking property propagation]
\label{thm:propagation}
    Let $d\geq 1$ be sufficiently large. Let $P\geq 2$. There exists a universal constant $C_b>0$ such that the following holds. Let $\Delta\in(0,1]$ and suppose that 
    $C_b\frac{M/N_p+P}{\tilde d}\cdot P \log(d/\Delta)\log P \leq k \leq \frac{\tilde d}{C_b P\log(d/\Delta)\log d}$.
    
    Fix $p\in\{2,\ldots,P\}$ and a layer-$(p-1)$ phase $v\in\Vcal_{p-1}$ and let 
    $\Delta'\in [\lambda_p,c\frac\Delta k\left(\frac{\tilde{d}}{d}\right)^{3/2}]$. Suppose that $v$ is well-behaved and $R_v$ is $\Delta$-$\Block_k(q\mid \texttt{Mem}_v,C_v,E_{>p-1})$.
    Then, for a majority of children $w\in\Vcal_p$ of $v$, $w$ is well-behaved and $R_w$ is $\Delta'$-$\Block_k(q-\tfrac1P\mid \texttt{Mem}_w,C_w,E_{>p})$.
\end{theorem}

For the initial layer $p=1$, no propagation is needed. Prior to some layer-$1$ phase $(i_1)$ for $i_1\in[N_1]$, the information gathered by the learner about the first subspace $E_1$ is exactly summarized by the lower-accuracy subspaces $E_{(1)},\ldots,E_{(i_1-1)}$. That is, the learner has only observed low-accuracy versions of $E_1$ compared to $E_{(i_1)}$; we can then show that the posterior on $E_{(i_1)}$ is in fact essentially uniform on a Grassmannian ball centered at $E_{(i_1-1)}$, and hence no subspace is perpendicular to $E_{(i_1)}$ with significant probability:

\begin{restatable}{proposition}{PropInitializationBlocking}
\label{prop:initialization_blocking}
    Let $d\geq 1$ be sufficiently large and $4k\leq \tilde d$. For any layer-$1$ phase $v\in\Vcal_1$, $v$ is well-behaved and $R_v$ is $\lambda_1$-$\Block_k(1-(2P)^{-1}\mid \texttt{Mem}_v,C_v,E_{>1})$. 
\end{restatable}

Combining the initialization from \cref{prop:initialization_blocking} and the induction step from \cref{thm:propagation} then shows that a large number of layer-$P$ phases---i.e., leaf phases---preserve the blocking property. 

\begin{corollary}\label{cor:induction_blocking}
    Let $d\geq1$ be sufficiently large and assume $d\log(1/\epsilon)\leq M\leq c_0\frac{d^2\log(1/\epsilon)}{\log^4 d}$ 
    for some universal constant $c_0>0$. Then, one can select $$P=1+\ceil{\left(\frac{c_0 d^2\log(1/\epsilon)}{M\log^5 d}\right)^{1/8}}.$$ 
    and some $k\geq 1$, for which there are at least
    $N_1\cdot\prod_{p=2}^P\paren{N_p/2}$
    layer-$P$ phases $\ell\in\Vcal_P$ such that $R_\ell$ is $\lambda_P$-$\Block_k((2P)^{-1}\mid \texttt{Mem}_\ell,C_\ell)$.
\end{corollary}
Focus on such a leaf phase $\ell\in\Vcal_P$. By definition, with probability at least $1/(2P)$, a widened $k$-dimensional subspace cannot lie in the successful region $R_\ell$, even having access to the memory state $\texttt{Mem}_\ell$. Hence, on that event, the learner cannot make a single query $x_t$ based only on its memory $\texttt{Mem}_\ell$ that would skip that leaf, since $x_t\in\Span(x_t)$ has dimension $1\leq k$. Plugging in the parameter choices for the numbers of phases $N_1,\ldots,N_P$ then proves our exponential query lower bound.

\subsection{Propagating the blocking property}

One of the main technical steps of the proof is to show \cref{thm:propagation}, which propagates the blocking property from one layer $p$ to the next layer $p+1$. We give in this section the main steps for this argument. We start by showing that the majority of children of a well-behaved phase are also well-behaved and satisfy an additional technical information constraint:
\begin{restatable}{lemma}{LemmaPigeonholePrinciple}
\label{lemma:pigeonhole_principle}
    Fix $p\in\{2,\ldots,P\}$ and a well-behaved layer-$(p-1)$ phase $v\in\Vcal_{p-1}$. A majority of child phases $w=(v\parallel i_p)$ of $v$ are well-behaved and additionally satisfy
    \begin{equation}\label{eq:second_technical_info_constraint}
        I(E_w;E_{v}\mid E_{v,p}^{(<i_p)} ,\texttt{Mem}_{v}, C_{v},E_{>p})\leq 16M/(N_pN_{p-1}).
    \end{equation}
\end{restatable}
The well-behavedness of a majority of child phases $w$ follows immediately from the information chain rule and Markov's inequality. The second technical information constraint uses the well-behavedness of the parent $v$, which in turn gives the extra factor $1/N_{p-1}$.

We now focus on phases $w\in\Vcal_p$ that satisfy the two properties from \cref{lemma:pigeonhole_principle}. We aim to show that these inherit the blocking property from their parent, via a reduction to a compression game. 

\vspace{4pt} \noindent\textbf{Reduction to a compression game.} At a high level, in this compression game, a player aims to output $k$ robustly independent vectors approximately perpendicular to some high-entropy $\tilde d$-dimensional subspace $E$, having access to some prior constrained memory. This game is heavily inspired by the orthogonal vector game of \cite{marsden22efficient} in which a player aims to output robustly independent vectors approximately perpendicular to a matrix with independent Rademacher entries, having additional access to an oracle for row-access to the matrix. An analogous game was also used in \cite{blanchard24gradient} in which the goal was to compress a Haar uniform random subspace, without access to an additional row oracle. 
Our compression game is essentially the same as in \cite{blanchard24gradient} except that the initial distribution of the subspace $E$ is not Haar uniform, but is instead the posterior distribution after observing lower-accuracy versions of this subspace. This specific choice naturally arises from our separation oracle construction. The compression game is formally defined in Game~\ref{game:compression_game}.

\begin{game}[htb]
\SetAlgoLined
\LinesNumbered
\caption{Compression game at layer $p\geq 2$}
\label{game:compression_game}
\KwIn{dimensions $d,\tilde{d},k$, map $\Tcal$, accuracies $\delta^{(\leq N)}$, index $i$, parameters $\beta,\gamma,\eta$}
\emph{Nature:} Sample $E\sim \Unif(\Gr(\tilde d,d))$ and $E^{(\leq N)}:=\NoiseAddition(\Tcal(E),\delta^{(\leq N)})$ (see Alg.~\ref{alg:noise_addition})

\emph{Encoder:} Observe $E^{(\leq i)}$ and store a message $\texttt{Message}$ about these

\emph{Decoder:} Based on $\texttt{Message}$ and $E^{(\leq i-1)}$ only, return nonzero vectors $x_1,\ldots,x_k$

Letting $\tilde E^{(i)}:=T_{E^{(i-1)},\eta} (E^{(i)})$, the compression is successful if for all $t\in[k]$

\begin{enumerate}[noitemsep, topsep=2pt, partopsep=4pt]
    \item $\|\text{Proj}_{\tilde E^{(i)}} (x_t)\|\leq \beta\|x_t\|$
    \item $\|\text{Proj}_{\Span(x_1,\dots,x_{t-1})^\perp}(x_t)\|\geq \gamma\|x_t\|.$
\end{enumerate}
\end{game}

The reduction is formally stated below: essentially, if the blocking property does not propagate from a layer-$(p-1)$ phase to some children satisfying \cref{lemma:pigeonhole_principle}, one may use this phase to implement a winning strategy for the compression game.

\begin{proposition}[Reduction to Game~\ref{game:compression_game}]\label{prop:reduction_to_compression_game}
    For $d$ sufficiently large, there is a universal constant $c>0$ such that the following holds. Fix $p\in\{2,\ldots,P\}$, some parameters $\lambda_p\leq\Delta'<\tfrac18\Delta\leq\tfrac18$. Fix a well-behaved layer-$(p-1)$ phase $v\in\Vcal_{p-1}$ such that $R_v$ is $\Delta$-$\Block_k(q\mid \texttt{Mem}_v,C_v,E_{>p-1})$.
    Let $w=(v\parallel i_p)$ be a well-behaved child phase satisfying \cref{eq:second_technical_info_constraint}. Then, either 
    \begin{enumerate}
        \item $R_w$ is $\Delta'$-$\Block_k(q-1/P\mid \texttt{Mem}_w,C_w,E_{>p})$,
        \item or there is a strategy to win at Game~\ref{game:compression_game} with some probability $\xi\geq (12P\ln(6P))^{-1}$, for dimensions $d,\tilde d,k$, accuracies $\delta^{(\leq N)}:=\delta_p^{(\leq N)}/\sqrt{3P}$, index $i:=i_p$, parameters $(\beta,\gamma):=(8\Delta',\Delta/(2\sqrt{P}))$, $\eta=\delta_p^{(i_p-1)}$, and some linear map $\Tcal$ with $\kappa(\Tcal)\leq 2/(d\delta_p^{(0)})$, such that the message satisfies the information constraint
        \begin{equation*}
             I(\texttt{Message};E^{(i)}\mid E^{(<i)})\leq \paren{12P\ln(6P)\paren{\frac{4M}{N_p}\paren{1+\frac{8k}{N_{p-1}}}+6k }+8k\log(6Pk)+4Pk^2 \log d }\xi.
        \end{equation*}
    \end{enumerate}
\end{proposition}

\vspace{4pt} \noindent\textbf{Hardness result for the compression game.} In this last step of the proof, we prove an upper bound on the success probability for the compression Game~\ref{game:compression_game}.

We give some intuitions when the linear map is simply the identity $\Tcal=I$. Due to the noise-addition process, we can then directly check that in this case, conditionally on lower-accuracy subspaces $E^{(<i)}$, the posterior of the subspace $E^{(i)}$ is exactly uniform over the Grassmannian ball $B(E^{(i-1)},\delta^{(i-1)})$. The transformation $T(E^{(i-1)},\delta^{(i-1)})$ then essentially stretches the radius of the ball to be of constant order. For simplicity, we may then consider that the posterior of $\tilde E^{(i)}$ is uniform over a ball $B(\tilde E^{(i-1)},\rho)$ for $\rho=\Theta(1)$. As a result, Game~\ref{game:compression_game} essentially asks to find a $k$-dimensional subspace approximately perpendicular to some $\tilde d$-dimensional subspace $F$ sampled uniformly from a constant-radius ball $F\sim\Unif(B(F_0,\rho))$. This naturally requires $O(k\tilde d)$ information; hence if the compression is successful with good probability, the message $\texttt{Message}$ must contain $\tilde\Omega(k\tilde d)$ information about the considered subspace. This gives the following result.

\comment{
When the linear map $\Tcal$ is not the identity, the posterior of $E^{(i)}$ may not be exactly uniform over $B(E^{(i-1)},\delta^{(i-1)})$. However, it remains approximately uniform provided that the linear map $\Tcal(E)$ does not significantly distort the distribution of the Haar uniform $E$ at the scale of the considered ball $\delta^{(i-1)}$. This gives the following result.
}

\begin{restatable}{proposition}{PropHardnessGame}
\label{prop:hardness_compression_game}
    There is a universal constant $C>0$ such that the following holds. Suppose that $\delta^{(i-1)}\leq 1/d$, $\eta\geq\delta^{(i-1)}$, and
    \begin{equation*}
        \kappa(\Tcal) \leq \frac{4}{d(\delta^{(i-1)}+\ldots+\delta^{(N)})} \quad \text{and} \quad \frac{\gamma}{\beta} \geq Ck\frac{\eta}{\delta^{(i-1)}}\sqrt{\frac{d}{\tilde d}}.
    \end{equation*}
    Then, any compression strategy for Game~\ref{game:compression_game} that succeeds with probability at least $\xi>0$ must satisfy
    \begin{equation*}
        I(\texttt{Message};E^{(i)}\mid E^{(<i)})\geq \xi \cdot \frac{k\tilde d}{1+ \log(\sqrt d/\gamma)}-\log 2.
    \end{equation*}
\end{restatable}
In words, this result shows that if the memory budget $M'$ is not enough to fully store a solution, $M'\ll k\tilde d$, then the compression cannot succeed with good probability.

\vspace{4pt} \noindent\textbf{Putting things together.} The hardness result from \cref{prop:hardness_compression_game} implies that the second alternative in \cref{prop:reduction_to_compression_game} cannot happen for appropriate parameter choices, in particular $\Delta'\leq c\frac\Delta k\left(\frac{\tilde{d}}{d}\right)^{3/2}$ with the universal constant $c>0$. 
This precisely gives the desired propagation for the blocking property: child layer-$p$ phases satisfying the constraints in \cref{lemma:pigeonhole_principle} inherit the blocking property from their layer-$(p-1)$ parent. This proves \cref{thm:propagation}.

\section{Missing proofs for Section~\ref{sec:proof_main_result}}
\label{sec:missing_proofs}

This section is organized as follows. In \cref{subsec:validity} we give some preliminary geometric properties of the constructed feasibility problem and prove its validity. We then turn to proving the result to propagate the blocking property: in \cref{subsec:proof_pigeonhole} we prove \cref{lemma:pigeonhole_principle} which in particular propagates well-behavedness to child phases. Next, in \cref{subsec:reduction_compression_proof} we prove the reduction to the compression game when the blocking property does not propagate (\cref{prop:reduction_to_compression_game}). In \cref{subsec:proof_hardness_compression}, we prove that the compression game is hard (\cref{prop:hardness_compression_game}). Finally, we put all these pieces together in \cref{subsec:putting_things_together} to prove our final memory/query-complexity result.

\subsection{Validity of the constructed feasibility problem}
\label{subsec:validity}

We first give some preliminary geometric results on our construction. First, although the successive transformations stretch the space, the phase regions remain bounded:

\begin{lemma}\label{lem:boundedness_of_regions-pre}
    Let $P\geq 2$ and let $R_0\subseteq B:=B_d(0,1)\subset \Rbb^d$ satisfy $\inf_{x\in R_0}\|x\|\geq 1/2$. Let $U_p,\tilde{U}_p\in \Gr(k,d)$ and $0<\tilde \delta_p<\delta_p<1/2$ for all $p\in [P]$. For $p\in[P]$, iteratively define
    \begin{equation}\label{eq:lem:boundedness_of_regions-pre-1}
        R_{p}:=T_{U_{p},\delta_{p}^{-1}} (R_{p-1}\cap(\tilde{U}_{p}^\perp +\tilde \delta_p B)).
    \end{equation}
    Suppose $\angle_{\max}(U_p,\tilde{U}_p)\leq \delta_p/(\sqrt{3P})$ and $\tilde \delta_p / \delta_p\leq 1/3$ for all $p\in[P]$. Then, provided $R_p\neq \emptyset$, we have
    \begin{equation*}
        \frac{1}{2}\leq \inf_{x\in R_p}\|x\|\leq \sup_{x\in R_p}\|x\| \leq \sqrt{p+1}.
    \end{equation*}
\end{lemma}
\begin{proof}
    By the recursive definition in \eqref{eq:lem:boundedness_of_regions-pre-1}, we have $R_p\subseteq T_{U_p,\delta_p^{-1}}(R_{p-1})$. Since $\delta_p<1$, we have that $T_{U_p,\delta_p^{-1}}$ is a stretching. Hence
    \[\inf_{x\in R_p}\|x\|\geq \inf\set{\|x\|:x\in T_{U_p,\delta_p^{-1}}(R_{p-1})}\geq \inf_{x\in R_{p-1}}\|x\|\geq \cdots\geq \inf_{x\in R_0}\|x\|\geq \frac12.\]

    Next, we inductively prove the $\sqrt{p+1}$ upper bound. The statement is clearly true for any $R_0\subseteq B$. Suppose $\sup_{x\in R_{p-1}}\|x\|\leq \sqrt{p}$ for some $p\in[P]$, and we want to show $\sup_{x\in R_{p}}\|x\|\leq \sqrt{p+1}$. 
    
    For any $x\in R_{p}$, there exists $y\in \tilde{U}_p ^\perp$ such that $x':=T_{U_{p},\delta_{p}}(x)\in R_{p-1} \cap ( y+\tilde \delta_p B)$. Note that given $x'\in R_{p-1}$, by the induction hypothesis we have $\|x'\|\leq \sqrt{p}\leq\sqrt{P}$.

    By $\|P_{U_p} P_{\tilde{U}_p^\perp}\|_{op} \leq \delta_p /(\sqrt{3P})$, we have $\lambda_{\max}(P_{U_p} P_{\tilde{U}_p^\perp})\leq \delta_p/(\sqrt{3P})$, hence
    $$\|P_{U_p}x'\|\leq \|P_{U_p}(x'-y)\| + \delta_p\|y\|/(\sqrt{3P})\leq \tilde \delta_p + 2\delta_p/3 \leq \delta_p.$$
    The second inequality is from $\|x'-y\|\leq \tilde\delta_p$ and $\|y\|\leq \|x'\|+\tilde \delta_p\leq (1+1/(6\sqrt{2}))\sqrt{P}$.

    Finally, with $x'=T_{U_p,\delta_p}(x)= P_{U_p^\perp} (x)+ \delta_p P_{U_p} (x)$, we have
    \[\|x\|^2= \|P_{U_p} x\|^2 + \|P_{U_p^\perp} x\|^2 = \delta_p ^{-2} \|P_{U_p} x'\|^2 + \|P_{U_p^\perp} x'\|^2\leq 1+ \|x'\|^2\leq p+1.\]
    This concludes the proof for $\sup_{x\in R_{p}}\|x\|\leq \sqrt{p+1}$.
\end{proof}

\begin{corollary}[Boundedness of regions]\label{lem:boundedness_of_regions}
    Let $P\geq 2$ and $d\geq 3$.
    Let $\ell=(i_1,\dots,i_P)\in\Vcal_P$ be a leaf with layer-$p$ ancestor $v_p=(i_1,...,i_p)$ for $p\in[P]$. Then, provided $R_{v_p}\neq\emptyset$, we have
    \begin{equation*}
        \frac{1}{2}\leq \inf_{x\in R_{v_p}}\|x\|\leq \sup_{x\in R_{v_p}}\|x\| \leq \sqrt{p+1}.
    \end{equation*}
\end{corollary}
\begin{proof}
    Let $R_{v_0}:=\set{\bm e^\top x\leq -\frac12}\cap B_d (0,1)$ and for $p\in[P]$, let $R_p:=R_{v_p}$, $U_p:=E_{v_{p-1},p}^{(i_p-1)}$, $\tilde U_p:=E_{v_{p-1},p}^{(i_p)}=E_{v_p}$ and $\delta_p:=\delta_p^{(i_p-1)}$, $\tilde \delta_p:=\delta_p^{(i_p)}$.
    One can check that conditions in \cref{lem:boundedness_of_regions-pre} are satisfied. In particular, by \cref{def:regions_contexts}, we indeed have
    \[R_{v_p}=T_{E_{v_{p-1},p}^{(i_p-1)},1/\delta_p^{(i_p-1)}}(R_{v_{p-1}}\cap (E_{v_{p}}^\perp +\delta_p^{(i_p)}B)),\quad\forall p\in[P].\]
    By $\NoiseAddition$, we also have $\angle_{\max}(E_{v_{p-1},p}^{(i_p-1)}, E_{v_{p-1},p}^{(i_p)})\leq\delta_p^{(i_p-1)}/\sqrt{3P}$.
\end{proof}

Next, we check that with high probability, all transformations simultaneously have appropriately bounded condition numbers.

\begin{lemma}[Condition number bound on transformations]\label{cor:the_uniform_singular_bound}
    Let $P\geq 2$ and $36P\tilde d\leq d$. With at least $1-2Pe^{-d/288}$ probability over $E_1,...,E_P$, simultaneously over all $p\in[P]$ and $v\in\Vcal_{p-1}$, we have 
    \[1\leq \lambda_{\min}(\Tcal_v)\leq \lambda_{\max}(\Tcal_v) \leq \frac{2}{d\delta_p^{(0)}}.\]
    regardless of the randomness used in $\NoiseAddition$ for sampling phases in each layer.
\end{lemma}
The main point is that the condition number $\kappa(\Tcal_v)$ does not grow as the \emph{product} of each successive stretching factor used to define $\Tcal_v$, but is essentially bounded by its \emph{last} stretching factor only---this improves the final exponent in \cref{thm:main_lower_bound_result}. We defer the proof of this technical result to \cref{app:proof_singular_bound}. 

We are now ready to prove the validity of our constructed separation oracle.

\begin{proof}[Proof of \cref{lem:oracle-consistency}]
    Using the same argument as in the proof of Lemma 7 in \cite{blanchard24gradient}, given that $d-\tilde d P \geq \frac{35}{36}d$, for $f:=P_{span(E_p,p\in[P])^\perp}(\bm e)$ we have 
    $$\Pbb\left(\|f\|\leq \frac56\right)\leq\Pbb\left(\|f\|\leq\sqrt{\frac{d-\tilde d P}{d}\left(1-\frac27\right)}\right)\leq e^{-(d-\tilde d P)/49} \leq e^{-d/60}.$$
    Let $\Ecal_1$ denote the event in which $\|f\|>5/6$. Let the $\epsilon$-ball $B_\epsilon$ be centered at $-\frac{3f}{4\|f\|}$, with radius $\epsilon$. By construction, $B_\epsilon$ belongs to $B_d(0,1)$. And under $\Ecal_1$, we have for any $x\in B_\epsilon$, 
    $$\bm e^\top x\leq -\frac{3\bm e^\top f}{4\|f\|} +\epsilon\leq -\frac58 +\epsilon<-\frac12.$$
    Hence with at least $1-e^{-d/60}$ probability, $Q_{E_1,...,E_P}$ does contain a ball of radius $\epsilon$. 

    By \cref{cor:the_uniform_singular_bound}, under an event $\Ecal_2$ on $E_1,...,E_P$ that happens with at least $1-2Pe^{-d/288}$ probability, simultaneously for all $p\in[P]$ and $v\in\Vcal_{p-1}$ we have $\lambda_{\max}(\Tcal_v)\leq 2(\delta_p^{(0)} d)^{-1}\leq 2\epsilon^{-\frac{P-1}{2P}}$. Combining the two events, we have 
    $$\Pbb(\Ecal_1\cap\Ecal_2)\geq 1-e^{-d/60}-2Pe^{-d/288}.$$
    We will assume event $\Ecal_2$ for the rest of the proof.

    In the transform separation oracle in \cref{alg:final_oracle}, a query is either answered by a path separation oracle $\Ocal_\ell$ with some $\ell\in\Vcal_P$, or by some arbitrary oracle that is valid as in line~\ref{line:transform-5}. Hence, it remains to show for each $\Ocal_\ell$ that, whenever it does not answer $\texttt{Success}$, it provides a valid separating hyperplane for $Q_{E_1,...,E_P}$.

    Fix $\ell$ and let $v_p$ be the layer-$p$ ancestor of $\ell$. 
    We first show by induction that, for each $p\in\set{1,2,\dots,P}$, we have $\Tcal_{v_p}(Q_{E_1,...,E_P})\subseteq  R_{v_p}$. Recall \cref{def:regions_contexts} for the definition of the phase regions $R_v$. We will proceed with a convention for the root phase as $R_\emptyset=\set{x\in B_d(0,1): \bm e^\top x\leq -\tfrac12}$. With this convention at hand, for the root phase we clearly have $\Tcal_{\emptyset}(Q_{E_1,...,E_P})=Q_{E_1,...,E_P}\subseteq  R_\emptyset$. Assume $\Tcal_{v_{p-1}}(Q_{E_1,...,E_P})\subseteq R_{v_{p-1}}$; then it suffices to show 
    \begin{equation}\label{eq:lem:oracle-consistency-1}
        \Tcal_{v_{p-1}}(Q_{E_1,...,E_P})\subseteq R_{v_{p-1}}\cap\set{z:\|P_{E_{v_p}}(z)\|\leq\delta_{v_p}}.
    \end{equation}
    Indeed, if this is true, then applying $\Tcal_{v_{p-1}}^{-1}$ on both sides gives $Q_{E_1,...,E_P}\subseteq\Tcal_{v_p}^{-1}( R_{v_p})$. To see why this holds, note that by \cref{lem:boundedness_of_regions}, we have $\|z\|\leq \sqrt{P}$ for any $z\in R_{v_{p-1}}$. In particular, for any $z\in\Tcal_{v_{p-1}}(Q_{E_1,...,E_P})$,
    \[\|P_{E_{v_p}}(z)\|\leq\|P_{\Hcal_{v_{p-1}}(E_p)}(z)\|+\sqrt{P}\angle_{\max}(\Hcal_{v_{p-1}}(E_p),E_{v_p})\]
    To show \eqref{eq:lem:oracle-consistency-1}, we need to bound $\|P_{E_{v_p}}(z)\|$ by $\delta_{v_p}$. For the first term, it is useful to think of 
    $$Q_{E_1,...,E_P}\subseteq\set{\|P_{E_p}(x)\|\leq \epsilon}=E_p^\perp+\epsilon B.$$
    Since $\lambda_{\max}(\Tcal_{v_{p-1}})\leq2\epsilon^{-\frac{P-1}{2P}}$, we have $$\Tcal_{v_{p-1}}(Q_{E_1,...,E_P})\subseteq \Tcal_{v_{p-1}}(E_p^\perp)+\epsilon\cdot 2\epsilon^{-\frac{P-1}{2P}} B=\set{z:\|P_{\Hcal_{v_{p-1}}(E_p)}(z)\|\leq 2\epsilon^{\frac{P+1}{2P}}}$$ 
    where the last inequality follows from the equivalence $\Tcal_{v_{p-1}}(E_p^\perp)=(\Hcal_{v_{p-1}}(E_p))^{\perp}$. Therefore, 
    \[\|P_{E_{v_p}}(z)\|\leq 2\epsilon^{\frac{P+1}{2P}}+\sqrt{P}\angle_{\max}(\Hcal_{v_{p-1}}(E_p),E_{v_p})\leq 
    2\epsilon^{\frac{P+1}{2P}}+\frac{1}{\sqrt{3}(1-d^{-1})}\delta_{v_p}\leq\delta_{v_p}.\]
    The second inequality follows from the definition of $\NoiseAddition$ along with a geometric sum of principal angle differences starting from $\delta_{v_p}$, with ratio $\lambda_p\leq d^{-1}$. In the last inequality we use $\delta_{v_p}\geq d^{-1}\epsilon^{1/2}$ and $d\geq 80 P\log d$.
    So far, we have shown the relation in \eqref{eq:lem:oracle-consistency-1} and by induction, for all $p\in[P]$ we have
    \begin{equation}\label{eq:lem:oracle-consistency-2}
        \Tcal_{v_{p-1}}(Q_{E_1,...,E_P})\subseteq R_{v_{p-1}}\cap\set{z:\|P_{E_{v_p}}(z)\|\leq\delta_{v_p}}\quad\text{ and }\quad\Tcal_{v_p}(Q_{E_1,...,E_P})\subseteq R_{v_p}.
    \end{equation}
    
    Now we are ready to check if $\Ocal_\ell$ returns a valid separating hyperplane for $Q_{E_1,...,E_P}$ whenever it does not return $\texttt{Success}$. If a query $x$ is answered by a gradient using layer $p$ subspace $E_{v_p}$, then $\Tcal_{v_{p-1}}(x)$ belongs to $ R_{v_{p-1}}$ but $\|P_{E_{v_p}}(\Tcal_{v_{p-1}}(x))\|>\delta_{v_p}$. Hence, $g:=P_{E_{v_p}}(\Tcal_{v_{p-1}}(x))$ separates $\Tcal_{v_{p-1}}(x)$ from the set $\set{z:\|P_{E_{v_p}}(z)\|\leq\delta_{v_p}}.$ In turn, $g$ provides a separation of $\Tcal_{v_{p-1}}(x)$ from $\Tcal_{v_{p-1}}(Q_{E_1,...,E_P})$ by \eqref{eq:lem:oracle-consistency-2}. This is a separation in the $\Tcal_{v_{p-1}}$-transformed space. To convert it back to the original space, notice that
    $${(\Tcal_{v_{p-1}}^\top g})^\top (y-x)=g^\top (\Tcal_{v_{p-1}} y-\Tcal_{v_{p-1}} x)<0,\quad\forall y\in Q_{E_1,...,E_P}.$$
    Hence, $\Tcal_{v_{p-1}}^\top g=\Tcal_{v_{p-1}}^{\top} P_{E_{v_p}}(\Tcal_{v_{p-1}}(x))$ provides a valid separation of $x$ from $Q_{E_1,\dots,E_P}$, which matches the choice in line~\ref{eq:alg:hard_oracle_explicit_single-4} of \cref{alg:leaf_oracle}.
\end{proof}

\subsection{Proof of Lemma~\ref{lemma:pigeonhole_principle}}
\label{subsec:proof_pigeonhole}

\LemmaPigeonholePrinciple*

\begin{proof}
    We start by analyzing whether children of $v$ are well-behaved. First, by definition,
    $C_{v\parallel i_p}=(C_{v},E_{v},E_{v,p}^{(<i_p)})$.
    By the chain rule,
    \begin{align*}
        \sum_{w:w=v\parallel i_p} I(E_w;\texttt{Mem}_{v}\mid E_{> p}, C_{w}) &=\sum_{i_p=1}^{N_p} I(E_{v,p}^{(i_p)};\texttt{Mem}_{v}\mid E_{v,p}^{(<i_p)} ,E_{> p}, C_{v},E_v) \\
        &= I(E_{v,p}^{(\leq N_p)};\texttt{Mem}_v\mid E_{> p},C_v,E_v) \leq M.
    \end{align*}
    Recall that $E_{v,p}^{(i_p)}=E_{v\parallel i_p}$ by construction.
    Hence, the set $\Acal_1$ of well-behaved children of $v$ satisfies
    \begin{equation*}
        M \geq \frac{4M}{N_p}(N_p-|\Acal_1|),
    \end{equation*}
    that is, $\Acal_1$ contains at least $3/4$ of all children of $v$. Next,
    \begin{equation}
        \sum_{i_p=1}^{N_p} I(E_{v,p}^{(i_p)};E_v\mid E_{v,p}^{(<i_p)} ,\texttt{Mem}_{v}, C_{v},E_{>p}) = I(E_{v,p}^{(\leq N_p)};E_v\mid \texttt{Mem}_v,C_v,E_{>p}) \overset{(i)}{\leq} I(E_p;E_v\mid \texttt{Mem}_v,C_v,E_{>p}) \label{eq:previous_sum_information}
    \end{equation}
    In $(i)$ we used the data processing inequality (DPI) recalling that (1) $E_{v,p}^{(\leq N_p)}$ are sampled using only $E_p$ and the transformation $\Tcal_v$ which is $C_v$-measurable, and (2) the learner selects the initial memory $\texttt{Mem}_v$ prior to interacting with the realization $E_{v,p}^{(\leq N_p)}$.

    Write $v=(v',i_{p-1})$ where $v'$ is the parent of $v$. Then,
    \begin{align*}
        I(E_p;E_v\mid \texttt{Mem}_v,C_v,E_{>p}) &\leq I(E_p,\texttt{Mem}_v;E_v\mid C_v,E_{>p})\\
        &\overset{(i)}{=} I(\texttt{Mem}_v;E_v\mid C_v,E_{\geq p})\overset{(ii)}{\leq } I(\texttt{Mem}_{v'};E_v\mid C_v,E_{\geq p})\overset{(iii)}{\leq} \frac{4M}{N_{p-1}}.
    \end{align*}
    In $(i)$ we used the fact that $C_v$ and $E_v$ can be generated using $E_1,\ldots,E_{p-1}$ alone, hence $I(E_p,E_v\mid C_v,E_{>p})=0$. In $(ii)$ we used the DPI, noting that the memory state $\texttt{Mem}_v$ can be generated from the initial memory state of its parent $\texttt{Mem}_{v'}$ together with the full context $C_v$, the deeper subspaces $E_{\geq p}$, and internal randomness independent from the sampling of $E_v$---recall that here, we assume the learner is deterministic. 
    Precisely, this internal randomness corresponds to the randomness used to generate the noise-addition process in deeper layers. In $(iii)$ we used the assumption that $v\in\Vcal_{p-1}$ is well-behaved.
\comment{
    In $(iii)$ we used the definition of the context from \cref{def:regions_contexts}. In $(iv)$ we used the fact that the initial memory state $\texttt{Mem}_{v'}$ is generated before interacting with the current phase subspace $E_{v'}$ (proxy for $E_{p-2}$). Hence, $\texttt{Mem}_{v'}$ is independent of $E_{v'}$ given the past context $C_{v'}$, deeper layers $E_{\geq p}$, and the subspaces $E_{v',p-1}^{(\leq N_{p-1})}$ which can be generated using the deeper layer $E_p$ and the transformation $\Tcal_{v'}$ which is $C_{v'}$-measurable.
    In $(v)$ we used the assumption that $v$ is well-behaved. As a remark, if $p=2$, $E_{v'}$ is void, hence we may directly go from $(iii)$ to $(iv)$.
}
    In summary, together with \cref{eq:previous_sum_information}, this gives
    \begin{equation*}
        \sum_{i_p=1}^{N_p} I(E_{v,p}^{(i_p)};E_v\mid E_{v,p}^{(<i_p)} ,\texttt{Mem}_{v}, C_{v},E_{>p}) \leq \frac{4M}{N_{p-1}}.
    \end{equation*}
    Then, the same arguments as for $\Acal_1$ show that the set $\Acal_2$ of children of $v$ satisfying the second information constraint contains at least $3/4$ of all children of $v$. Therefore, $\Acal_1\cap\Acal_2$ contains at least half of the children of $v$, ending the proof.
\end{proof}

\comment{

\begin{proposition}[(Temp)Reduction to Game~\ref{game:compression_game}]
    There is a universal constant $c>0$ such that the following holds. Assume $(3P\ln 3P)^{-1}\geq 6Pe^{-d/288}$. Fix $p\in\{2,\ldots,P\}$, and $\Delta,\Delta'>0$ satisfying {\color{purple}$\lambda_p<\Delta' < \tfrac18 \Delta\leq \tfrac18$}.
    
    $\Delta,\Delta'\geq \lambda_p$ and a well-behaved layer-$(p-1)$ phase $v\in\Vcal_{p-1}$ such that $R_v$ is $\Delta$-$\Block_k(q\mid \texttt{Mem}_v,C_v,E_{>p-1})$.
    Let $w=(v\parallel i_p)$ be a well-behaved children phase satisfying \cref{eq:second_technical_info_constraint}. Then, either 
    \begin{enumerate}
        \item $R_w$ is $\Delta'$-$\Block_k(q-1/P\mid \texttt{Mem}_w,C_w,E_{>p})$,
        \item or there is a strategy to win at Game~\ref{game:compression_game} with some probability $\xi>0$, for dimensions $d,\tilde d,k$, accuracies $\delta^{(\leq N)}:=\delta_p^{(\leq N)}$, index $i:=i_p$, parameters $(\beta,\gamma):=(8\Delta',\frac{\Delta}{2\sqrt{P}})$, and some linear map $\Tcal$ with $\kappa(\Tcal)\leq 2/(d\delta_p^{(0)})$, such that the message satisfies the information constraint
        \begin{equation*}
             I(\texttt{Message};E^{(i)}\mid E^{(<i)})\leq \paren{6P\ln(6P)\paren{\frac{4M}{N_p}\paren{1+\frac{8k}{N_{p-1}}}+5k }+2k\log(6Pk)+2Pk^2 \log d } \xi.
        \end{equation*}
    \end{enumerate}
\end{proposition}
}

\subsection{Reduction to the compression game: Proof of Proposition~\ref{prop:reduction_to_compression_game}}
\label{subsec:reduction_compression_proof}

In this section, we prove \cref{prop:reduction_to_compression_game}, which reduces the blocking property propagation to the compression Game~\ref{game:compression_game}. To this end, we suppose that all assumptions from \cref{prop:reduction_to_compression_game} hold for some fixed layer $p\in\{2,\ldots,P\}$ and layer-$p$ phase $w=(v\parallel i_p)$.
Specifically, we suppose that $R_w$ is not $\Delta'$-$\Block_k(q-P^{-1}\mid \texttt{Mem}_w,C_w,E_{>p})$ and aim to construct a good encoder/decoder strategy for the compression Game~\ref{game:compression_game}. For simplicity, throughout this section, we use the shorthands 
\begin{equation*}
    i:= i_p,\quad T_{i-1}:=T_{E_{v,p}^{(i-1)},1/\delta_p^{(i-1)}},\quad \widetilde C_v:=(E_{v,p}^{(<i)},C_v,E_{>p}),\quad \text{and} \quad B:=B_d(0,1).
\end{equation*}
In particular, note that by construction, one has
    \begin{equation*}\label{eq:decomposition_R_w}
        R_w=T_{i-1}\set{R_v\cap(E_w^\perp+\delta_p^{(i)} B)}.
    \end{equation*}
We recall that the algorithm is fixed throughout, hence, all variables arising during the learning procedure form a fixed stochastic process, referred to as $\Scal$.
Before describing the encoder/decoder strategy for Game~\ref{game:compression_game}, we first introduce a few preliminaries that will be useful throughout the construction.

\begin{lemma}\label{lemma:construction_W_P}
    If $R_w$ is not $\Delta'$-$\Block_k(q-P^{-1}\mid \texttt{Mem}_{w},C_{w},E_{>p})$, then there is a $(E_v,\texttt{Mem}_v,\widetilde C_v)$-measurable random subspace $W_p\in\Gr(k,d)$ such that
    \begin{equation}%
        \Pbb (R_w\cap (W_p +\Delta' B) =\emptyset)<q-P^{-1}.
    \end{equation}
\end{lemma}
\begin{proof}
    By construction of $C_w=(E_v,E_{v,p}^{(<i)},C_v)$, the claim is equivalent to $R_w$ not being $\Delta'$-$\Block_k(q-P^{-1}\mid \texttt{Mem}_{v},C_{w},E_{>p})$. We prove it by contrapositive.
    Suppose that $R_w$ is $\Delta'$-$\Block_k(q-P^{-1}\mid \texttt{Mem}_{v},C_{w},E_{>p})$. Note that $R_v$ and $T_{i-1}$ are already $C_w$-measurable, hence the only additional randomness in $R_w$ comes from $E_w$. However, $\texttt{Mem}_w$ and $E_w$ are conditionally independent given $\texttt{Mem}_v,C_w,E_{>p}$. Indeed, as we assume the learner is deterministic, the memory state $\texttt{Mem}_w$ can be generated from the initial memory state of its parent $\texttt{Mem}_{v}$ together with the full context $C_v$, the deeper subspaces $E_{> p}$, and internal randomness used for $\NoiseAddition$ calls in deeper layers. This proves that $R_w$ is also $\Delta'$-$\Block_k(q-P^{-1}\mid \texttt{Mem}_w,C_w,E_{>p})$.
\end{proof}
From now on, we fix a mapping $W_p(E_v,\texttt{Mem}_v,\widetilde C_v)$ satisfying \cref{lemma:construction_W_P} throughout the proof.

    Second, our decoding strategy will involve recovering i.i.d.\ samples from the conditional distribution
    \begin{equation}
        \label{eq:target_distribution}
        P_\Scal(E_{v}|E_w,\texttt{Mem}_{v},\widetilde C_v)
    \end{equation}
    Importantly, in the decoding part of Game~\ref{game:compression_game}, $E_w$ will not be available to the decoder. However, the decoder will have access to the distribution $P_\Scal(E_{v}|\texttt{Mem}_{v},\widetilde C_v)$. Using the one-shot channel simulation result from \cite{Li2017ContinuousCommu}, the decoder can still recover i.i.d.\ samples from \eqref{eq:target_distribution} using extra information.  
    
    \begin{lemma}\label{lemma:construct_additional_message}
        A decoder can reconstruct an i.i.d. sample from $P_\Scal(E_{v}|E_w,\texttt{Mem}_{v},\widetilde C_v)$, having access to $(\texttt{Mem}_{v},\widetilde C_v)$, some shared randomness, and an additional message $\texttt{Mes}\in\Nbb$ such that
        \begin{equation*}
            \Ebb[\length(\texttt{Mes})] \leq \frac{32M}{N_pN_{p-1}}+6.
        \end{equation*}
    \end{lemma}
    \begin{proof}
        From \cref{thm:entropy_of_Mj}, conditional on $\texttt{Mem}_{v},\widetilde C_v$, the message only requires $2I(E_v;E_w \mid \texttt{Mem}_{v},\widetilde C_v)+6$ bits on average, bounding its average length. Since by assumption, $w$ satisfies \cref{eq:second_technical_info_constraint}, this gives the desired expected length bound.
    \end{proof}

    From now on, we denote the $j$-th i.i.d.\ sample from \eqref{eq:target_distribution} by $E_{vj}$ and its corresponding message, as given by \cref{lemma:construct_additional_message}, by $\texttt{Mes}_j$.
    Finally, for every subspace $W\in \Gr(k,d)$, we fix \emph{a priori} a $\Delta'$-net $\Sigma_W$ for $W\cap \sqrt{2P}\cdot B$ of size at most $(3\sqrt{2P}/ \Delta')^{k}$. For every $x\in\Sigma_W$ let $[x]_W$ denote its encoding given $W$, which can therefore be encoded in at most $k \log (3\sqrt{2P}/\Delta')$ bits.

    We are now ready to construct the encoding/decoding strategy for Game~\ref{game:compression_game}. For now, we consider that the linear map $\Tcal$ may be chosen by the learner at the very beginning of the game:

    \begin{assumption}\label{ass:Tcal_chosen}
        We consider a slight modification of Game~\ref{game:compression_game} in which at the very beginning (before nature samples $E,E^{(\leq N)}$), the encoder may select the linear map $\Tcal$ to be used in the game.
    \end{assumption}

    \paragraph{Encoding scheme.}
    \begin{enumerate}
        \item\label{item:selecting_Tcal} Sample $(\Tcal_v,C_v,E_{>p})$ from $P_\Scal(\Tcal_v,C_v,E_{>p})$ and let $\Hcal_v=\Tcal_v^{-\top}$. If $\kappa(\Hcal_v)> \tfrac{2}{d\delta_p^{(0)}}$, the compression fails. Otherwise, select $\Tcal:=\Hcal_v$ and continue. 
        
        \item\label{item:encoding-1} Observe $E ^{(\leq i)}$ and use those as $E_{v,p}^{(\leq i)}$ in the process $\Scal$. In particular, $E_w:=E^{(i)}$. This also defines $\widetilde C_v$. Sample $\texttt{Mem}_{v}$ from
        $P_\Scal(\texttt{Mem}_{v}| E_w,\widetilde C_v)$ using private randomness. Next, check whether some event $\Ecal$ happened, where $\Ecal$ depends only on $(\texttt{Mem}_{v}, E_w,\widetilde C_v)$ and will be specified later. If $\Ecal$ did not happen, the compression fails and ends.
        
        \item\label{item:encoding-2} For any $j\geq 1$, sample $E_{vj}$ from \eqref{eq:target_distribution} along with its additional description $\texttt{Mes}_j${ \color{blue} using fresh randomness $\Rcal$.} Construct the region $R_{vj}$ using $E_{vj}$ and $C_v$ as in \cref{def:regions_contexts}.
        
        \item\label{item:encoding-3} Define $W_{pj}:=W_p(E_{vj},\texttt{Mem}_{v},\widetilde C_v)$ and $R_{wj} :=T_{i-1}\set{R_{vj}\cap(E_w^\perp+\delta_p^{(i)} B)}$. These correspond to the variables $(W_p,R_w)$ had the subspace $E_v$ been $E_{vj}$.
        
        Let $S=\set{j:R_{wj}\cap (W_{pj}+\Delta'B)\neq\emptyset}$; we refer to these as \emph{successful} samples. 
        By construction, for each successful sample $j\in S$, there is some $z_j\in R_{wj}\cap (W_{pj}+\Delta' B)$. Then, let $x_j$ be the nearest neighbor of $z_j$ within the $\Delta'$-net $\Sigma_{W_{pj}}$: this will serve as a compressed vector close to $R_{wj}$.
        
        \item\label{item:encoding-4} We aim to find a sequence of $k$ robustly linearly independent successful vectors $x_j$. To do so, initialize $J=\emptyset$. Loop through $j\in S$ by increasing index, and add $j$ to $J$ if $R_{wj}\cap (\Span(x\in J)+\Delta B)=\emptyset$. Repeat until $|J|=k$. The compression fails if this target is never reached.
        \item\label{item:encoding-5} If the compression failed, select $\texttt{Message}:=\texttt{Mem}_v$, otherwise select $$\texttt{Message}:=(\texttt{Mem}_{v},(j,\texttt{Mes}_j,[x_j]_{W_{pj}})_{j\in J},{\color{blue}C_v,E_{>p},\Rcal}).$$
    \end{enumerate}

    \paragraph{Decoding scheme.}
    \begin{enumerate}
        \item 
        {\color{blue} Receive $\texttt{Message}.$}
        {\color{blue} Recover $(C_v,E_{>p})$ from \texttt{Message}.}
        Observe $E^{(<i)}$ and use these as $E_{v,p}^{(<i)}$. This gives $\widetilde C_v$.
        
        \item For each $j\in J$, recover $E_{vj}$ using $\texttt{Mes}_j$, $(\texttt{Mem}_{v},\widetilde C_v)$, and {\color{blue}$\Rcal$.}
        Then, recover $W_{pj}=W_p(E_{vj},\texttt{Mem}_{v},\widetilde C_v)$. Given $W_{pj}$ and the encoding $[x_j]_{W_{pj}}$, compute $x_j$.
        
        \item Output $\set{x_j,j\in J}$.
    \end{enumerate}

    We provide some intuitions on this compression strategy. We recall that the goal of the decoder is to output robustly linearly independent vectors that are approximately within $(\tilde E^{(i)})^\perp=T_{i-1}((E^{(i)})^\perp)$. Here, the player selects $\Tcal=\Hcal_v$ so that $E^{(i)}$ is generated exactly as $E_w$. Therefore, vectors in $R_w=T_{i-1}\{R_v\cap (E_w^\perp + \delta_p^{(i)}B)\}$ are natural solution candidates, since they are almost perpendicular to $\tilde E^{(i)}$---i.e., close to $T_{i-1}((E^{(i)})^\perp)$---by construction. It remains to see how we can find $k$ robustly independent solutions using this observation. First, note that vectors in $R_w$ are also within $T_{i-1}(R_v)$. The main idea is to use a resampling strategy: we consider several potential choices of $R_v$---denoted $R_{vj}$---leading to several potential values of $R_w$---denoted $R_{wj}$. Since $R_v$ has the blocking property, eventually this will yield robustly independent solution vectors.

    More precisely, conditioned on $(\texttt{Mem}_v,\widetilde C_v)$, the encoder samples different $R_{vj}$'s, in turn, defining different $R_{wj}$'s. Since we assumed that $R_w$ does not have the blocking property, for each sample $j$, there is a good chance of finding a relatively low-dimensional subspace $W_{pj}$ that cuts $R_{wj}$ at distance $\Delta'$, using only information from $(\texttt{Mem}_{v},\widetilde C_v)$. If we pick a vector $x_j$ from the intersection of this cut, it (1) would be in $R_{wj}$ and hence sufficiently perpendicular to $\tilde E^{(i)}$ as discussed above and (2) could be compressed using $O(k\log P)$ bits using the net $\Sigma_{W_{pj}}$ of $W_{pj}$.

    The encoder then accepts the vectors $x_j$ when these explore a new direction sufficiently far from previous solution vectors $\Span(x_j:j\in J)$, ensuring these are robustly independent. Intuitively, this is doable because we assumed $R_{v}$ has the blocking property given the information in $(E_p^{(\leq N_p)},\texttt{Mem}_{v},C_v,E_{>p})$. This blocking property controls intersections of $R_v$ with subspaces of dimension less than $k$. When resampling several $R_{vj}$'s, this ensures that $x_j$ is not in the span of any $\leq k$ previous solution vectors with good probability.

\comment{
    As a technical remark, this argument requires the blocking property of $R_{v}$ in the $T_{i-1}$-transformed space which is slightly different from the blocking property of $R_v$; in fact, the blocking property does not directly hold for $T_{i-1}(R_v)$. However, we will show that the blocking property does hold for the restricted region $T_{i-1}(R_{v})\cap T_{i-1}(E_w^\perp+\delta_p^{(i)} B)$, which will be sufficient for our purposes since it is contained within $R_w$.

    \todo{This paragraph is not super clear... will come back to this later.
    
    "Finally, as a reminder, we actually need the blocking property of $R_{v}$ in the $T_{i-1}$-transformed space, which does not simply hold for $T_{i-1}(R_{v})$. Hence, our proof considers a restricted region---when we condition on $(E_{v,p}^{(\leq i)},\texttt{Mem}_{v},C_v,E_{>p})$, the region $T_{i-1}({E_{v,p}^{(i)}}^\perp+\delta_p^{(i)} B)$ is fixed, and the blocking property does hold for the restricted region $T_{i-1}(R_{v})\cap T_{i-1}({E_{v,p}^{(i)}}^\perp+\delta_p^{(i)} B)$."}

}

    \comment{
    In the following, we will first focus on analyzing the behavior of this encoding-decoding strategy when the input $\Tcal$ follows the exact same distribution as $\Tcal_v$ in $\Scal$. 
    \begin{assumption}\label{ass:Tcal_is_Tcal_v}
        We assume the input $\Tcal$ of \cref{game:compression_game} follows the same distribution as $\Tcal_v$ in $\Scal$. 
    \end{assumption}
    }

    \paragraph{Validity of the compression scheme.} We start by checking that the simulation procedure worked as intended: the procedure from the encoding scheme indeed emulates the original process $\Scal$.

    \begin{lemma}\label{lemma:validity_scheme}
        All variables constructed within the encoder match their counterparts in $\Scal$ in distribution. In particular, selecting $E_{v,p}^{(\leq i)}$ as $E^{(\leq i)}$ was valid. Also, $(\tilde E^{(i)})^\perp = T_{i-1}(E_w^\perp)$
    \end{lemma}
    \begin{proof}
        By construction, $(\Tcal_v,C_v,E_{>p})$ are sampled as in the original process $\Scal$. More precisely, $v\in\Vcal_{p-1}$ is a layer-$(p-1)$ phase, hence $(\Tcal_v,C_v)$ can be generated using $E_{\leq p-1}$ only. As a result, $(\Tcal_v,C_v,E_{>p})$ are fully independent of $E_p$.
        Next, by construction of the skeleton tree (see \cref{alg:construct_tree_structure}), the subspaces $E_{v,p}^{(\leq N_p)}$ are sampled using only $\Hcal_v=\Tcal_v^{-\top}$, $E_p$, and fresh randomness. We recall from the above discussion that conditionally on the sampled variables $(\Tcal_v,C_v,E_{>p})$, the layer-$p$ subspace is still distributed as $E_p\sim\mathrm{Unif}(\Gr(\tilde d,d))$. In summary, since the map $\Tcal$ was chosen as $\Hcal_v$, the sampling of $E^{(\leq N_p)}$ in Game~\ref{game:compression_game} exactly matches that of $E_{v,p}^{(\leq N_p)}$.
        For all other variables, the encoder uses the same sampling distributions as in the original process $\Scal$. This proves the first claim.

        For the second claim, we have
        \begin{equation*}
            (\tilde E^{(i)})^\perp = (T_{E^{(i-1)},\delta_p^{(i-1)}} (E^{(i)}))^\perp = T_{E^{(i-1)},1/\delta_p^{(i-1)}} ((E^{(i)})^\perp) = T_{i-1}(E_w^\perp).
        \end{equation*}
        This ends the proof.
    \end{proof}

    Next, we check that provided the encoder did not fail to generate its message, then the compression would be successful.
    \begin{lemma}\label{lemma:valid_compression_success}
        Suppose that the encoder did not fail in either step~\ref{item:encoding-1} or step~\ref{item:encoding-4} of the encoding scheme. Then, the compression is successful, i.e., the decoder returns vectors $x_1,\ldots,x_k$ such that for all $t\in[k]$, 
        $\|P_{\tilde E^{(i)}}(x_t)\| \leq 8\Delta' \|x_t\|$ and $\|P_{\Span(x_{t'}:t'<t)^\perp}(x_t)\| \geq \frac{\Delta}{2\sqrt P}  \|x_t\|$.
    \end{lemma}
    \begin{proof}
        By construction of the decoder, the output $\{x_j:j\in J\}$ is exactly the same as in the encoding scheme. Fix $j\in J$ and denote by $S_{<j}$ the span of the previous solution vectors. First, $x_j$ is the compression of $z_j\in R_{wj}\cap (W_{pj}+\Delta'B)$, hence
        \begin{equation*}
            d(x_j,R_{wj}) \leq d(x_j,z_j) \leq \Delta' + d(x_j,P_{W_{pj}}(z_j)) \leq 2\Delta'.
        \end{equation*}
        In the last inequality, we used the definition of $\Sigma_{W_{pj}}$ as a $\Delta'$-net of $W_{pj}\cap \sqrt{2P} \cdot B$, and the fact that $z_j\in R_{wj}\subseteq \sqrt {2P}\cdot B$ from \cref{lem:boundedness_of_regions}.
        Next, by construction, $j$ was selected so that $R_{wj}\cap (S_{<j}+\Delta B)=\emptyset$. However, $z_j\in R_{wj}$, therefore the previous identity gives
        \begin{equation*}
            \|P_{S_{<j}^\perp}(x_t)\| = d(x_j,S_{<j}) \geq d(z_j,S_{<j})-d(x_j,z_j) \geq \Delta-2\Delta' \geq \frac{3\Delta}{4} \geq \frac{\Delta}{2\sqrt{P}} \|x_t\|.
        \end{equation*}
        In the last inequality we used the fact that $x_t\in \Sigma_{W_{pj}}$ which is contained within $\sqrt{2P}\cdot B$.

        We next turn to the first identity:
        \begin{equation*}
            \|P_{\tilde E^{(i)}} (x_j)\|=d(x_j,(\tilde E^{(i)})^\perp) \overset{(i)}{=} d(x_j,T_{i-1}(E_w^\perp))\leq d(x_j,T_{i-1}(E_w^\perp+\delta_p^{(i)} B))+\frac{\delta_p^{(i)}}{\delta_p^{(i-1)}} \leq d(x_j,R_{wj})+\lambda_p \overset{(i)}{\leq} 3\Delta'.
        \end{equation*}
        In $(i)$ we used \cref{lemma:validity_scheme} and in $(ii)$ we used the previous bound $d(x_j,R_{wj})\leq 2\Delta'$ and $\lambda_p\leq\Delta'$. We conclude by noting that $\|x_j\|\geq \|z_j\|-\Delta' \geq 1/2-\Delta' \geq 3/8$ where we used $z_j\in R_{wj}$ and \cref{lem:boundedness_of_regions}. This proves $\|P_{\tilde E^{(i)}} (x_j)\| \leq 8\Delta'\|x_j\|$ as desired.
    \end{proof}

\paragraph{Bounding the required sample size for the encoding scheme.}
Let $j_{k}$ denote the total number of $E_{vj}$ samples until getting $k$ indices in $J$ at step \ref{item:encoding-4} of the encoding scheme. We aim to bound its expected size.
Starting from step~\ref{item:encoding-2} of the encoding scheme, the encoder conditions on 
\begin{equation*}
    \texttt{Context}:=(\texttt{Mem}_{v},E_w,\widetilde C_v)
\end{equation*}
and repeatedly samples i.i.d.\ copies $(R_{vj},W_{pj},R_{wj})\overset{iid}{\sim}P_\Scal(R_v,W_p,R_w\mid \texttt{Context})$. We reason conditionally on $\texttt{Context}$ in the following argument and aim to lower bound the probability that a new sample $j$ gets accepted within $J$.

To do so, we first analyze the distribution $P_\Scal(R_v,W_p,R_w\mid \texttt{Context})$. We start by considering the event that the encoding scheme is successful in both steps~\ref{item:selecting_Tcal} and \ref{item:encoding-3} of the encoding scheme:
$$\Ecal_{\text{cut}}:= \set{R_{w}\cap (W_p+\Delta'B)\neq \emptyset}\cap \set{\kappa(\Hcal_v)\leq 2(d\delta_p^{(0)})^{-1}}.$$
The first event measures the probability of $W_p$ successfully cutting $R_{w}$ at $\Delta'$ distance. 
It follows from \cref{lemma:construction_W_P} that this first event has probability at least $1-q+P^{-1}$. 

On the other hand, since $R_v$ has the blocking property and $R_w\subseteq T_{i-1}(R_v)$, we will show that for any subspace $W\in\Gr(k,d)$ that is $\texttt{Context}$-measurable, with good probability we simultaneously have (1) $R_{w} \cap (W+\Delta B)=\emptyset$ and (2) $\Ecal_{\text{cut}}$ holds.
This is formalized below.

\comment{

When we condition on $\text{Context}$, the region $T_{i-1}({E_{v,p}^{(i)}}^\perp+\delta_p^{(i)} B)$ is fixed, and roughly speaking, a copy gets accepted when 1. $W_p$ successfully cuts $R_{w}$ and 2. $T_{i-1}(R_{v})$ has the blocking property against $span(x\in J)$ when restricted to $T_{i-1}({E_{v,p}^{(i)}}^\perp+\delta_p^{(i)} B)$.

We will show something stronger for the second condition. In fact, for any $W$ that potentially depends on $(\text{Context})$,
$$R_{w} \cap (W(\text{Context})+\Delta B)=\emptyset\quad\text{ happens with a good chance.}$$
Note that $R_{w}= T_{i-1}(R_{v})\cap T_{i-1}({E_{v,p}^{(i)}}^\perp+\delta_p^{(i)} B)$. The following lemma lower bounds the overall probability of this happening. 
}

\begin{lemma}\label{lem:R1_RLIS_implies_R2win_RLIS}
    If $R_{v}$ is $\Delta$-$\Block_{k}(q|\texttt{Mem}_v,C_v,E_{>p-1})$, then for any $\texttt{Context}$-measurable subspace $W\in\Gr(k,d)$,
    $$\Pbb (R_{w} \cap (W+\Delta B)=\emptyset,\Ecal_{\text{cut}})> (2P)^{-1}.$$
\end{lemma}
\begin{proof}
    First, we show that $R_v$ is $\Delta$-$\Block_{k}(q|\texttt{Context})$. Recall that by definition, $\texttt{Context}:=(\texttt{Mem}_{v},E_w,\widetilde C_v) = (\texttt{Mem}_{v},E_{v,p}^{(\leq i)},C_v,E_{>p})$. Hence, compared to $(\texttt{Mem}_v,C_v,E_{>p-1})$, the context \texttt{Context} may additionally depend on the fresh randomness from $\NoiseAddition$ when generating $E_{v,p}^{(\leq N_p)}$ from $E_p$---recall that $\Tcal_v$ is already $C_v$-measurable. In particular, this fresh randomness is independent of $R_v$. In summary, $R_v$ is also $\Delta$-$\Block_{k}(q|\texttt{Context})$. 
    
    As a result, for the considered \texttt{Context}-measurable subspace $W$, since $T_{i-1}$ is \texttt{Context}-measurable as well, we have
    \begin{align*}
        q \leq \Pbb(R_v \cap (T_{i-1}^{-1} (W)+\Delta B)=\emptyset) & \overset{(i)}{=} \Pbb(T_{i-1}(R_v) \cap (W+\Delta T_{i-1}(B))=\emptyset)\\
        &\overset{(ii)}{\leq} \Pbb(R_w \cap (W+\Delta B)=\emptyset).
    \end{align*}
    In $(i)$ we applied the transformation $T_{i-1}$. In $(ii)$ we used the fact $B\subseteq T_{i-1}(B)$ since $T_{i-1}$ is obtained by a succession of stretchings of the space, and $R_{w} \subseteq T_{i-1}(R_{v})$ from \eqref{eq:decomposition_R_w}.
    On the other hand, recall that from \cref{lemma:construction_W_P} we have $\Pbb (R_v\cap (W_p+\Delta'B)=\emptyset)< q-P^{-1}$. Combining these two inequalities yields
    \begin{align*}
        \Pbb(R_w \cap (W+\Delta B)=\emptyset,\Ecal_{cut}) &\geq \Pbb(R_w \cap (W+\Delta B)=\emptyset) -\Pbb (\Ecal_{\text{cut}}^c) \\
        &\overset{(i)}{>} q - (q-P^{-1}) - 2Pe^{-d/288} > P^{-1} - 2Pe^{-d/288} \overset{(i)}{>} (2P)^{-1}.
    \end{align*}
    In $(i)$ we also used \cref{cor:the_uniform_singular_bound} to bound the probability that $\kappa(\Hcal_v)> 2(d\delta_p^{(0)})^{-1}$. In $(ii)$ we took $d$ sufficiently large, recalling that $P\leq d$.
    This ends the proof. 
\end{proof}

Since we assumed $R_{v}$ is $\Delta$-$\Block_{k}(q|\texttt{Mem}_v,C_v,E_{>p-1})$, \cref{lem:R1_RLIS_implies_R2win_RLIS} precisely shows that
\begin{equation}\label{eq:lower_bound_success_proba}
    \Ebb[P(\texttt{Context})]>\frac{1}{2P}\quad\text{where}\quad P(\texttt{Context}):=\min_{W\in \Gr(k,d)}\Pbb (R_{w} \cap (W +\Delta B)=\emptyset,\Ecal_{\text{cut}}|\texttt{Context}).
\end{equation}
This result is useful because $P(\texttt{Context})$ precisely lower-bounds the probability that a new sample $j$ gets accepted within $J$ throughout steps~\ref{item:encoding-2}, \ref{item:encoding-3}, and \ref{item:encoding-4} of the encoding scheme.

\begin{lemma}\label{lemma:lower_bound_proba_accept}
    Consider a stage $j$ of the encoding scheme such that after sampling $(E_{vj'},W_{pj'},R_{wj'})_{j'<j}$, the collection $J$ has at most $|J|< k$ accepted samples. Then, conditionally on these past samples and the complete past history, $j$ is accepted within $J$ with probability at least $P(\texttt{Context})$. In particular,
    \begin{equation*}
        \Ebb[j_k\mid \texttt{Context}] \leq \frac{k}{P(\texttt{context})}.
    \end{equation*}
\end{lemma}

\begin{proof}
    Consider a stage $j$ of the encoding process such that at the beginning of this round there are $l<k$ accepted samples: $J=\{j_1,\ldots,j_l\}$. Since $(R_{vj},W_{pj},R_{wj})$ is a fresh sample of $P_\Scal(R_v,W_p,R_w\mid\texttt{Context})$ and $\Span(x_i:i\in J)$ has dimension at most $l<k$, we precisely have
    \begin{equation*}
        \Pbb (R_{wj} \cap (\Span(x_i:i\in J) +\Delta B)=\emptyset,\Ecal_{\text{cut},j}|\texttt{Context}) \geq P(\texttt{Context}).
    \end{equation*}
    Here, we used the notation $\Ecal_{\text{cut},j}$ for the corresponding event for the variables $R_{vj}$ and $W_{pj}$. This exactly gives the desired result: $j$ is accepted within $J$ precisely when it is successful (see step~\ref{item:encoding-3})---i.e., $\Ecal_{\text{cut},j}$ holds---and when $R_{wj} \cap (\Span(x_i:i\in J) +\Delta B)=\emptyset$ (see step~\ref{item:encoding-4}).

    In particular, conditional on the past history, the waiting time for a new accepted sample is stochastically dominated by a geometric random variable $\Gcal(P(\texttt{Context}))$, which has expected value $1/P(\texttt{Context})$. Since $j_k$ is the index after $k$ successful vectors, this directly gives the desired expected bound on $j_k$.
\end{proof}

\cref{lemma:lower_bound_proba_accept} shows that the length of the encoding depends inversely on $P(\texttt{Context})$. To avoid bad tail probabilities when this probability is very small, we introduce the thresholding event
\begin{equation*}
    \Ecal:=\set{P(\texttt{Context}) \geq p^\star},
\end{equation*}
for some fixed probability threshold $p^\star\in(0,1]$ selected such that
$\Pbb[\Ecal] \geq (6Pp^\star\ln(6P))^{-1} $. 
This is possible as guaranteed by \cref{lem:expected_inverse} since $\Ebb[P(\texttt{Context})]>(2P)^{-1}$ from \cref{eq:lower_bound_success_proba}. In summary, we obtain the following result:
\begin{lemma}\label{lem:bounding_number_of_samples_in_compression_scheme}
    $\Ecal$ is $\texttt{Context}$-measurable and satisfies $p^\star\Pbb[\Ecal] \geq (6P\ln(6P))^{-1}$ for some $p^\star\geq(6P\ln(6P))^{-1}$. 
    Finally, conditionally on $\Ecal$, the compression succeeds almost surely.
\end{lemma}
\begin{proof}
    It only remains to check the last claim. By \cref{lemma:lower_bound_proba_accept},
    \begin{equation}\label{eq:conditional_expected_num_jk}
        \Ebb[j_k\mid\Ecal]\leq \Ebb_{\texttt{Context}\mid\Ecal}[\Ebb[j_k\mid\texttt{Context}]]\leq \frac{k}{p^\star}\leq 6kP\ln(6P)<\infty,
    \end{equation} 
    almost surely conditionally on $\Ecal$, the encoding scheme does not fail. Hence, the encoding scheme does not fail with probability at least $\Pbb[\Ecal]$. By \cref{lemma:valid_compression_success}, this shows that the compression succeeds with the same probability.
\end{proof}

\paragraph{Information bound on the $\texttt{Message}$.} We have already shown the desired success probability in \cref{lem:bounding_number_of_samples_in_compression_scheme}. To complete the proof it still remains to bound the information between $\texttt{Message}$ and $E_w=E^{(i)}$, conditionally on the previous subspaces $E^{(<i)}$ and the transformation $\Tcal$. This is done by combining the bound on the number of needed samples in \cref{lem:bounding_number_of_samples_in_compression_scheme} together with the information bounds from \cref{lemma:construct_additional_message} and the well-behavedness of phase $v$.

\begin{lemma}\label{lemma:information_upper_bound_for_game}
    We have
    \begin{equation*}\label{eq:information_final_bound}
    I(E^{(i)};\texttt{Message}| E^{(<i)},\Tcal) \leq \paren{6P\ln(6P)\paren{\frac{4M}{N_p}\paren{1+\frac{8k}{N_{p-1}}}+6k }+4k\log(6Pk)+2Pk^2 \log d }\Pbb(\Ecal)
\end{equation*}
\end{lemma}
\begin{proof}
We start by bounding the encoding lengths for the additional messages $\texttt{Mes}_j$. By Wald's identity for stopping times, we have 
\begin{align}  
    \Ebb\left[\sum_{j\in J}\text{length}(\texttt{Mes}_j)\mid \Ecal\right]&\leq \Ebb\left[\sum_{j=1}^{j_{k}}\text{length}(\texttt{Mes}_j) \mid \Ecal\right] \notag\\
    &\leq \Ebb_{\text{Context}|\Ecal}\left[\Ebb[j_k\mid\text{Context}] \cdot \Ebb[\length(\texttt{Mes}_1)|\text{Context}]\right] \notag\\
    &\leq \Ebb_{\text{Context}|\Ecal}\left[\frac{k}{P(\text{Context})} \cdot \Ebb[\length(\texttt{Mes}_1)|\text{Context}]\right] \notag\\
    &\leq \frac{k}{p^\star}\Ebb\left[\text{length}(\texttt{Mes}_j) \mid \Ecal\right] \notag\\
    &\leq \frac{k\Ebb[\length(\texttt{Mes}_1)]}{p^\star\Pbb[\Ecal]} \leq 6Pk\ln(6P)\paren{\frac{32M}{N_pN_{p-1}}+6}. \label{eq:lem:pass_on_RLIS-Mj}
\end{align}
The third inequality follows from \cref{lemma:lower_bound_proba_accept}. In the last inequality we used \cref{lem:bounding_number_of_samples_in_compression_scheme} to lower bound $\Pbb[\Ecal]$ and we used \cref{lemma:construct_additional_message} to bound the expected length of the additional message. 
\comment{
We want to bound $I(E_w;\texttt{Message} | E_w,\Tcal_{v})$, which denotes the information of $\texttt{Message}$ on $E_w$. 

We will use event $\Ecal$ for encoder as the event in \cref{lem:bounding_number_of_samples_in_compression_scheme}. Under event $\Ecal$, we have $\Ebb[j_{k}|\Ecal]<\infty$ by the same Lemma. Hence almost surely the encoder is not trapped in the loop in step \ref{item:encoding-4} under this event. Therefore, the decoder outputs the desired set of vectors with probability at least $\xi:=\Pbb(\Ecal)\geq (3P\ln(3P))^{-1}$.

By the result on one-shot channel simulation in \cite{Li2017ContinuousCommu}, we have the binary encoding length for each fixed $j$, 
$$\Ebb(\text{length}(M_j))\leq 2I(E_w;E_{v}|E_{v,p}^{(<i)},\texttt{Mem}_{v},C_v,E_{>p}) +5\leq \frac{32M}{N_p N_{p-1}}+5.$$
The last inequality follows from our assumption that phase $w$ satisfies \cref{eq:second_technical_info_constraint}. \cref{lem:bounding_number_of_samples_in_compression_scheme} implies $\Ebb[j_{k}|\Ecal]\leq\Pbb(\Ecal)\cdot 3Pk\ln(3P)$. 

By Wald's identity for stopping times, we have 
\begin{equation}\label{eq:lem:pass_on_RLIS-Mj}   \Ebb\left[\left.\sum_{j\in J}\text{length}(M_j)\right|\Ecal\right]\leq \Ebb\left[\left.\sum_{j=1}^{j_{k}}\text{length}(M_j)\right|\Ecal\right]\leq \Pbb(\Ecal)\cdot 3Pk\ln(3P)\cdot \Ebb[\text{length}(M_1)|\Ecal]\leq 3Pk\ln(3P) (2kd/N +5).
\end{equation}
}
Next, using a prefix-free encoding, the size of the set of the at most $k$ indices in $J$ is bounded by
\begin{equation}\label{eq:lem:pass_on_RLIS-j}   \Ebb\left[ \sum_{j\in J}\text{length}(j)\mid \Ecal\right] \leq k (2\Ebb [\log j_{k}|\Ecal]+1)\overset{(i)}{\leq} k(2\log(\Ebb[j_k\mid \Ecal])+1)\overset{(ii)}{\leq} 4k\log(6Pk).
\end{equation}
In $(i)$ we used the concavity of $\log$ and in $(ii)$ we used \eqref{eq:conditional_expected_num_jk}.
Finally, by construction, recall that $\text{length}([x]_W)\leq k \log (3\sqrt{2P}/\Delta')\leq 2Pk\log d$, where the last inequality uses $\Delta'\geq \lambda_p$.
Combining this with \cref{eq:lem:pass_on_RLIS-Mj,eq:lem:pass_on_RLIS-j}, we obtain 
\begin{align}\label{eq:prop:reduction_to_compression_game-final-bound}
        I(E_w;\texttt{Message}\mid {\color{blue}\widetilde C_v}) &\leq I(E_w;\texttt{Mem}_{v}\mid {\color{blue}\widetilde C_v})+\Ebb\left[\sum_{j\in J}(\length(\texttt{Mes}_j)+\length(j) + \length([x_j]_{W_{pj}}))\right] \notag\\ 
        &{\color{blue}\leq I(E_w;\texttt{Mem}_{v}\mid E_{> p}, C_{w})+\Ebb\left[\sum_{j\in J}(\length(\texttt{Mes}_j)+\length(j) + \length([x_j]_{W_{pj}}))\right] \notag}\\
    &\leq \paren{6P\ln(6P)\paren{\frac{4M}{N_p}\paren{1+\frac{8k}{N_{p-1}}}+6k }+4k\log(6Pk)+2Pk^2 \log d }\Pbb(\Ecal).
\end{align}
{\color{blue}The first inequality follows from $C_v,E_{>p}$ being $\widetilde C_v$-measurable, and $\Rcal$ being independent of all quantities sampled before step~\ref{item:encoding-2} of encoder. The second inequality follows from $I(E_w;E_v|\widetilde C_v)=0$ and $(\widetilde C_v,E_v)=(E_{>p},C_w)$.}
The last inequality follows from the well-behavedness of phase $w$ and the fact that $\Pbb[\Ecal] \geq (6P\ln(6P))^{-1}$ from \cref{lem:bounding_number_of_samples_in_compression_scheme}.
Finally, since $\Tcal=\Tcal_v^{-\top}$ is ${\color{blue}\widetilde C_v}$-measurable,
\begin{align*}
    I(E_w;\texttt{Message} | {\color{blue}\widetilde C_v})&=I(E_w;\texttt{Message} | {\color{blue} E_{v,p}^{(<i)},\Tcal,C_v,E_{>p}})\\
        &=I(E_w;\texttt{Message},{\color{blue}C_v,E_{>p}} | E_{v,p}^{(<i)},\Tcal)\\
        &\geq I(E_w;\texttt{Message}| E_{v,p}^{(<i)},\Tcal) = I(E^{(i)};\texttt{Message}| E^{(<i)},\Tcal)
\end{align*}
In $(i)$ we used the fact that  $E_w=E^{(i)}$ is independent of ${\color{blue}(C_v,E_{>p})}$ conditionally on  $E_{v,p}^{(<i)}=E^{(<i)}$ and $\Tcal_{v}=\Tcal^{-\top}$. This ends the proof. 
\end{proof}

\comment{
Hence $I(E_p ^{(i)};\texttt{Message}| E_p^{(<i)},\Tcal_{v})$ is bounded by the right hand side of last inequality in \eqref{}.

The same mutual information bound for $(E^{(\leq i)},\Tcal)$ follows, since we are actually using the $(E^{(\leq i)},\Tcal)$ provided by Game~\ref{game:compression_game} as $(E_{v,p}^{(\leq i)},\Tcal_v)$ here. 
}

\paragraph{Removing the choice of $\Tcal$.} The only remaining step to prove \cref{prop:reduction_to_compression_game} is to remove \cref{ass:Tcal_chosen} in which the player may select the linear map $\Tcal$. First, note that by construction of the compression scheme, $\Tcal$ always satisfies the desired condition number bound (see step~\ref{item:selecting_Tcal} of the encoding scheme). We now aim to select a specific realization of $\Tcal$ that enjoys the same information guarantee. 

Specifically, denote by $\widetilde M$ the factor on the right-hand side of \cref{lemma:information_upper_bound_for_game} and $\xi:=(12P\ln(6P))^{-1}$.  Since $\Pbb[\Ecal]\geq 2\xi$, by \cref{lemma:information_upper_bound_for_game}, we have
\begin{equation*}
    I(E^{(i)};\texttt{Message}| E^{(<i)},\Tcal) \leq 2\widetilde M(\Pbb[\Ecal]-\xi).
\end{equation*}
Rearranging terms, this precisely gives
\begin{equation*}
    \Ebb_{\Tcal_0\sim\Tcal}\sqb{I(E^{(i)};\texttt{Message}| E^{(<i)},\Tcal=\Tcal_0)  - 2\widetilde M (\Pbb(\Ecal\mid \Tcal=\Tcal_0) - \xi)} \leq 0.
\end{equation*}
Hence, there must be a fixed realization of $\Tcal=\Tcal_0$ for which the quantity within the expectation is non-positive. Since the mutual information is always non-negative, this implies 
\begin{equation*}
    I(E^{(i)};\texttt{Message}| E^{(<i)},\Tcal=\Tcal_0) \leq 2\widetilde M (\Pbb(\Ecal\mid \Tcal=\Tcal_0)-\xi) \leq 2\widetilde M \Pbb(\Ecal\mid \Tcal=\Tcal_0)
\end{equation*}
and $\Pbb(\Ecal\mid \Tcal=\Tcal_0)\geq\xi$.
Finally, on this realization of $\Tcal=\Tcal_0$, \cref{lem:bounding_number_of_samples_in_compression_scheme} shows that the success probability is at least $\Pbb(\Ecal\mid \Tcal=\Tcal_0)$. As a remark, since this probability is non-zero, this implies $\kappa(\Tcal_0)\leq 2/(d\delta_p^{(0)})$ (see step~\ref{item:selecting_Tcal} of the encoding scheme). This completes the proof of \cref{prop:reduction_to_compression_game}.

\subsection{Hardness of the compression game: Proof of Proposition~\ref{prop:hardness_compression_game}}
\label{subsec:proof_hardness_compression}
In this section, we prove \cref{prop:hardness_compression_game}. We recall that in this game, the goal is to find an approximately perpendicular subspace to some transformed version of $E^{(i)}$ given some message $\texttt{Message}$ and lower-accuracy versions $E^{(1)},\ldots,E^{(i-1)}$, where all these versions are taken from the noise addition procedure from \cref{alg:noise_addition}. We start by showing that if we ignore the message, the posterior of the next version $E^{(i)}$ is essentially uniform in a Grassmannian ball centered at $E^{(i-1)}$.

\begin{lemma}\label{lemma:reduce_to_uniform}
    Fix an invertible linear map $\Tcal$, accuracies $\delta^{(\leq N)}$, and $\eta\in(0,1]$. Sample $E\sim\mathrm{Unif}(\Gr(\tilde d,d))$ and $E^{(\leq N)}:=\NoiseAddition(\Tcal(E),\delta^{(\leq N)})$ (as in Game~\ref{game:compression_game}). Then,
    \begin{equation*}
        p_{E^{(i)}\mid E^{(\leq i-1)}} \leq e^{2d\tilde d(\delta^{(i-1)}+\ldots+\delta^{(N)}) \kappa(\Tcal)}\cdot  p_{\mathrm{Unif}(B(E^{(i-1)},\delta^{(i-1)}))},
    \end{equation*}
    where $p_A$ denotes the density of the distribution of some random variable $A$ with respect to the Haar uniform measure.
\end{lemma}

\begin{proof}
    We start by studying the posterior distribution of $E^{(i)}$ given $E^{(\leq i-1)}$. Since by construction the noise addition process is Markovian from $\Tcal(E)\to E^{(N)}\to\ldots\to E^{(0)}$, this posterior only depends on $E^{(i-1)}$, i.e., $p_{E^{(i)}\mid E^{(\leq i-1)}} = p_{E^{(i)}\mid E^{( i-1)}}$. Then, for any $G\in\Gr(\tilde d,d)$, by Bayes' rule,
    \begin{align}
        p_{E^{(i)}\mid E^{( i-1)}}(G) \propto p_{(E^{(i)},E^{(i-1)})}(G,E^{(i-1)}) &\overset{(i)}{\propto} p_{E^{(i)}}(G) \1[E^{(i-1)}\in B(G,\delta^{(i-1)})] \notag \\
        &= p_{E^{(i)}}(G) \1[G\in B(E^{(i-1)},\delta^{(i-1)})]. \label{eq:posterior_computation_1}
    \end{align}
    In $(i)$ we used the definition of the noise addition process. Next, we analyze the prior $p_{E^{(i)}}$ on the ball $B(E^{(i-1)},\delta^{(i-1)})$. By construction, the noise addition process constructs $E^{(i)}$ from $\Tcal(E)$ by taking a centered random walk starting from $\Tcal(E)$. Importantly, all steps $E^{(j)}\to E^{(j-1)}$ of this random walk are independent and isotropic given their center $E^{(j)}$. Further, to construct $E^{(i)}$ from $E$ it takes sequential steps of size at most $\delta^{(N)},\ldots,\delta^{(i)}$. In particular, the distance between $\Tcal(E)$ and $E^{(i)}$ in principal angle is at most $\delta^{(N)}+\ldots+\delta^{(i)}$ by the triangle inequality. In summary, there is a kernel $K(G,A)$ such that (1)
    \begin{equation*}
        p_{E^{(i)}} (G) = \int_A K(G,A) \cdot p_{\Tcal(E)}(A)\, dA,
    \end{equation*}
    where the integral is with respect to the Haar uniform measure, (2) for any $G_1,G_2\in\Gr(\tilde d,d)$, $K(G_1,\cdot)$ can be obtained from $K(G_2,\cdot)$ by applying some rotation from $G_1$ to $G_2$, and (3) $K(G,\cdot)$ is supported on $B(G,\delta^{(N)}+\ldots+\delta^{(i)})$.

    Therefore, for any $G_1,G_2\in B(E^{(i-1)},\delta^{(i-1)})$, we have
    \begin{equation}\label{eq:reduction_to_density_ratio}
        \frac{p_{E^{(i)}}(G_1)}{p_{E^{(i)}}(G_2)} \leq \sup_{\substack{A_1\in B(G_1,\delta^{(N)}+\ldots+\delta^{(i)})\\ A_2\in B(G_2,\delta^{(N)}+\ldots+\delta^{(i)}) }} \frac{p_{\Tcal(E)}(A_1)}{p_{\Tcal(E)}(A_2)} \leq \sup_{\substack{A_1,A_2\in\Gr(\tilde d,d) \\ \theta_{\max}(A_1,A_2) \leq 2(\delta^{(i-1)}+\ldots +\delta^{(N)})}} \frac{p_{\Tcal(E)}(A_1)}{p_{\Tcal(E)}(A_2)}.
    \end{equation}

    \paragraph{Bounding the marginal density ratio.} In this paragraph, we focus on bounding the right-hand-side density ratio for the marginal of $\Tcal(E)$. We recall that $E$ is Haar uniform. If $Q\in\mathbb R^{d\times \tilde d}$ denotes an orthonormal basis matrix of $A\in\Gr(\tilde d,d)$, then by the density formula for the matrix angular central Gaussian distribution \cite{chikuse1990matrix}
\[
p_{\Tcal(E)}(A)
=
|\det\Tcal|^{-\tilde d}
\det\bigl(
Q^\top\Tcal^{-\top}\Tcal^{-1}Q
\bigr)^{-d/2}.
\]
Fix $A_1,A_2$, and write
$\theta=\theta_{\max}(A_1,A_2)$.
There exists a
skew-symmetric matrix $\Omega$ such that
$e^\Omega A_1=A_2$ and $\|\Omega\|_{ op }=\theta$.
Consider the path of this rotation
\[
A(s)=e^{s\Omega}A_1,
\qquad
Q(s)=e^{s\Omega}Q,
\qquad s\in[0,1],
\]
and define $M(s)=\Tcal^{-1}Q(s)$. Differentiating the log-density gives
\[
\frac{d}{ds}\ln p_{\Tcal(E)}(A(s))
=
-d\,\operatorname{tr}
\bigl(M(s)^\dagger M'(s)\bigr).
\]
Since $\|\Tcal\|_{ op }\le \lambda_{\max}(\Tcal)$, we have $\lambda_{\min}(M(s))\geq 1/\lambda_{\max}(\Tcal)$ and $\|M(s)^\dagger\|_{ op }\le \lambda_{\max}(\Tcal).$
Furthermore,
\[
\|M'(s)\|_{ op }
=
\|\Tcal^{-1}\Omega Q(s)\|_{ op }
\le\frac{\theta}{\lambda_{\min}(\Tcal)}.
\]
Consequently,
\[
\left|
\frac{d}{ds}\ln p_{\Tcal(E)}(A(s))
\right|
\le d\tilde d \|M(s)^\dagger M'(s)\|_{op}\leq d\tilde d\cdot \theta \kappa(\Tcal).
\]
Integrating over $s\in[0,1]$ proves
\begin{equation*}\label{eq:useful_comparison}
    \abs{\ln \frac{p_{\Tcal(E)}(A_1)}{p_{\Tcal(E)}(A_2)}} \leq d\tilde d\cdot \theta \kappa(\Tcal).
\end{equation*}

We recall from \cref{eq:posterior_computation_1} that $E^{(i)}\mid E^{(\leq i-1)}$ is supported on $B(E^{(i-1)},\delta^{(i-1)})$. Hence, there must be some $G_2$ on this support such that the density $p_{E^{(i)}\mid E^{( i-1)}}(G_2)$ is at most that of the uniform on $B(E^{(i-1)},\delta^{(i-1)})$. Then, plugging the previous identity within \cref{eq:posterior_computation_1,eq:reduction_to_density_ratio} shows that
\begin{equation}\label{eq:first_comparison_no_trasnform}
    \frac{p_{E^{(i)}\mid E^{(\leq i-1)}}(G)}{p_{\mathrm{Unif}(B(E^{(i-1)},\delta^{(i-1)}))}(G)} \leq \frac{p_{E^{(i)}\mid E^{(\leq i-1)}}(G)}{p_{E^{(i)}\mid E^{(\leq i-1)}}(G_2)} = \frac{p_{E^{(i)}}(G_1)}{p_{E^{(i)}}(G_2)} \leq e^{2d\tilde d  (\delta^{(i-1)}+\ldots+\delta^{(N)})\kappa(\Tcal)}.
\end{equation}
This ends the proof.
\end{proof}

\cref{lemma:reduce_to_uniform} essentially reduces Game~\ref{game:compression_game} to the case when $\tilde E^{(i)}$ is a transformation $T_{E^{(i-1)},\eta}(F)$ from a uniform subspace $F$ in $B(E^{(i-1)},\delta^{(i-1)})$. Further, finding $k$ perpendicular vectors to a uniform subspace in a Grassmannian ball requires $\tilde O(k\tilde d)$ information, as formalized below.

\begin{restatable}{lemma}{LemUnifCaseTranslation}
\label{lemma:Unif_case_plus_transformation}
    There is a universal constant $C_0>0$ such that the following holds. Fix $E_0\in\Gr(\tilde d,d)$ and some parameters $\delta\in(0,1/d]$ and $\rho\geq 1$. Sample $E\sim\mathrm{Unif}(B(E_0,\delta))$ and let $F:=T_{E_0,\rho\delta}(E)$. Then, for any fixed subspace $W\in\Gr(k,d)$ and $\epsilon>0$, we have
    \begin{equation*}
        \Pbb(\|P_WP_F\|_{op} \leq \epsilon) \leq \left(
                C_0
                \epsilon\rho\sqrt{\frac{d}{\tilde d}}
            \right)^{k\tilde d}.
    \end{equation*}
\end{restatable}

\comment{
\begin{restatable}{lemma}{LemmaUnifCase}
\label{lemma:success_proba_for_uniform}
    There is a universal constant $C_0>0$ such that the following holds.
    Let $E_0\in\Gr(\tilde d,d)$ and $\rho\in[0,1/d]$. Then, for any fixed subspace $W\in\Gr(k,d)$ for $k\leq d$, letting $E\sim \mathrm{Unif}(B(E_0,\rho))$:
    \begin{equation*}
        \Pbb(\|P_WP_E\|_{op} \leq \epsilon) \leq \left(
                C_0
                \frac{\epsilon }{\tan \rho}\sqrt{\frac{d}{\tilde d}}
            \right)^{k\tilde d}.
    \end{equation*}
\end{restatable}
}

We are now ready to prove \cref{prop:hardness_compression_game}: if the encoding strategy worked with some probability $\xi$ for Game~\ref{game:compression_game}, the message \texttt{Message} should have provided the necessary $\tilde O(\xi k\tilde d)$ information.

\begin{proof}[Proof of \cref{prop:hardness_compression_game}]
 Fix an encoding-decoding strategy for Game~\ref{game:compression_game} and let $\Ecal$ denote the winning event. 
    Write $\tilde{E}^{(i)}=T_{E^{(i-1)},\eta}(E^{(i)})$. 
    On event $\Ecal$, the decoder outputs $x_1,...,x_{k}$ with
    \[\|\text{Proj}_{\tilde{E}^{(i)}} (x_t)\|\leq \beta\|x_t\|\quad\text{ and }\quad\|\text{Proj}_{span(x_1,\dots,x_{t-1})^\perp}(x_t)\|\geq \gamma\|x_t\|,\quad{\forall t\in[k]}.\]
    Let $s:=1+\log\frac{\sqrt{d}}{\gamma}$. Using the same argument as in the proof of Lemma 13 in \cite{blanchard24gradient}, we can construct an orthonormal sequence $z_1,...,z_r$ with $r=\ceil{k/s}$ using only $x_1,...,x_{k}$, such that for all $j\in[r]$, $$\|\text{Proj}_{\tilde{E}^{(i)}}(z_j)\|\leq\frac{e\beta\sqrt{k}}{\gamma}.$$
    Let $W=\Span(z_1,...,z_r)$. Hence, on event $\Ecal$, $\|P_W P_{\tilde{E}^{(i)}}\|_{op}\leq e\beta\sqrt{rk}/\gamma$. We extend the definition of $W$ to $\Ecal^c$ as well. 
    The decision of the decoder in Game~\ref{game:compression_game} only uses $\texttt{Message}$ and $E^{(<i)}$. Hence, by the data processing inequality,
    \begin{equation}\label{eq:usef_1}
    \begin{aligned}
        I(E^{(i)};\texttt{Message}|E^{(<i)})=I(\tilde{E};\texttt{Message}|E^{(<i)}) &\geq I(\tilde{E}^{(i)};W|E^{(<i)})\\
        &=\Ebb[\text{KL}(\Dcal(\tilde{E}^{(i)},W|E^{(<i)}),\Dcal(\tilde{E}^{(i)}|E^{(<i)})\otimes \Dcal(W|E^{(<i)}))]
    \end{aligned}    
    \end{equation}
    Let $Q(\tilde{E}^{(i)},W|E^{(<i)}):=\Dcal(\tilde{E}^{(i)}|E^{(<i)})\otimes \Dcal(W|E^{(<i)})$ denote the marginal product law and consider the event $\Ecal_\perp(\tilde{E}^{(i)},W):=\set{\|P_W P_{\tilde{E}^{(i)}}\|_{op}\leq e\beta\sqrt{rk}/\gamma} \supseteq\Ecal$. 
    Again, by the data processing inequality,
    \begin{equation}\label{eq:usef_2}
    \begin{aligned}
        \text{KL}(\Dcal( E^{(i)},W|E^{(<i)}),Q(E^{(i)},W|E^{(<i)}))&\geq \text{KL}(\Dcal(\1[\Ecal_\perp]|E^{(<i)}),Q(\1[\Ecal_\perp]|E^{(<i)}))\\
        &\geq \Pbb(\Ecal_\perp|E^{(<i)})\log\frac{1}{\Pbb_Q(\Ecal_\perp|E^{(<i)})}-\log 2, 
    \end{aligned}    
    \end{equation}
    where $\Pbb_Q(\Ecal_\perp|E^{(<i)})$ denotes the conditional probability of event $\Ecal_\perp$ under the marginal product law. For simplicity, let $F\sim\mathrm{Unif}(B(E^{(i-1)},\delta^{(i-1)}))$ and $\widetilde F:=T_{E^{(i-1)},\eta}(F)$.
    Then, we have
    \begin{align*}
        \Pbb_Q(\Ecal_\perp|E^{(<i)})&=\Pbb\paren{\|P_W P_{\tilde{E}^{(i)}}\|_{op}\leq \frac{e\beta\sqrt{rk}}{\gamma}\mid E^{(<i)}}\\
        &\overset{(i)}{\leq} e^{8\tilde d} \cdot \Pbb\paren{\|P_W P_{\tilde F}\|_{op}\leq \frac{e\beta\sqrt{rk}}{\gamma}}\\
        &\overset{(ii)}{\leq} \paren{C'\frac{\beta\sqrt{rk}}{\gamma} \frac{\eta}{\delta^{(i-1)}}\sqrt{\frac{d}{\tilde d}}}^{r\tilde d} \overset{(iii)}{\leq} e^{-r\tilde d},
    \end{align*}
    for some universal constant $C'>0$.
    In $(i)$ we used \cref{lemma:reduce_to_uniform} and the assumptions on $\kappa(\Tcal)$. In $(ii)$ we used \cref{lemma:Unif_case_plus_transformation}, recalling that $\delta^{(i-1)}\leq 1/d$ and $\eta\geq\delta^{(i-1)}$. In $(iii)$ we used the assumption on $\beta/\gamma$ for $C$ sufficiently large.
    Therefore, taking the expectation over $E^{(<i)}$ on both sides of \eqref{eq:usef_2} and then plugging the result within \eqref{eq:usef_1}, we obtain
    \[I(E^{(i)},\texttt{Message}|E^{(<i)})\geq r\tilde d \cdot \Ebb[\Pbb(\Ecal_\perp|E^{(<i)})]-\log 2\overset{(i)}{\geq} r\tilde d \xi-\log 2.\]
    In $(i)$ we used the fact that $\Ecal\subseteq\Ecal_\perp$, hence $\Pbb(\Ecal_{\perp})\geq\Pbb(\Ecal)\geq \xi$. This ends the proof.
\end{proof}

\subsection{Putting things together}
\label{subsec:putting_things_together}

We start by proving the blocking property propagation from \cref{thm:propagation}. This is a direct consequence of the reduction to the compression game from \cref{prop:reduction_to_compression_game} and its hardness result in \cref{prop:hardness_compression_game}.

\begin{proof}[Proof of \cref{thm:propagation}] 

    Taking $c<1/8$ ensures that by assumption $\lambda_p\leq \Delta'< \tfrac18 \Delta\leq \tfrac18$. 
    
    Fix a well-behaved layer-$(p-1)$ phase $v$ with the $\Delta$-$\Block_k(q\mid \texttt{Mem}_v,C_v,E_{>p-1})$ property.
    By \cref{lemma:pigeonhole_principle} a majority of its children are also well-behaved and satisfy the information constraint \eqref{eq:second_technical_info_constraint}. If any of these children $w$ is not $\Delta'$-$\Block_k(q-P^{-1}\mid \texttt{Mem}_w,C_w,E_{>p})$, then \cref{prop:reduction_to_compression_game} shows that one can win at the compression Game~\ref{game:compression_game} with the parameters described therein. We recall that the sequence $\delta_p^{(0)},\delta_p^{(1)},\ldots$ is geometrically decreasing by factors $\lambda_p\leq 1/d\leq 1/8$ for $d$ sufficiently large. Hence, the chosen linear map $\Tcal$ satisfies
    \begin{equation*}
        \kappa(\Tcal) \leq \frac{2}{d\delta_p^{(0)}} \leq \frac{4}{d(\delta_p^{(0)}+\delta_p^{(1)}+\ldots+\delta_p^{(N_p)})}.
    \end{equation*}
    Also, for the chosen parameters we have $(\delta_p^{(i_p-1)}/\sqrt{3P}) \leq 1/d$ and $\eta=\delta^{(i_p-1)}$. Hence
    \begin{equation*}
        \frac{\gamma}{\beta} = \frac{\Delta}{16\Delta'\sqrt{P}} {\geq}  \frac{k}{16c\sqrt{P}}\paren{\frac{d}{\tilde d}}^{3/2} \geq Ck\sqrt{3P\frac{d}{\tilde d}} = Ck\frac{\eta}{\delta_p^{(i_p-1)}/\sqrt{3P}}\sqrt{\frac{d}{\tilde d}},
    \end{equation*}
    for a universal constant $c>0$ chosen sufficiently small, where $C$ is the universal constant from the assumptions in \cref{prop:hardness_compression_game}. Here, we used the definition of $\Delta'$ and $\tilde d \leq d/(2P)$.

    We have checked all assumptions for \cref{prop:hardness_compression_game}. Then, combining the information $I(\texttt{Message};E^{(i)}\mid E^{(<i)})$ upper bound from \cref{prop:reduction_to_compression_game} with the information lower bound from \cref{prop:hardness_compression_game}, this implies
    \begin{equation}\label{eq:thm:propagation-final}
        12P\ln(6P)\paren{\frac{4M}{N_p}\paren{1+\frac{8k}{N_{p-1}}}+6k }+8k\log(6Pk)+4Pk^2 \log d \geq \frac{k\tilde d}{1+ \log(d/\Delta)} - \frac{\log 2}{\xi}
    \end{equation}
    We recall that the success probability is at least $\xi\geq (12P\ln(6P))^{-1}$ from \cref{prop:reduction_to_compression_game}. Since 
    \[N_{p-1}\geq \frac{\log(1/\epsilon)}{2P^2\log d}\geq\frac{d}{2P^2\log d}=\Omega\left(\frac{\tilde{d}}{P\log(d/\Delta)\log d}\right) \]
    and with $$k\leq \frac{\tilde d}{C_bP\log(d/\Delta)\log d},$$ 
    \eqref{eq:thm:propagation-final} becomes impossible for $C_b$ a sufficiently large universal constant. In summary, all well-behaved children of $v$ satisfying \cref{eq:second_technical_info_constraint} are $\Delta'$-$\Block_k(q-P^{-1}\mid \texttt{Mem}_w,C_w,E_{>p})$. 
    
\end{proof}

We now check that the first-layer phases are well-behaved and already have the blocking property. This is essentially immediate from the fact that before a given layer-$1$ phase $(i_1)$, the learner has only interacted with the versions $E_{(0)},\ldots,E_{(i_1-1)}$ of the subspace $E_1$ and has no additional information about it.

\PropInitializationBlocking*
\begin{proof}
    Fix some layer-$1$ phase $v=(i_1)\in\Vcal_1$. Its parent is the root phase $\emptyset$ for which the initial memory state $\texttt{Mem}_\emptyset=\texttt{Memory}_1=0$ is empty. Hence, $v$ is directly well-behaved.

    It remains to check that $v$ has the blocking property conditional on $\texttt{Mem}_v,C_v,E_{>1}$, where $C_v = E_{\emptyset,1}^{(<i_1)} = (E_{(0)},\ldots,E_{(i_1-1)})$.
    We recall that the subspaces $E_{(0)},\ldots,E_{(N_1)}$ are obtained as the result of the $\NoiseAddition$ procedure: $E_{(0)},\ldots,E_{(N_1)} := \NoiseAddition(E_1,\frac1{\sqrt{3P}}\delta_1^{(\leq N_1)})$.
    The main observation is that before phase $(i_1)$, the oracle responses were constructed using only lower-accuracy versions $E_{(0)},\ldots,E_{(i_1-1)}$, the subspaces $E_2,\ldots,E_P$ which are independent of $E_1$, as well as fresh randomness in the $\NoiseAddition$ procedure. As a result, the initial memory state $\texttt{Mem}_v$ could be reconstructed using only $E_{(0)},\ldots,E_{(i_1-1)}$, the subspaces $E_2,\ldots,E_P$, and fresh randomness for the $\NoiseAddition$ process on deeper layers $E_2,\ldots,E_P$. 
    
    In summary, $E_v$ is independent of $\texttt{Mem}_v,E_{>1}$ conditionally on $C_v=E_{(<i_1)}$. We note that since this is the first layer, $R_v$ only depends on $E_v=E_{(i_1)}$ and $E_{(i-1)}$. In summary, it suffices to show that $R_v$ is $\lambda_1$-$\Block_k(1-(2P)^{-1}\mid E_{(<i_1)})$. 
    To do so, fix an $E_{(<i_1)}$-measurable subspace $W\in\Gr(k,d)$. Let $\tilde E_v:=\Hcal_v(E_v)$ be the resulting subspace after applying the transformation $\Hcal_v$.
    In the following, let $\Sigma_W$ denote a $\lambda_1$-net of $W\cap S^{d-1}$ of size at most $(3/\lambda_1)^k$.
    Then, by definition of $R_v$, we have
    \begin{align*}
        R_v \cap (W+\lambda_1 B)\neq\emptyset 
        &\overset{(i)}{\implies} \set{x:\|x\|\geq \tfrac12} \cap (\tilde E_v^\perp + \lambda_1 B) \cap (W+\lambda_1 B) \neq \emptyset\\
        &\overset{(ii)}{\implies} \exists x\in \Sigma_W:\, \|P_{\tilde E_v}(x)\| \leq  10\lambda_1.
    \end{align*}
    In $(i)$ we noted that $\Tcal_v(B)\subseteq B/\delta_1^{(i_1-1)}$ and that $\Tcal_v$ stretches the space. In $(ii)$ we use $\lambda_1\leq 1/8$ for $d$ sufficiently large.
    We now bound the probability of the right-hand side event conditioned on $E_{(<i_1)}$. As a result,
    \begin{align*}
        \Pbb(R_v \cap (W+\lambda_1 B)\neq\emptyset\mid E_{(<i_1)}) &\leq \Pbb\paren{\exists x\in \Sigma_W: \|P_{\tilde E_v}(x)\| \leq  10\lambda_1 \mid E_{<(i_1)}}\\
        &\overset{(i)}{\leq} \abs{\Sigma_W}e^{d\tilde d(\delta_1^{(i_1-1)}+\ldots+\delta_1^{(N_1)})/\sqrt{3P}} \paren{10C_0\lambda_1\sqrt{3P\frac{d}{\tilde d}}}^{\tilde d}\\
        &\overset{(ii)}{\leq} \paren{\frac{3}{\lambda_1}}^k \paren{10e^2C_0\lambda_1\sqrt{3P\frac{d}{\tilde d}}}^{\tilde d} \leq \frac{1}{2P}.
    \end{align*}
    In $(i)$ we used the same arguments as in the proof of \cref{prop:hardness_compression_game}: first we use \cref{lemma:reduce_to_uniform} to reduce to the case when $E_v\mid E_{(<i_1)}$ is uniform in the Grassmannian ball $B(E_{(i_1-1)},\tfrac{1}{\sqrt{3P}}\delta_1^{(i_1-1)})$, second we use \cref{lemma:Unif_case_plus_transformation} for $1$-dimensional subspaces. In $(ii)$ we used the definition of the accuracies $\delta_1^{(\leq N_1)}$ for $d$ sufficiently large. The last inequality uses $4k\leq \tilde d$ and $d$ being sufficiently large.
    Taking the expectation over $E_{(<i_1)}$ ends the proof that $v$ is $\lambda_1$-$\Block_k(1-(2P)^{-1}\mid \texttt{Mem}_v,C_v,E_{>1})$.
\end{proof}

Combining the two previous results, we therefore show that the blocking property is propagated to a large number of leaf phases. This is a direct implication of the initialization in layer-$1$ phases from \cref{prop:initialization_blocking} and the inductive propagation result from \cref{thm:propagation}.

\begin{proof}[Proof of \cref{cor:induction_blocking}]

    We first select $P$ and $k$. Given $d^{1+\delta}\log\frac1{\epsilon}=M\leq c_0 \frac{d^{2}}{\log^4 d}\log\frac1\epsilon$, take 
    \begin{equation}\label{eq:cor:induction_blocking-choice-of-P}
        P=1+\ceil{\frac{(c_0 d^{1-\delta})^{1/8}}{\log^{5/8}d}}=1+\ceil{\left(\frac{c_0 d^2\log(1/\epsilon)}{M\log^5 d}\right)^{1/8}},\qquad k=\floor{\frac{d}{72C_b P^3\log^2 d}}
    \end{equation}
    For a sufficiently small universal constant $c_0>0$ and sufficiently large $d$, 
    one can verify that \[
    \widetilde d\ge\frac{d}{72P},
    \qquad
    N_p\ge\frac{\log(1/\epsilon)}{4Pp\log d},
    \qquad
    k\ge\frac{d}{144C_bP^3\log^2 d}\ge1,
    \]
    and that
    \[C_b \frac{M/N_p + P}{\tilde d} \cdot P\log P \cdot\log\frac{d}{\lambda_{p-1}}\leq k\leq \frac{\tilde d}{C_b P \log(d/\lambda_{p-1})\log d},\quad\forall p\in\set{2,\dots,P}.\]
    We also have $\lambda_p=d^{-p}\leq c\frac{\lambda_{p-1}}{k}\left(\frac{\tilde d}{d}\right)^{3/2}$  when $d$ is sufficiently large. 
    
    The remaining proof recursively applies \cref{thm:propagation}, with $\Delta=\lambda_{p-1}$ and $\Delta'=\lambda_p$ at layer $p$.
    By \cref{prop:initialization_blocking}, with $4k\leq \tilde d$, initially, we have for each of the $N_1$ phases $v\in\Vcal_1$, the corresponding $R_v$ is $\lambda_1$-$\Block_k(1-\tfrac{1}{2P}\mid \texttt{Mem}_v,C_v,E_{>1})$. By the induction hypothesis, assume for $p\in\set{2,\dots,P}$ there are at least 
    \[N_1\cdot\prod_{s=2}^{p-1}\left(\frac{N_s}{2}\right)\]
    layer-$(p-1)$ phases $v$ that are well-behaved and whose regions $R_v$ are $\lambda_{p-1}$-$\Block_k(1-\tfrac{2p-3}{2P}\mid \texttt{Mem}_v,C_v,E_{>p-1})$. Then, by \cref{thm:propagation}, for at least $\frac{N_p}{2}$ children $w\in\Vcal_p$ of $v$, $w$ is well-behaved and $R_w$ is $\lambda_p$-$\Block_k(1-\frac{2p-1}{2P}\mid \texttt{Mem}_w,C_w,E_{>p})$. Summing over all $v$ gives a total of at least
    $N_1\cdot\prod_{s=2}^{p}\left(\frac{N_s}{2}\right)$ such layer-$p$ phases. This completes the induction. In particular, the desired result is obtained at layer $P$.
\end{proof}

It only remains to prove that leaf phases with the blocking property cannot be skipped with significant probability: since there is an exponential number of them, this will prove the desired exponential sample-complexity lower bound.

\begin{proof}[Proof of \cref{thm:main_lower_bound_result}]
    Suppose $M\leq c_0 \frac{d^2}{\log^4 d}\log\frac1\epsilon$ as requested by \cref{cor:induction_blocking}. Following the same choice of $P$ and $k$ as in the proof of \cref{cor:induction_blocking}, there are at least $N_1\cdot\prod_{p=2}^P\paren{\frac{N_p}{2}}$
    layer-$P$ phases $\ell\in\Vcal_P$ such that $R_\ell$ is $\lambda_P$-$\Block_k(\tfrac{1}{2P}\mid \texttt{Mem}_\ell,C_\ell)$, which again follows from \cref{cor:induction_blocking}. Let $\Ucal_P$ denote the set of all such phases. We then have 
    $$|\Ucal_P|\geq N_1\cdot\prod_{p=2}^P\paren{\frac{N_p}{2}}.$$
    
    For each $\ell\in\Ucal_P$, recall that $\texttt{Mem}_\ell$ is the memory state of the learner at time $t_\ell$, which corresponds to the first round when the transform separation oracle had its internal leaf equal to $\ell$. If the oracle ever reaches this internal leaf, there are two possibilities: either the query $x_{t_\ell}$ is answered by the path separation oracle $\Ocal_\ell$ in \cref{alg:leaf_oracle} that corresponds to this internal leaf, or $\Ocal_\ell(x_{t_\ell})=\texttt{Success}$ which means the query cannot be answered by this $\Ocal_\ell$. Define the event
    \[\Ecal_\ell:=\set{\text{The oracle reaches leaf }\ell\text{ and }\Ocal_\ell(x_{t_\ell})=\texttt{Success}}.\]
    Importantly, when $\Ocal_\ell(x_{t_\ell})=\texttt{Success}$, we have $\Tcal_{\ell}(x_{t_\ell})\in R_\ell$.
    We then use the observation that the next query $x_{t_\ell}$ is $\texttt{Mem}_\ell$-measurable, and that the transformation $\Tcal_{\ell}$ is $C_\ell$-measurable. Hence $\Tcal_{\ell}(x_{t_\ell})$ is $(\texttt{Mem}_\ell,C_\ell)$-measurable. Then, the blocking property of $R_\ell$ implies
    \[\Pbb(\Ecal_\ell)\leq\Pbb(\Tcal_{\ell}(x_{t_\ell})\in R_\ell)\leq\Pbb(R_\ell\cap (\Tcal_{\ell}(x_{t_\ell})+\lambda_P B)\neq\emptyset)< 1-\frac1{2P}.\]
    Meanwhile, it is possible that the oracle is invalid from the beginning, which corresponds to event $\Ecal_{\text{bad}}$. When this event happens, no query should be counted.
    By \cref{lem:oracle-consistency}, we have 
    \[\Pbb(\Ecal_\ell^c \cap\Ecal_{\text{bad}}^c)\geq \frac1{2P}-e^{-d/300}\geq \frac1{4P}.\]
    In expectation, the total number of queries during the run is at least
    \begin{equation}\label{eq:thm:main_lower_bound_result}
        \Ebb\left[\sum_{\ell\in\Ucal_P}\1(\Ecal_{\ell}^c\cap\Ecal_{\text{bad}}^c)\right]\geq \frac{|\Ucal_P|}{4P}\geq \frac1{2P}\left(\frac{\ln(1/\epsilon)}{8P^2\log d}\right)^P \geq \left(d\ln (1/\epsilon)\right)^{P/4}=(d\ln(1/\epsilon))^{\widetilde{\Omega}(d^{(1-\delta)/8})}.
    \end{equation}
    The last equality follows from $\epsilon\leq e^{-d}$, $N_p=\floor{\frac{\log(1/\epsilon)}{2Pp\log d}}$ and the choice of $P$ in \eqref{eq:cor:induction_blocking-choice-of-P}. 

    Next, when $M> c_0 \frac{d^2}{\log^4 d}\log\frac1\epsilon$, \cref{cor:induction_blocking} cannot be applied, and we have $\frac{d^{(1-\delta)/8}}{\log^{5/8}d}<1$
    for $d$ sufficiently large. Here, instead of using the transform separation oracle, we simply apply the resisting separation oracle in \cite{Nemirovsky1983} that achieves the $d\log \frac1{\epsilon}$ query complexity. In particular, this also achieves the desired exponent.

    The transform separation oracle and resisting oracle are adaptive to queries. 
    Nevertheless, for deterministic learners, non-adaptive (oblivious) oracles are as powerful as adaptive ones. This gives, for each deterministic learner, an oblivious oracle that incurs at least the same oracle complexity with probability one.

    Finally, we extend our result to randomized learners by Yao's minimax principle-- any such learner that succeeds within $T$ steps can be considered as a $M+\ceil{\log T}$-bit deterministic learner with an internalized random string that is independently sampled. Without loss of generality one can assume $T=O(1/{\epsilon^2})$, as we can cap the runtime of any algorithm and run gradient descent. Conditioned on this random string, the previous argument holds. Averaging over this random string then gives the same expected lower bound as in \eqref{eq:thm:main_lower_bound_result}. Importantly, the oracle we use here is still fully adaptive. 

    The expected query lower bound extends to randomized algorithms succeeding with probability at least \(2/3\) by a standard restarting argument (see \cite{AdaProcaccia2017}): a \(T\)-query algorithm with success probability at least \(2/3\) against every valid adaptive separation oracle can be repeatedly restarted with fresh randomness (without resetting the oracle), yielding an algorithm that succeeds almost surely with expected query complexity at most \(3(T+1)/2\).
\end{proof}

\subsection*{Statement on the use of AI}
The authors only used the help of AI in bounding the marginal density ratio for \cref{lemma:reduce_to_uniform}, in part of the proof for \cref{lem:transform_process}, and in the analysis of Grassmannian density for \cref{lemma:Unif_case_plus_transformation}. The rest of the proof, including the construction of the transform separation oracle, the complete structure of the proof and all the reductions and information-theoretic arguments, was carried out by humans.

\appendix

\section{Other related work and potential applications}\label{app:additional-related-work}

\paragraph{Other works on memory.} 
Understanding the role of memory has also been a central topic in many other learning problems. For parity learning, \cite{Raz2017} first proved that any algorithm with sub-quadratic memory suffers exponentially in sample complexity. Similar results for other parity problems were then obtained by \cite{Beame2018,Garg2018,garg2019time,garg2021memory,Kol2017,Moshkovitz2017,Moshkovitz2018,raz2018fast}. Tradeoffs between memory and sample complexity have also been studied in many other topics, including linear regression \cite{Sharan2019,Steinhardt15}, principal component analysis (PCA) \cite{Mitliagkas2013}, online prediction with expert advice \cite{peng2023near,peng2023online,Srinivas2022}, learning under the statistical query model \cite{steinhardt16} and other general learning models \cite{Brown2021,brown22a}.

\paragraph{Potential applications.}
Our hard instance construction (see \cref{subsec:construction_of_oracle}) can potentially be insightful for other problems--- it provides a heterogeneous information hiding structure that attempts to maximize the noise of observed gradients under the geometric constraint of separation. When optimizing an objective with multiple scales \cite{kelner2021bigstep}, this heterogeneous noise also arises naturally from the different scales in the objective. Another possibility lies in showing negative results for the problem of estimating the maximum of a linear objective over some unknown convex body, where the learner has access to a membership oracle, which can be viewed as a zeroth-order separation oracle. The recent paper \cite{vempala2026nearlyquadratic} approaches this problem by constructing a nested pair of potential convex bodies that are indistinguishable within a certain number of membership queries. Intuitively, this involves hiding some regions that belong only to the larger body, where our construction ideas could be useful.

\input{transform_necessity}

\section{One-shot channel simulation.}
    We will use the following one-shot channel simulation model \cite{Harsha2010Communication,Li2017ContinuousCommu}:
    Let $X,Y$ be two potentially correlated random variables taking values in compact Polish spaces. Let Alice and Bob be two parties with a shared random string $r$. Even without direct access to the realization $x$ of the random variable $X$, Bob can obtain a sample from $P(Y|X=x)$ with a message $\texttt{Mes}=M(x,r)$ from Alice, who has access to $x$. A formal description of this is summarized in \cref{alg:one-shot-channel-commu}.
    
    \begin{algorithm}[H]
    \SetAlgoLined
    \LinesNumbered
    \caption{One-Shot Channel Simulation}
    \label{alg:one-shot-channel-commu}
    \KwIn{Random variables $(X,Y)$, realization $x$ of $X$, independent random string $r$}
    \KwData{Both observe $r$ and the distribution of $(X,Y)$; only Alice has access to $x$, Bob does not}
    
    Alice observes $x$ and sends a message $\texttt{Mes}= M(x,r)$ to Bob

    Bob outputs a sample $Y' = g(\texttt{Mes},r)$ 
    \end{algorithm}
    The message $\text{Mes}$ doesn't need to be large enough to store $x$ or $Y$. In fact, it can be described with $O(I(X;Y))$ bits on average, using a prefix-free encoding: \cite{Li2017ContinuousCommu} provides the following guarantee.
    \begin{theorem}\label{thm:entropy_of_Mj}
        There exist a function $g(\cdot)$ and a prefix-free mapping $M(\cdot)$ such that in \cref{alg:one-shot-channel-commu}, Bob's output $Y'$ follows $P(Y|X=x)$ almost surely in $x\sim P_X$, with $H(\texttt{Mes})\leq I(X;Y)+\log(I(X;Y)+1)+4$.
    \end{theorem}
    We will take $2I(X;Y)+6$ as a simpler bound on the expected length $\Ebb[\text{length}(\texttt{Mes})]$, which is enough for our purposes. As a direct generalization, if additionally, Alice and Bob have common knowledge of another random variable $Z$, then Bob can sample from $P(Y|X,Z)$ without knowing $X$, using on average $2I(X;Y|Z)+6$ bits from Alice.

\section{Helper results}

\subsection{Proof of Lemma~\ref{cor:the_uniform_singular_bound}}
\label{app:proof_singular_bound}

In this section, we prove the singular value bounds for $\Tcal_v$ in \cref{cor:the_uniform_singular_bound}. We will mostly focus on the related mapping $\Hcal_v=\Tcal_v^{-\top}$.
For each layer-$p$ phase $v$, one can verify that $\Hcal_v$ can  be written as a concatenation of the form
\[T_{U_{p},\delta_{p}}\circ\cdots\circ T_{U_2,\delta_2}\circ T_{U_1,\delta_1}.\]
with $T_{U,\delta}=P_{U^\perp}+\delta P_U$ being a contraction by a factor of $1/\delta$ along the subspace $U$.
Here, $U$ corresponds to ancestor phases of $v$ from earlier layers. Importantly, as in the construction of the skeleton tree, we are using phases obtained by adding noise \emph{after} applying the previous concatenated mapping to a Haar-uniform subspace $V_p$, or $E_p$ in our oracle construction. Therefore, the directions of the next stretching in $U_p$ also depend on the previous concatenated mapping using $U_1,\dots,U_{p-1}$. 

We formulate this process as the \emph{Transform Process} in \cref{def:transform_process} below.

\begin{definition}[Transform Process]\label{def:transform_process}
Let $V_1,\dots,V_P\in \Gr(\tilde d,d)$ be i.i.d.\ Haar-uniform $\tilde{d}$-dimensional subspaces. Correspondingly, let the contraction factors be $\delta_1,\dots,\delta_P\in (0,1]$. Define the following \emph{Transform Process}: Let $U_1$ be any subspace such that $\angle_{\max}(U_1,V_1)\leq \delta_1 /2$, let $U_2$ be any subspace such that $\angle_{\max}(U_2,T_{U_1,\delta_1} (V_2))\leq \delta_2 /2$, let $U_3$ be any subspace such that $\angle_{\max}(U_3,T_{U_2,\delta_2}\circ T_{U_1,\delta_1} (V_3))\leq \delta_3 /2$, ..., let $U_P$ be any subspace such that $\angle_{\max}(U_P,T_{U_{P-1},\delta_{P-1}}\circ\cdots\circ T_{U_2,\delta_2}\circ T_{U_1,\delta_1} (V_P))\leq \delta_P /2$.   

For simplicity, we use $\Hcal_{p}$ to denote the combined transformation $T_{U_p,\delta_p}\circ\cdots\circ T_{U_2,\delta_2}\circ T_{U_1,\delta_1}$. 
\end{definition}
The above transform process definition retains the flexibility in selecting $U_1,...,U_P$. In our oracle, this corresponds to the choice given by $\NoiseAddition$. Ideally, if we manage to control the singular values of $\Hcal_p$ regardless of this flexibility, then we will obtain uniform control that doesn't depend on the randomness used in $\NoiseAddition$ for sampling phases in each layer, as desired in \cref{cor:the_uniform_singular_bound}.

We observe that $\Hcal_p$ only takes effect on $span(U_1,...,U_p)$ and leaves the perpendicular component unchanged. This leads to the following characterization, which essentially controls the difference between $span(U_1,...,U_P)$ and $span(V_1,...,V_P)$ regardless of the flexibility in selecting $U_1,...,U_P$.

\begin{lemma}\label{lem:cumulative_span_identity}
In the transform process, let $V_1,..,V_P$ have orthonormal basis matrices $Q_1,\dots,Q_P\in\Rbb^{d\times \tilde{d}}$.
Then there exist matrices $D_p\in\Rbb^{d\times\tilde{d}}$ with
$\|D_p\|_{op}\leq\delta_p/2$ for $p\in[P]$, such that
\begin{equation}\label{eq:lem:cumulative_span_identity}
span(U_1,\dots,U_p)=span\left([Q_1+D_1,\dots,Q_p+D_p]\right).\qquad\forall p\in[P]
\end{equation}
\end{lemma}

\begin{proof}
For each $p$, we have $\angle_{\max}(U_p,\Hcal_{p-1}(Q_p))\leq\delta_p/2<\pi/2$. Let $D_p:=-P_{U_p^\perp}\Hcal_{p-1}Q_p$. Hence,
\[U_p=span(P_{U_p}(\Hcal_{p-1}Q_p))=span(\Hcal_{p-1}Q_p+D_p).\]
\
Since $\Hcal_{p-1}$ is a contraction,
\[
\|D_p\|_{op}
\leq
\|P_{U_p^\perp}P_{\Hcal_{p-1}Q_p}\|_{op}
\|\Hcal_{p-1}Q_p\|_{op}
\leq\delta_p/2.
\]
We prove the claim in \eqref{eq:lem:cumulative_span_identity} by induction. For $p=0$ this is trivially true. Assume 
\[span(U_1,\dots,U_{p-1})=span\left([Q_1+D_1,\dots,Q_{p-1}+D_{p-1}]\right),\]
then we have
\begin{align*}
    span(U_1,\dots,U_{p-1},U_p)
    &=span\left([Q_1+D_1,\dots,Q_{p-1}+D_{p-1},\Hcal_{p-1} Q_p+D_p]\right)\\
    &=span\left([Q_1+D_1,\dots,Q_{p-1}+D_{p-1},Q_p+D_p]\right)
\end{align*}
where the last equality follows from $\text{range}(I-\Hcal_{p-1})\subseteq span(U_1,...,U_{p-1})$.
\end{proof}

We are ready to prove the following lemma which uniformly controls the minimum singular value of $\Hcal_p$. Intuitively, the concatenated transformation $\Hcal_{p-1}$ makes $\Hcal_{p-1}(V_p)$ nearly perpendicular to $span(U_1,...,U_{p-1})$, which leads to $U_p$ being nearly perpendicular to the same subspace as well. In turn, the direction of the next contraction $T_{U_p,\delta_p}$ is also close to being perpendicular to the directions of all previous contractions in $span(U_1,...,U_{p-1})$. Roughly speaking, given that the contractions are nearly perpendicular to each other, and provided that the sequence $\delta_1,...,\delta_p$ decreases fast enough, the minimum singular value of $\Hcal_p$ should correspond to the contraction with the smallest $\delta$ factor, which would simply be $\delta_p$.
\begin{lemma}\label{lem:transform_process}
    Let $36P\tilde{d}\leq d$. Then with at least $1-2Pe^{-d/288}$ probability over $V_1,\dots,V_P$, simultaneously for all factors $\delta_1,\dots,\delta_P\in(0,1]$ satisfying $\delta_1<1/7$ and $\delta_{p+1}\leq\delta_p/36$ for all $p\in[P-1]$, and all choices of $U_1,\dots,U_P$ satisfying the principal-angle constraints of the corresponding transform process, the combined transformation
    $$\Hcal_p:= T_{U_{p},\delta_{p}}\circ\cdots\circ T_{U_2,\delta_2}\circ T_{U_1,\delta_1}
    $$ 
    satisfies $\|\Hcal_p\|_{\min}\geq \delta_p /2$ for all $p\in[P]$.
\end{lemma}
\begin{proof}
    As in \cref{lem:cumulative_span_identity}, let $Q_1,...,Q_P$ be orthonormal basis matrices of $V_1,...,V_P$. By the proof of \cref{lem:cumulative_span_identity}, we have
    \[U_p=span(\Hcal_{p-1}Q_p+D_p)\]
    for some matrices $D_p\in\Rbb^{d\times \tilde{d}}$ with $\|D_p\|_{op}\leq\delta_p/2$. Define the event
    \begin{equation}\label{eq:lem:transform_process-event}
        \Ecal:=\set{\lambda_{\min}([Q_1,...,Q_P])\geq\frac12,\quad\|P_{L_{p-1}}Q_p\|_{op}\leq\frac12\text{ for every }p\in[P]}
    \end{equation}
    where $L_{p}:=span(V_1,...,V_p)$ with $L_0=\set{0}$. This event only depends on $V_1,...,V_P$. Next, we inductively show the singular value bounds given $\Ecal$. 
    
    The statement is clearly true when $p=1$. 
    For $p>1$, let $B_p:= \Hcal_p ^\top \Hcal_p$. Since $T_{U,\delta} = T_{U,\delta}^\top$, we have
    \[B_p=\Hcal_{p-1} ^\top T_{U_p,\delta_p}^2 \Hcal_{p-1} = \Hcal_{p-1} ^\top (I-(1-\delta_p^2 )P_{U_p}) \Hcal_{p-1} = B_{p-1} - (1-\delta_p^2)\Hcal_{p-1} ^\top P_{U_p} \Hcal_{p-1}.\]
    The proof proceeds by giving singular value bounds for those $B_p$'s. Let $\widetilde{Q}_p:=Q_p + \Hcal_{p-1} ^{-1} D_p$. We have
    \comment{

}
    $$P_{U_{p}} = (\Hcal_{p-1} \widetilde{Q}_p) [(\Hcal_{p-1} \widetilde{Q}_p)^\top (\Hcal_{p-1} \widetilde{Q}_p)]^{-1} (\Hcal_{p-1} \widetilde{Q}_p)^\top = \Hcal_{p-1} \widetilde{Q}_p(\widetilde{Q}_p^\top B_{p-1} \widetilde{Q}_p)^{-1}\widetilde{Q}_p^\top \Hcal_{p-1} ^\top.$$
    Therefore,
    $$B_{p} = B_{p-1} - (1-\delta_{p}^2) B_{p-1} \widetilde{Q}_p(\widetilde{Q}_p^\top B_{p-1} \widetilde{Q}_p)^{-1}\widetilde{Q}_p^\top B_{p-1}.$$
    Inverting with the Woodbury formula gives
    \begin{equation}\label{eq:lem:transform_process-1}
        B_{p}^{-1}=B_{p-1} ^{-1} + (\delta_{p}^{-2}-1)\widetilde{Q}_p(\widetilde{Q}_p^\top B_{p-1} \widetilde{Q}_p)^{-1} \widetilde{Q}_p^\top.
    \end{equation}
    Since $\|D_p\|_{op}\leq\delta_p/2$, we have 
    \begin{equation}\label{eq:lem:transform_process-2}
        \|\widetilde{Q}_p\|_{op}\leq 1+\delta_p\|\Hcal_{p-1}\|_{\min}^{-1}/2.
    \end{equation}
    
    \comment{
    Since $\|E\|_{op}\leq \sqrt{\tilde{d}}\delta_{p}$ and $\|\Lambda\|_{op}\leq\|\Hcal_{p-1} Q_p\|_F$, we have 
    \begin{equation}\label{eq:lem:transform_process-2}
        \|\widetilde{Q}_p\|_{op}\leq 1 + \|\Hcal_{p-1}\|_{\min}^{-1}\cdot\sqrt{\tilde{d}}\delta_{p}\cdot \|\Hcal_{p-1} Q_p\|_F \leq 1 + \|\Hcal_{p-1}\|_{\min}^{-1}\cdot\sqrt{\tilde{d}}\delta_{p}\cdot \|Q_p\|_F = 1 + \tilde{d}\|\Hcal_{p-1}\|_{\min}^{-1}\delta_{p}
    \end{equation}
    where the second inequality is because $\Hcal_{p-1}$ is a composition of contraction mappings.
    }
    To control $\|B_p^{-1}\|_{op}$ using \eqref{eq:lem:transform_process-1}, it remains to lower-bound $\|\widetilde{Q}_p^\top B_{p-1} \widetilde{Q}_p\|_{\min}$.
    Consider
    \[\widetilde{Q}_p^\top B_{p-1} \widetilde{Q}_p \succeq Q_p^\top B_{p-1} Q_p + (\Hcal_{p-1} ^{-1} D_p) ^\top B_{p-1} Q_p + Q_p^\top B_{p-1} \Hcal_{p-1} ^{-1} D_p.\]
    For the first term, let $\widetilde L_{p-1}=span(U_1,\dots,U_{p-1})$. We have $B_{p-1}(x)=x$ for all $x\in \widetilde L_{p-1}^\perp$ as well as $\text{dim}(\widetilde L_{p-1})\leq (p-1)\tilde{d}\leq d/36$. Hence,
    \begin{equation}\label{eq:em:transform_process-2.5}
        \|Q_p^\top B_{p-1} Q_p \|_{\min}\geq \|P_{\widetilde L_{p-1}^\perp}Q_p\|_{\min}^2 = 1- \|P_{\widetilde L_{p-1}} P_{V_{p}}\|_{op}^2
    \end{equation}
    We then use $L_{p-1}$ to approximate $\widetilde L_{p-1}$. If we manage to control $\angle_{\max}(L_{p-1},\widetilde L_{p-1})$, then we would be able to substitute the $P_{\widetilde L_{p-1}}$ in \cref{eq:em:transform_process-2.5} by $P_{L_{p-1}}$, which is a quantity that only depends on $V_1,...,V_{p-1}$ as desired.
    
    To obtain this control on $\angle_{\max}(L_{p-1},\widetilde L_{p-1})$, first, by \cref{lem:cumulative_span_identity}, we have
    \[\widetilde L_{p-1}=span(Q_1+D_1,\dots,Q_{p-1}+D_{p-1}).\]
    Then, with $\delta_1<1/7$, we have
    $$(\sum_{j=1}^{p-1}\|D_j\|_{op}^2)^{1/2}\leq \frac{\delta_1}{2\sqrt{1-(36)^{-2}}}<\frac1{12}.$$
    Let $x\in \Rbb^{(p-1)\tilde d}$. For each $y=[Q_1,...,Q_{p-1}]x=\sum_{i=1}^{p-1}Q_i x_i\in L_{p-1}$, we have
    \[d(y,\widetilde L_{p-1})\leq \left\|y-\sum_{i=1}^{p-1}(Q_i+D_i)x_i\right\|=\left\|\sum_{i=1}^{p-1}D_i x_i\right\|< \frac1{12} \|x\|\leq \frac{1}{6}\|y\|\]
    The last inequality follows from $\lambda_{\min}([Q_1,...,Q_P])\geq 1/2$ by event $\Ecal$. In particular, $y$ can be any vector in $L_{p-1}$. And
    \[\sin\angle_{\max}(L_{p-1},\widetilde L_{p-1})= \sup_{\|y\|=1,y\in L_{p-1}}d(y,\widetilde L_{p-1})\leq \frac16.\]
    In summary, we have controlled $\sin\angle_{\max}(L_{p-1},\widetilde L_{p-1})$. We now go back to \eqref{eq:em:transform_process-2.5}. First, we immediately have
    \[\|P_{\widetilde L_{p-1}} P_{V_{p}}\|_{op}\leq \|P_{L_{p-1}} P_{V_{p}}\|_{op}+\frac16.\]
    On event $\Ecal$, we have $\|P_{L_{p-1}} P_{V_{p}}\|_{op}=\|P_{L_{p-1}} Q_p\|_{op}\leq 1/2$, which implies
    \[\|Q_p^\top B_{p-1} Q_p \|_{\min}\geq \frac59.\]
    Under event $\Ecal$, we have
    $$\|\widetilde{Q}_p^\top B_{p-1} \widetilde{Q}_p\|_{\min}\geq \frac59-2\|Q_p^\top B_{p-1} \Hcal_{p-1} ^{-1} D_p\|_{op}\geq \frac59-\|\Hcal_{p-1}\|_{\min}^{-1}\delta_{p}.$$

    Combining this with \cref{eq:lem:transform_process-1,eq:lem:transform_process-2}, we have
    \[\|B_{p}^{-1}\|_{op}\leq\|B_{p-1} ^{-1}\|_{op} + (\delta_{p}^{-2}-1)\frac{(1 + \delta_{p}\|\Hcal_{p-1}\|_{\min}^{-1}/2)^2}{5/9-\|\Hcal_{p-1}\|_{\min}^{-1}\delta_{p}}\]
    By the induction hypothesis, $\|\Hcal_{p-1}\|_{\min}\geq\delta_{p-1} /2$, hence
    \[\|\Hcal_{p}\|_{\min}^{-2} = \|B_{p}^{-1}\|_{op}\leq\frac{4}{\delta_{p-1} ^2} + \frac{(1 + \delta_{p}/\delta_{p-1})^2}{\delta_{p}^2 (5/9-2\delta_{p}/\delta_{p-1})}\leq \frac{4}{\delta_{p}^2}.\]
    The last inequality follows from $\delta_{p}\leq \delta_{p-1} / 36$. This completes the inductive proof for $\|\Hcal_p\|_{\min}\geq \delta_p /2$. 

    The above proof conditions on $\Ecal$, which only depends on $V_1,...,V_P$. Therefore, whenever event $\Ecal$ happens, the result holds simultaneously for all admissible $\delta_1,...,\delta_P$ and $U_1,...,U_P$. 
    
    It remains to control the failure probability. By basis invariance, we may take $Q_p=G_p(G_p^\top G_p)^{-1/2}$ for independent standard normal
    matrices $G_p\in\Rbb^{d\times \tilde{d}}$, and write $G=[G_1,\dots,G_P]$.
    Applying \cite[Corollary~5.35]{vershynin2012introduction}
    with $t=\sqrt d/12$ to $G$ and, conditionally on
    $L_{p-1}=span([G_1,\dots,G_{p-1}])$, to each $P_{L_{p-1}}G_{p}$ expressed in
    orthonormal coordinates on $L_{p-1}$, gives
    \begin{equation}\label{eq:lem:transform_process-3}
         \lambda_{\min}(G)\geq\frac34\sqrt d,
    \qquad
    \|G\|_{op}\leq\frac54\sqrt d,
    \qquad
    \|P_{L_{p-1}}G_{p}\|_{op}
    \leq\sqrt{(p-1)\tilde{d}}+\sqrt{\tilde{d}}+\frac{\sqrt d}{12}
    \quad\text{for every }p\in[P].
    \end{equation}
    with at least $1-2Pe^{-d/288}$ probability. On the event in \eqref{eq:lem:transform_process-3},
    \[
    \lambda_{\min}([Q_1,\dots,Q_P])
    \geq
    \frac{\lambda_{\min}(G)}{\max_p\|G_p\|_{op}}
    \geq
    \frac{\lambda_{\min}(G)}{\|G\|_{op}}
    \geq\frac35.
    \]
    Moreover, with $\lambda_{\min}(G_{p})\geq \lambda_{\min}(G)$
    and $36P\tilde{d}\leq d$, we obtain for every $p\in[P]$,
    \[
    \begin{aligned}
    \|P_{L_{p-1}}Q_{p}\|_{op}
    &\leq
    \frac{\|P_{L_{p-1}}G_{p}\|_{op}}{\lambda_{\min}(G_{p})}\\
    &\leq
    \frac{\sqrt{(p-1)\tilde{d}}+\sqrt{\tilde{d}}+\sqrt d/12}{3\sqrt d/4}
    \leq\frac{2\sqrt2+1}{9}<\frac12.
    \end{aligned}
    \]
    Thus $\Pbb(\Ecal^c)\leq2Pe^{-d/288}$.

\end{proof}
\comment{
\begin{corollary}\label{cor:the_uniform_singular_bound}
    Let $P\geq 2$ and $36P\tilde d\leq d$. With at least $1-2Pe^{-d/288}$ probability over $E_1,...,E_P$, simultaneously over all $p\in[P]$ and $v\in\Vcal_{p-1}$, we have 
    \[\lambda_{\max}(\Tcal_v),\text{ }\kappa(\Tcal_v)\leq \frac{2}{d\delta_p^{(0)}}.\]
    regardless of the randomness used in $\NoiseAddition$ for sampling phases in each layer.
\end{corollary}
}

We are now ready to prove \cref{cor:the_uniform_singular_bound}.
\begin{proof}[Proof of \cref{cor:the_uniform_singular_bound}]
    $\Tcal_\emptyset =I_d$, so the statement is true for $p=1$. In the skeleton tree construction of our oracle, each path $\ell=(i_1,...,i_P)$ from root to leaf can be viewed as a transform process. To be specific, let $v_p=(i_1,...,i_p)$ denote the layer-$p$ ancestor of $\ell$; then $(E_{v_0,1}^{(i_1-1)},E_{v_1,2}^{(i_2-1)},\dots,E_{v_{P-1},P}^{(i_P-1)})$ follows a transform process as $(U_1,...,U_P)$, together with $(V_1,\dots,V_P)=(E_1,\dots,E_P)$ and $(\delta_1,\dots,\delta_P)=(\delta_1^{(i_1-1)},\dots,\delta_P^{(i_P-1)})$. Indeed, one can verify that for each $p\in\set{2,\dots,P}$,
    $$\Hcal_{v_{p-1}}=T_{U_{p-1},\delta_{p-1}}\circ\cdots\circ T_{U_1,\delta_1},$$
    and that the $\NoiseAddition$ process we use to sample the phases with parameters $\delta_p^{(\leq N_p)}/\sqrt{3P}$ satisfies
    $$\angle_{\max}(E_{v_{p-1},p}^{(i_p-1)},\Hcal_{v_{p-1}}(E_p))\leq\frac{1}{\sqrt{3P}}(\delta_p^{(i_p-1)}+\delta_p^{(i_p)}+\cdots+\delta_p^{(N_p)})\leq \frac{\delta_p^{(i_p-1)}}{2}.$$ 
    The last inequality follows from a geometric sum with ratio $\lambda_p\leq d^{-p}$ and $P\geq 2$, $d\geq 72$. One can also verify the condition $$\delta_{p}^{(i_{p}-1)}\leq d^{-1}\epsilon^{\frac{p-1}{2P}}\leq\delta_{p-1}^{(N_{p-1})}\leq\delta_{p-1}^{(i_{p-1}-1)}/d\leq\delta_{p-1}^{(i_{p-1}-1)}/36.$$ 
    Applying \cref{lem:transform_process} then gives $$\lambda_{\min}(\Hcal_{v_{p-1}})\geq \delta_{p-1}^{(i_{p-1}-1)}/2\geq d\delta_{p}^{(0)}/2.$$ 
    Since we are using the same choice $(V_1,...,V_P)=(E_1,...,E_P)$ for all paths, which corresponds to exactly the same event of success, we then have simultaneously over $p\in[P]$ and $v\in\Vcal_{p-1}$,
    \[\lambda_{\max}(\Tcal_{v})=\lambda_{\min}(\Hcal_{v})^{-1}\leq \frac{2}{d\delta_p^{(0)}}.\]
    Since $\Hcal_v$ is always a contraction mapping, we have $\lambda_{\min}(\Tcal_v)\geq 1$, hence $\kappa(\Tcal_v)\leq \frac{2}{d\delta_p^{(0)}}$ as well.
\end{proof}

\subsection{Expected inverse threshold}

\begin{lemma}\label{lem:expected_inverse}
    Let $X\in [0,1]$ be a random variable with $\Ebb[X]\geq \alpha\in (0,1]$. Then there exists $x^* \in (0,1]$, such that $\Pbb(X\geq x^*)\geq \alpha/(3x^*\ln(3/\alpha))$ and $x^*\geq\alpha/(3\ln(3/\alpha))$.
\end{lemma}
\begin{proof}
    Let $x^*\in\arg\max_{x\in (0,1]} x\Pbb(X\geq x)$, $\Ecal:=\set{X\geq x^*}$ and put $r:=x^* \Pbb(X\geq x^*)$. Then,
    $$\Ebb[X^{-1}|\Ecal]\leq 1/x^* = \Pbb (\Ecal)/ r.$$
    Since $\Pbb(X\geq x)\leq \min\set{r/x,1}$, we have
    $$\alpha\leq\Ebb[X]=\int_{[0,1]} \Pbb(X\geq x)dx\leq r+r\ln (1/r).$$
    This implies $r\geq \alpha/(3\ln(3/\alpha))$. In turn, $\Pbb (\Ecal) = r/x^* \geq \alpha/(3x^*\ln(3/\alpha))$. The last claim follows from $\Pbb(\Ecal)\leq 1$.
\end{proof}

\section{Concentration inequalities}

In this section, we prove that appropriate transformations of uniform subspaces in Grassmannian balls in $\Gr(\tilde d,d)$ are perpendicular to $k$-dimensional subspaces with probability exponentially small in the entropy $k\tilde d$.

\LemUnifCaseTranslation*

\begin{proof}
By rotational invariance, we may assume without loss of generality that $E_0=\operatorname{Span}(e_1,\ldots,e_{\tilde d})$ is simply the span of the first $\tilde d$ standard basis vectors.
If \(k+\tilde d>d\), then \(W\cap F\ne\{0\}\), so \(\|P_WP_F\|_{\mathrm{op}}=1\) and the result is immediate. We therefore assume \(k\le d-\tilde d\). Up to taking $C_0\geq 10$, we may also assume without loss of generality that $\epsilon\rho< 1/8$.

First, any subspace $E\in B(E_0,\delta)$ can be written as
\[
E=\Span
\begin{pmatrix}
I_{\tilde d}\\
(\tan\delta)\,Z
\end{pmatrix},
\qquad
Z\in\mathbb R^{(d-\tilde d)\times\tilde d},
\quad \|Z\|_{\mathrm{op}}\le1.
\]
We first reduce to the case when $Z$ is sampled uniformly with respect to Lebesgue measure from the operator-norm ball as above. From \cite[Section~6.2]{zhang2009radon}, the density of $Z$ when $E\sim\mathrm{Unif}(B(E_0,\delta))$ is given by
\begin{equation*}
    p_Z(Z)\propto\det(I_{\tilde d}+\tan^2\delta\,Z^\top Z)^{-d/2}.
\end{equation*}
Note that
\[
0\le
\frac d2\ln\det(I_{\tilde d}+\tan^2\delta\,Z^\top Z)
\le\frac{d\tilde d}{2}\tan^2\delta
\le2,
\]
where in the last inequality we used $\delta\leq 1/d$.
Consequently, if \(Z_0\) is uniform with respect to Lebesgue measure on the operator-norm unit ball, then
\begin{equation}\label{eq:comparison_Z_Z0}
    p_Z\le e^2p_{Z_0}.
\end{equation}
We now analyze the desired event for the uniform $Z_0$ instead of $Z$.
Choose an orthonormal basis matrix for \(W\), written as
$
\begin{pmatrix}A\\D\end{pmatrix}$ where $A\in\mathbb R^{\tilde d\times k}$ and $D\in\mathbb R^{(d-\tilde d)\times k}$.
By definition of the transformation,
\[
F=\Span
\begin{pmatrix}
\rho\delta I_{\tilde d}\\
\tan\delta\,Z
\end{pmatrix}
=
\Span
\begin{pmatrix}
I_{\tilde d}\\
\frac1\rho\frac{\tan\delta}{\delta}\,Z
\end{pmatrix}.
\]
Since \(1\le\tan\delta/\delta\le2\) and \(\rho\ge1\), the event
\(\|P_WP_F\|_{\mathrm{op}}\le\epsilon\) implies
\begin{equation}\label{eq:op_norm_bound}
\left\|A^\top+\frac{\tan\delta}{\rho\delta}D^\top Z\right\|_{\mathrm{op}}
\le
\epsilon\sqrt{1+\left(\frac{\tan\delta}{\rho\delta}\right)^2}
\le\sqrt5\,\epsilon.
\end{equation}
If this event is empty, the desired bound is immediate. Otherwise, for every unit vector \(x\in\mathbb R^k\), \eqref{eq:op_norm_bound} gives
\[
\sqrt{1-\|Dx\|^2}
=\|Ax\|
\le
\frac{\tan\delta}{\rho\delta}\|Dx\|
+\sqrt5\,\epsilon
\le2\|Dx\|+\sqrt5\,\epsilon.
\]
If \(\|Dx\|<1/4\), the left-hand side is at least \(\sqrt{15}/4\), whereas the right-hand side is less than \(1/2+\sqrt5/8\), since \(\epsilon\le\rho\epsilon<1/8\). This contradiction proves
\[
\lambda_{\min}(D)\ge\frac14.
\]
Take a thin QR decomposition \(D=VR\). The preceding bound gives
\(\|(R^\top)^{-1}\|_{\mathrm{op}}\le4\). Therefore, the first inequality in \eqref{eq:op_norm_bound} implies
\[
\begin{aligned}
\left\|
V^\top Z+
\frac{\rho\delta}{\tan\delta}(R^\top)^{-1}A^\top
\right\|_{\mathrm{op}}
&\le
\|(R^\top)^{-1}\|_{\mathrm{op}}\epsilon
\sqrt{1+\left(\frac{\rho\delta}{\tan\delta}\right)^2}\\
&\le4\epsilon\sqrt{1+\rho^2}
\le6\rho\epsilon.
\end{aligned}
\]

It remains to bound the probability of this translated matrix-ball event for \(Z_0\). By orthogonal invariance, \(V^\top Z_0\) has the same distribution as the first \(k\) rows \(Y\) of \(Z_0\). We recall that $Z_0$ is uniform on the unit operator-norm ball.

By the density formula for Haar orthogonal matrix blocks \cite[Lemma 2.5]{jiang2006entries}, the upper-left \((d-\tilde d)\times\tilde d\) block of a Haar-distributed matrix in \(O(d+1)\) is uniformly distributed on the operator-norm unit ball. Thus \(Y\) has the same law as its upper-left \(k\times\tilde d\) block. Applying the same formula then gives
\begin{equation}\label{eq:density_bound}
p_Y(Y)
=
\frac{1}{\pi^{k\tilde d/2}}
\prod_{j=1}^{k}
\frac{\Gamma\!\left(\frac{d+2-j}{2}\right)}
     {\Gamma\!\left(\frac{d-\tilde d+2-j}{2}\right)}
\det(I_{\tilde d}-Y^\top Y)^{(d-\tilde d-k)/2}
\mathbf1_{\{\|Y\|_{\mathrm{op}}<1\}} \leq (C\sqrt d)^{k\tilde d},
\end{equation}
for some universal constant $C>0$. Here we used the fact that $k\leq d-\tilde d$, hence the determinant factor is at most \(1\).
Finally, a translated operator-norm ball of radius \(6\rho\epsilon\) in \(\mathbb R^{k\times\tilde d}\) is contained in a product of \(k\) Euclidean \(\tilde d\)-balls of that radius. Its volume is therefore at most
\[
\left(\frac{C\rho\epsilon}{\sqrt{\tilde d}}\right)^{k\tilde d}.
\]
Combining this volume bound with the density bound \eqref{eq:density_bound}, and the comparison between \(Z\) and \(Z_0\) from \eqref{eq:comparison_Z_Z0}, we obtain
\[
\mathbb P\!\left(\|P_WP_F\|_{\mathrm{op}}\le\epsilon\right)
\le
e^2(C\sqrt d)^{k\tilde d}
\left(\frac{C\rho\epsilon}{\sqrt{\tilde d}}\right)^{k\tilde d}
\le
\left(C_0\rho\epsilon\sqrt{\frac d{\tilde d}}\right)^{k\tilde d}.
\]
This proves the lemma.
\end{proof}

\clearpage
\bibliographystyle{alpha}
\bibliography{refs}

\end{document}

%% file: shortcuts.tex
\newcommand{\Acal}{\mathcal{A}}

\newcommand{\Dcal}{\mathcal{D}}
\newcommand{\Ecal}{\mathcal{E}}
\newcommand{\Fcal}{\mathcal{F}}
\newcommand{\Gcal}{\mathcal{G}}
\newcommand{\Hcal}{\mathcal{H}}

\newcommand{\Ncal}{\mathcal{N}}
\newcommand{\Pcal}{\mathcal{P}}
\newcommand{\Scal}{\mathcal{S}}
\newcommand{\Tcal}{\mathcal{T}}
\newcommand{\Ucal}{\mathcal{U}}
\newcommand{\Vcal}{\mathcal{V}}

\newcommand{\Ocal}{\mathcal{O}}

\newcommand{\Rcal}{\mathcal{R}}

\newcommand{\Ebb}{\mathbb{E}}
\newcommand{\Nbb}{\mathbb{N}}
\newcommand{\Pbb}{\mathbb{P}}

\newcommand{\Rbb}{\mathbb{R}}
\newcommand{\Sbb}{\mathbb{S}}

\newcommand{\1}{\mathbbm{1}}%

\definecolor{dark_red}{rgb}{0.2,0,0}

\newcommand{\Unif}{\text{Unif}}

\newcommand{\paren}[1]{\left( #1 \right)}
\newcommand{\sqb}[1]{\left[ #1 \right]}
\newcommand{\set}[1]{\left\{ #1 \right\}}
\newcommand{\floor}[1]{\left\lfloor #1 \right\rfloor}
\newcommand{\ceil}[1]{\left\lceil #1 \right\rceil}
\newcommand{\abs}[1]{\left|#1\right|}

\newcommand{\poly}{\textsf{poly}}
\newcommand{\polylog}{\textsf{polylog}}
\newcommand{\NoiseAddition}{\mathrm{NoiseAddition}}
\newcommand{\Succ}{\mathrm{Succ}}
\newcommand{\Gr}{\mathrm{Gr}}
\newcommand{\Block}{\mathrm{Block}}
\newcommand{\Span}{\mathrm{Span}}
\newcommand{\length}{\mathrm{length}}

%% file: transform_necessity.tex
\section{Failure of the naive construction}\label{app:transform_necessity}

We follow the same notations in \cref{sec:technical_overview}. Informally, we will provide some intuitions on this strategy in $\Rbb^3$, using Figure~\ref{fig:naive-construction-geometry}. The left-hand diagram magnifies a small region of the unit sphere in $\Rbb^3$, shown on the right; the gray cone connects the two views. For simplicity, we ignore the constraint $\bm e^\top x\leq -1/2$ as we are working with vectors with norms close to one.
\input{geometric-configuration}

In the 2-layer oracle with $d=3$ and $\tilde d=1$, suppose we just finished phase $i_1$ at layer $1$ and managed to pass the check of every accuracy version of $E_2$ in the blocking oracle--- Essentially, we have observed $E_2$ itself, and then we store a vector $x_*$ that belongs to $E_2^\perp$ up to high accuracy.

Next, the first layer proceeds to phase $i_1+1$, and we forget everything from the end of phase $i_1$, depicted as dashed objects in Figure~\ref{fig:naive-construction-geometry}, except for the point $x_*$ that we chose to keep in memory. And again, we interact with layer 2 phases in the blocking oracle. Suppose that we just observed $E_2^{(i_2)}$. Then, the geometric constraint for the blocking oracle ensures that $B_d(0,1)\cap E_2^\perp\subseteq \set{\|P_{E_2^{(i_2)}}(x)\|\leq \delta^{(i_2)}}$, and hence we know the true $E_2^\perp$ must be within $\alpha=O(\delta^{(i_2)})$ angular distance from $(E_2^{(i_2)})^\perp$, and must pass through $x_*$. In other words, we know that $E_2^\perp$ must lie between the two black lines in Figure~\ref{fig:naive-construction-geometry}.

We now consider the intersection of $E_2^\perp$ and $(E_1^{(i_1+1)})^\perp$. With access to $E_1^{(i_1+1)}$ at the current first-layer-phase, we know that $E_2^\perp\cap (E_1^{(i_1+1)})^\perp$ must be somewhere within the blue segment of length $\delta_*$ in Figure~\ref{fig:naive-construction-geometry}. 
And given both $x_*$ and $B_d(0,1)\cap (E_1^{(i_1+1)})^\perp$ belong to $\set{\|P_{E_1^{(i_1)}}(x)\|\leq \delta^{(i_1)}}$, we have the distance between $x_*$ and $E_2^\perp\cap (E_1^{(i_1+1)})^\perp$ being $O(\delta^{(i_1)})$.\footnote{Technically, this argument only holds when $E_2$ and $E_1$ are in general positions, meaning the minimum principal angle between $E_1$ and $E_2$ is, say, at least $d^{-10}$ --- when $E_1$ and $E_2$ are both Haar random, this is indeed the case, and geometrically, it is also better to have different layers being relatively far from each other as otherwise they may get jointly learned.}
Combining this with the angle $\alpha=O(\delta^{(i_2)})$ then gives, roughly speaking, $\delta^*=O(\delta^{(i_1)}\delta^{(i_2)})=O(\delta^{(i_1+i_2)})$. As a result, any query on the blue segment would be at most $O(\delta^{(i_1+i_2)})$ distance from $E_2^\perp$. This allows us to directly jump from phase $i_2$ to, say, phase $i_1+i_2-10$ within one step, skipping all phases in between. We can repeat this strategy for every layer 1 phase and succeed with essentially
$10N+\sum_{i_1=11}^N N/i_1 = O(N\ln N)\ll N^2$ query complexity.

Finally, we note that this kind of strategy can be extended to more than two layers with a slight twist. In the two-layer strategy above, we store a vector $x^*\in( E_1^{(i_1)})^\perp\cap E_2^\perp$ so that we can skip layer-2 phases quickly. However, given $P$ is potentially large, we cannot afford to keep a vector in memory for each of the $P-1$ subsequent layers. The idea here is to share the same vector across layers. For instance, if there are $3$ layers, then we wait until both layer 2 and 3 are at the highest precision -- Using the 2-layer argument, this takes only $O(N\ln N)$ time. Once we are in this stage, we will be able to memorize a vector $x_*$ that belongs to, essentially, $(E_1^{(i_1)})^\perp\cap E_2^\perp\cap E_3^\perp$. This enables us to skip both layers in the following way: Each time we enter phase $(i_1,i_2,N)$, which corresponds to the third layer being at highest precision and $E_3$ being observed, we query a point in $E_3^\perp\cap (E_1^{(i_1)})^\perp\cap (x_* +(E_2^{(i_2)})^\perp)$. Following the same argument as the 2-layer case, we can see that this query skips $\Omega(i_1)$ layer-2 phases while not getting blocked by layer-3 phases at all. Next, this query is blocked by some layer-2 phase, it will then take only $\ln N$ queries to pass all layer-3 phases using the same argument for two layers. Now that we are restored to some phase of the form $(i_1,i_2',N)$, we can repeat the above argument until the second layer also reaches its highest precision.
This way, it takes only $(\ln N)^2$ queries to pass each layer-1 phase, which gives a total query complexity of $O(N(\ln N)^2)$. By extending the strategy we will eventually get $O(N(\ln N)^{P-1})$ for $P$ layers.

%% file: geometric-configuration.tex
\begin{figure}[htbp]
\centering
\begingroup
\definecolor{figpink}{HTML}{C34870}
\definecolor{figteal}{HTML}{229E95}
\resizebox{0.98\linewidth}{!}{%
\providecommand{\TealDirectionLabel}
  {\bigl(E_1^{(i_1+1)}\bigr)^\perp}
\providecommand{\PinkStripLabel}
  {\bigl\{\|P_{E_2^{(i_2)}}(x)\|\leq\delta^{(i_2)}\bigr\}}
\providecommand{\TealStripLabel}
  {\bigl\{\|P_{E_1^{(i_1)}}(x)\|\leq\delta^{(i_1)}\bigr\}}

\begin{tikzpicture}[baseline=(O),
  x=1cm,y=1cm,
  font=\Large,
  line cap=round,line join=round,
  boundary/.style={line width=1.1pt},
  guide/.style={line width=1.05pt,dash pattern=on 5pt off 5pt},
  blackline/.style={black,line width=1.35pt},
  dimension/.style={
    line width=1pt,
    {Latex[length=2.6mm,width=1.8mm]}-{Latex[length=2.6mm,width=1.8mm]}
  }
]
  \def\PinkAngle{17}
  \def\PinkGuideAngle{22}
  \def\LowerAngle{6}
  \def\UpperAngle{33}
  \def\TealAngle{-76}
  \def\PinkHalfWidth{1.53}
  \def\BlackBoundaryGap{0.10}
  \def\PinkExtent{5.15}
  \pgfmathsetmacro{\LowerLength}
    {min((\PinkHalfWidth-\BlackBoundaryGap)/abs(sin(\LowerAngle-\PinkAngle)),
         \PinkExtent/cos(\LowerAngle-\PinkAngle))}
  \pgfmathsetmacro{\UpperLength}
    {min((\PinkHalfWidth-\BlackBoundaryGap)/abs(sin(\UpperAngle-\PinkAngle)),
         \PinkExtent/cos(\UpperAngle-\PinkAngle))}
  \pgfmathsetmacro{\PinkGuideLength}
    {(\PinkExtent-0.10)/cos(\PinkGuideAngle-\PinkAngle)}

  \coordinate (O) at (0,0);

  \begin{scope}[rotate=\PinkAngle]
    \draw[figpink,boundary]
      (-\PinkExtent,\PinkHalfWidth)--(\PinkExtent,\PinkHalfWidth);
    \draw[figpink,boundary]
      (-\PinkExtent,-\PinkHalfWidth)--(\PinkExtent,-\PinkHalfWidth);
    \draw[figpink,dimension]
      (-\PinkExtent,-\PinkHalfWidth)--(-\PinkExtent,\PinkHalfWidth);
    \coordinate (PinkWidthMid) at (-\PinkExtent,0);
  \end{scope}
  \coordinate (PinkGuideEnd) at (\PinkGuideAngle:\PinkGuideLength);
  \draw[figpink,guide]
    ($(O)+(\PinkGuideAngle:-\PinkGuideLength)$)--(PinkGuideEnd);
  \node[figpink,anchor=west] at ($(PinkGuideEnd)+(0.16,0)$)
    {$E_2^\perp$};
  \node[figpink,anchor=east,inner sep=0pt] (PinkStripText)
    at ($(PinkWidthMid)+(2.4,-3)$) {$\PinkStripLabel$};
  \draw[figpink,line width=0.8pt,-{Latex[length=2mm,width=1.4mm]}]
    ($(PinkStripText.north)+(0,0.12)$)
    to[out=90,in=210] (PinkWidthMid);

  \draw[figteal,guide] (-2.45,3.15)--(-2.45,-5.35);
  \draw[figteal,guide] (0.45,3.15)--(0.45,-5.35);
  \coordinate (T) at (-1.30,{-1.30*tan(-3)});
  \path[name path=teal-axis]
    ($(T)+(\TealAngle:-3.35)$)--
    ($(T)+(\TealAngle:5.55)$);
  \draw[figteal!70,boundary]
    ($(T)+(\TealAngle:-3.35)$)--
    ($(T)+(\TealAngle:5.55)$);
  \node[figteal,anchor=south] at (-1.95,3.42)
    {$\TealDirectionLabel$};
  \draw[figteal,dimension] (-2.45,-5.55)--(0.45,-5.55);
  \node[figteal,anchor=north,inner sep=0pt] at (-1.0,-5.83)
    {$\TealStripLabel$};

  \draw[blackline,name path=lower-line]
    (\LowerAngle:-\LowerLength)--(\LowerAngle:\LowerLength);
  \draw[blackline,name path=upper-line]
    (\UpperAngle:-\UpperLength)--(\UpperAngle:\UpperLength);

  \draw[blackline]
    (\LowerAngle:1.8)
    arc[start angle=\LowerAngle,end angle=\UpperAngle,radius=1.8cm];
  \node at ({(\LowerAngle+\PinkGuideAngle)/2}:2.30) {$\alpha$};

  \path[name intersections={of=teal-axis and lower-line,by=A}];
  \path[name intersections={of=teal-axis and upper-line,by=B}];
  \draw[blue!85!black,dimension,
    {Latex[length=2mm,width=1.5mm]}-{Latex[length=2mm,width=1.5mm]}]
    (A)--(B)
    node[midway,left=12pt,yshift=-2pt] {$\delta_*$};
  \foreach \pt in {A,B}{
    \draw[blue!85!black,line width=1.15pt]
      ($(\pt)+(14:0.19)$)--($(\pt)+(14:-0.19)$);
  }

  \filldraw[draw=red!85!black,fill=red!85!black,line width=0.6pt]
    (O) circle[radius=2.5pt];
  \node[red!85!black,anchor=north,inner sep=1pt] at (0,-0.23)
    {$x_*$};

  \coordinate (ConeApex) at (12.99,-4.11);
  \begin{scope}[shift={(12.25,-3.85)},scale=0.4,
    solid curve/.style={line width=0.8pt},
    dashed curve/.style={line width=0.65pt,dash pattern=on 2.8pt off 2.4pt}]
    \definecolor{curvered}{RGB}{211,91,111}
    \definecolor{curvegreen}{RGB}{47,155,140}
    \def\R{5}
    \coordinate (SphereP) at (1.85,-0.65);
    \coordinate (SphereO) at (0.75,0.40);
    \fill[white] (0,0) circle (\R);
    \begin{scope}
      \clip (0,0) circle (\R);
      \draw[dashed curve,curvegreen]
        (109:\R)
        .. controls (-0.70,3.15) and (0.30,1.45) .. (0.57,-0.45)
        .. controls (0.89,-2.10) and (0.65,-3.75) .. (272:\R);
      \draw[dashed curve,curvegreen]
        (91:\R)
        .. controls (1.10,3.25) and (2.05,1.40) .. (2.17,-0.55)
        .. controls (2.28,-2.15) and (2.54,-3.55) .. (294:\R);
      \draw[solid curve,curvered]
        (176:\R)
        .. controls (-4.10,-1.25) and (1.00,-0.80) .. (18:\R);
      \draw[solid curve,curvered]
        (198:\R)
        .. controls (-3.65,-3.20) and (1.15,-1.65) .. (354:\R);
      \draw[solid curve]
        (178:\R)
        .. controls (-4.25,-1.00) and (-0.30,-1.10) .. (SphereP)
        .. controls (2.60,-0.52) and (3.90,-0.42) .. (355.8:\R);
      \draw[solid curve]
        (195:\R)
        .. controls (-3.10,-2.45) and (-0.30,-1.60) .. (SphereP)
        .. controls (2.60,-0.318) and (3.88,0.68) .. (15.8:\R);
      \draw[solid curve,curvegreen]
        (108:\R)
        .. controls (-0.941043,4.951548) and (0.480504,2.987149) .. (1.195104,0.320225)
        .. controls (1.909704,-2.346699) and (2.306925,-4.481237) .. (290:\R);
    \end{scope}
    \draw[dashed curve,black] (SphereO)--(SphereP);
    \draw[line width=0.8pt] (0,0) circle (\R);
    \fill (SphereO) circle (2.1pt);
    \node[font=\large] at (-0.20,0.65) {$O$};
  \end{scope}

  \coordinate (ConeUpper) at (3.80,1.45);
  \coordinate (ConeLower) at (2.95,-0.90);
  \path[fill=gray,fill opacity=0.18,draw=none]
    (ConeApex)--(ConeUpper)
    .. controls (3.20,0.95) and (2.85,-0.20) .. (ConeLower)
    --cycle;
  \draw[gray!65,line width=0.55pt]
    (ConeApex) ellipse[x radius=0.22cm,y radius=0.14cm];
\end{tikzpicture}
}

\endgroup
\caption{Failure of the naive two-layer construction. The diagram on the left magnifies a small region of the unit sphere shown on the right.}
\label{fig:naive-construction-geometry}
\label{fig:naive-left}
\label{fig:naive-right}
\end{figure}
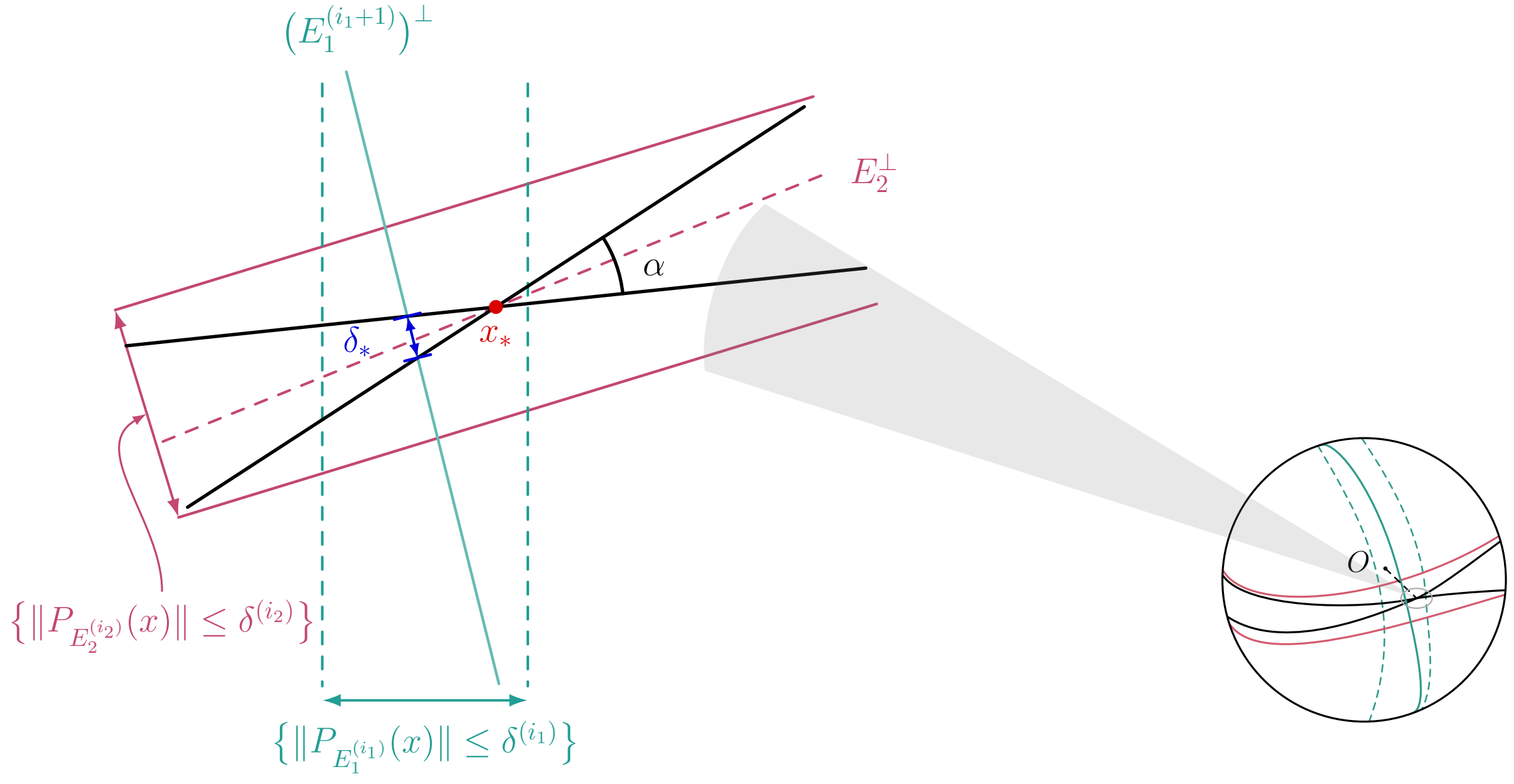